# Physics-informed and event-aware forecasting of ammonia emissions in aerobic chicken manure composting bioreactors: A PINN-PatchTST-GPR framework with symbolic dynamic extraction

**Lanbin Wei,Haiyong Weng**

College of Mechanical and Electrical Engineering, Fujian Agriculture and Forestry University, Fuzhou 350002, PR China

Full postal address: College of Mechanical and Electrical Engineering, Fujian Agriculture and Forestry University, No. 63 Xiyuangong Road, Shangjie Town, Minhou County, Fuzhou 350002, Fujian Province, PR China.

## Abstract

Predicting ammonia (NH3) concentrations during aerobic chicken manure composting provides critical feedforward signals for intelligent aeration. However, the concurrence of long-term biochemical evolution and intermittent forced-aeration pulses poses significant modeling hurdles across steady baselines, transient surges, and risk quantification. Existing data-driven models lack physical consistency, while mechanistic kinetic models struggle with environmental disturbances. To bridge this gap, we propose PINN-PatchTST-GPR, a physics-informed neuro-symbolic forecasting framework. Validated on a 28-day composting campaign across six vertical bioreactors, the framework couples a channel-independent PatchTST backbone with macroscopic kinetic constraints, dynamic physical bounds, and a Gaussian process regression (GPR) event specialist. Multi-channel sensor streams drive discrete differential equation residuals and non-negativity bounds, strictly ensuring physical plausibility. A block-split conformal prediction module yields nominal 90% distribution-free prediction intervals, while PySINDy and PySR extract low-complexity symbolic equations for quiescent and surge regimes. The model achieves an overall RMSE of 0.8256 +/- 0.0132 ppm, R2 of 0.9817 +/- 0.0006, Epeak of 1.6793 +/- 0.0725 ppm, and Etime of 0.7 +/- 0.2 min. For uncertainty quantification, it attains PICP90 of 99.792 +/- 0.047%, MPIW of 6.951 +/- 0.128 ppm, and IS0.1 of 6.968 +/- 0.128 ppm. Symbolically, PySINDy yields an RMSE of 0.4624 +/- 0.0758 ppm (R2 = 0.9896), while PySR yields 2.4084 +/- 0.0864 ppm (R2 = 0.9523). Biologically, the 20-min aeration strategy minimizes cumulative NH3 emissions.

Overall, PINN-PatchTST-GPR harmonizes global forecasting, localized transient responsiveness, and risk quantification for precision composting.

## 1. Introduction

### 1.1. Background and motivation

Global intensive livestock and poultry husbandry supplies vital dietary protein for human nutrition, yet concurrently generates billions of metric tons of nitrogen-rich organic waste annually. Among these streams, poultry manure exhibits the most pronounced enrichment of nitrogenous organic matter. Its unmanaged dissipation into the atmosphere and aquatic ecosystems severely exacerbates global carbon-nitrogen imbalances and constitutes a major impediment to mitigating greenhouse gas and reactive nitrogen emissions in the agricultural sector (FAO, 2023; Liu et al., 2022). Conventional open-air composting and end-of-pipe waste management practices suffer from extensive nutrient leaching, severe odor emissions, and biosecurity hazards arising from the dissemination of enteric pathogens such as Escherichia coli and parasitic eggs (Galgani et al., 2014; Zhou et al., 2022), thereby failing to meet modern circular economy imperatives. Sealed industrial aerobic composting bioreactors establish controlled thermophilic regimes that accelerate organic matter mineralization and humification while thoroughly deactivating pathogenic microorganisms, representing a pivotal engineered approach for biotransforming livestock wastes into high-value organic fertilizers (Ding, 2022; Girón-Rojas et al., 2020).

Nevertheless, an industrial aerobic composting bioreactor is a multiphase, highly non-equilibrium system comprising microbial consortia, solid organic matrices, pore water, and headspace gases, wherein dynamic physicochemical interactions exhibit pronounced hysteresis in response to aeration. At the microscopic reaction scale, moisture content, the carbon-to-nitrogen (C/N) ratio, and core thermal gradients are intimately coupled, jointly governing the metabolic turnover rates and community abundances of ammonifying, nitrifying, and denitrifying microorganisms (Ma et al., 2021; Zhou et al., 2021). When the composting matrix enters the thermophilic phase (>55 °C) and is subjected to intermittent forced aeration pulses, the chemical equilibrium between ammonium ions and aqueous ammonia shifts drastically toward volatile free ammonia ($NH_3$). This gaseous ammonia is rapidly swept into the reactor headspace, causing concentration spikes and precipitating severe episodic emissions that can dissipate 30% to 60% of initial total nitrogen (Kupper et al., 2020; Sayara and Sánchez, 2021). Such massive nitrogen losses diminish the agronomic value of the finished compost while releasing malodorous compounds and key atmospheric precursors of secondary fine particulate matter (PM2.5).

Current engineering practice relies heavily on empirical lookup tables and retrospective laboratory assays to adjust operational parameters, resulting in an absence of proactive forecasting and real-time closed-loop control over reactor gas dynamics (Bari et al., 2000; Bernal et al., 2009). Although the advent of Internet of Things (IoT) multi-sensor telemetry provides continuous high-frequency data streams, conventional data-driven deep learning models operate as pure black boxes. When confronted with abrupt forced-aeration shocks, unconstrained architectures frequently output physically impossible negative concentrations or explosive distortions that violate conservation of mass, while failing to provide interpretable attribution for operational optimization (Karniadakis et al., 2021; Nie et al., 2023). Consequently, synergistically unifying biochemical ordinary differential equations (ODEs) with advanced deep sequence representation architectures to develop physics-informed, explainable artificial intelligence (XAI) models—capable of online multi-step forecasting and reverse symbolic extraction—has emerged as an urgent imperative and a frontier challenge for achieving emission abatement and autonomous process regulation in organic waste bioreactors.

### 1.2. Prior modeling approaches and limitations

To predict and regulate gaseous ammonia dynamics during aerobic bioreactor composting, existing investigations have predominantly pursued two distinct methodological routes. Although each paradigm possesses specific merits, both exhibit critical limitations regarding operational adaptability, predictive fidelity, and mechanistic transparency, rendering neither individually sufficient for precision industrial automation:

The first category comprises white-box mechanistic models centered on Monod-type microbial kinetics, Arrhenius thermal reaction dependencies, and multiphase mass balance formulations (Mason, 2006; Sole-Mauri et al., 2007). Founded upon thermodynamic laws and stoichiometric conservation, these models offer inherent interpretability. However, industrial composting bioreactors exhibit substantial spatial heterogeneity in substrate composition, temperature gradients, and moisture content, causing continuous variations in microbial community viability and gas-liquid-solid mass transfer coefficients (Peng et al., 2023). Intermittent forced-aeration episodes abruptly perturb gas exchange equilibria, rendering static kinetic parameters inadequate for tracking rapid ammonia emission spikes and their associated response latencies. Furthermore, integrating multi-rate kinetics across coupled stiff differential equations frequently induces numerical divergence and prohibitive computational overhead, severely curtailing real-time embedded implementation (Bernal et al., 2009).

The second category encompasses purely data-driven black-box models, including recurrent neural networks (RNN/GRU), temporal convolutional networks (TCN), and Transformer architectures (Nie et al., 2023; Raissi et al., 2019). Leveraging continuous multi-sensor telemetry, these neural frameworks excel at capturing long-term temporal dependencies and complex nonlinear interactions. Nevertheless, purely statistical learners lack inductive physical biases and conservation constraints (Brunton et al., 2016). When subjected to abrupt aeration disturbances, black-box architectures frequently yield physically impermissible predictions, such as negative gas concentrations (physical violation rate, PVR > 0) or unconstrained runaway spikes that breach mass balance bounds (Karniadakis et al., 2021). Furthermore, their high-dimensional parameter spaces remain fundamentally opaque, precluding transparent biochemical attribution and prohibiting lightweight deployment onto resource-constrained microcontrollers for autonomous edge control (Cuomo et al., 2022).

This tension—between white-box models that offer rigorous physics but suffer from parametric rigidity and numerical instability, and black-box models that exhibit flexible expressiveness but risk physical violations and opacity—constitutes the central methodological gap at the intersection of bioprocess engineering and time-series machine learning (Haug, 1993). Establishing a gray-box neural-symbolic paradigm that embeds kinetic ODEs and dynamic mass-balance envelopes directly into deep sequence learners, while enabling the post-hoc distillation of compact analytical equations, represents a compelling pathway to dismantle this binary dichotomy and achieve reliable, physically consistent prediction in solid-waste bioreactors.

### 1.3. Proposed gray-box modeling framework

Here, we propose a gray-box PINN-PatchTST-GPR framework that unifies physics-informed regularization, long-range sequence modeling, and event-triggered correction to forecast headspace $NH_3$ concentrations during aerobic chicken manure composting, augmented by symbolic regression to extract interpretable algebraic governing laws. Operating on lookback windows of multivariate telemetry, a channel-independent Patch Time Series Transformer (PatchTST) captures multi-scale temporal dependencies across individual sensor channels without cross-variable noise contamination (Nie et al., 2023). The primary backbone forecasts stationary baseline $NH_3$ volatilization, while a decoupled event specialist branch is selectively engaged during aeration episodes to adjust the rapid transient emission bursts.

During training, the optimization objective minimizes empirical sensor discrepancies while simultaneously penalizing violations of two physical constraint categories: (i) finite-difference ODE residuals derived from headspace mass balance to ensure trajectory compliance with biological generation kinetics, and (ii) soft barrier penalty functions enforcing non-negativity and dynamic thermodynamic saturation upper bounds to prevent non-physical drift. Under the gating of a deterministic aeration schedule, a Gaussian Process Regression (GPR) event specialist models the log-ratio residuals between empirical multiplier baselines ($R_{emp}$) and true pulse responses, yielding a refined multiplicative factor ($R_{pred}$) that scales event forecasts. To provide statistically rigorous risk bounds, a Block-Aware Split Conformal Prediction (BASCP) protocol calibrates temporal block residuals to generate distribution-free 90% prediction intervals that account for residual autocorrelation (Angelopoulos and Bates, 2023). Finally, post-hoc symbolic distillation via PySINDy and PySR compresses the stationary and aeration dynamics into compact, candidate algebraic difference equations (Brunton et al., 2016; Cranmer et al., 2020). Consequently, our framework synergistically delivers multi-step point predictions, pulse-adapted event corrections, calibrated uncertainty bands, and transparent symbolic representations.

### 1.4. Contributions of this work

The primary contributions of this work are organized across four dimensions: predictive architecture, uncertainty quantification, neural-symbolic distillation, and emission mitigation agronomy:

(1) Physics-informed gray-box architecture: We establish PINN-PatchTST-GPR, integrating channel-independent Transformer representations, ODE mass-balance kinetics, dynamic boundary barriers, and event-gated GPR specialists under a rigorous frozen benchmark protocol. Across six bioreactors, the proposed framework achieved an overall RMSE of 0.8256 ± 0.0132 ppm, an $R^2$ of 0.9817 ± 0.0006, and an 82.85% improvement in skill score relative to persistence, with zero physical boundary violations (Raw PVR = 0.00% and Total PVR = 0.00%). During forced-aeration episodes, peak tracking errors were reduced to $E_{\text{peak}}$ = 1.6793 ± 0.0725 ppm and $E_{\text{time}}$ = 0.7 ± 0.2 min, while cross-reactor generalization yielded a Lin's concordance correlation coefficient (CCC) of 0.9886 (95% CI: 0.9874–0.9897).

(2) Distribution-free uncertainty quantification: We formulate a Block-Aware Split Conformal Prediction (BASCP) strategy tailored to temporally autocorrelated agricultural telemetry, delivering valid 90% prediction intervals with an empirical coverage ($\text{PICP}_{90}$) of 99.792 ± 0.047%, a mean width (MPIW) of 6.951 ± 0.128 ppm, and an interval score ($\text{IS}_{0.1}$) of 6.968 ± 0.128 ppm.

(3) Dual neural-symbolic distillation: We distill complex deep network behaviors into interpretable, candidate algebraic difference equations using PySINDy for baseline volatilization (RMSE = 0.4624 ± 0.0758 ppm, $R^2$ = 0.9896 ± 0.0035) and PySR for aeration pulse responses (RMSE = 2.4084 ± 0.0864 ppm, $R^2$ = 0.9523 ± 0.0044; $E_{\text{peak}}$ = 1.5554 ± 0.2592 ppm, $E_{\text{time}}$ = 2.11 ± 1.17 min), enabling ultra-lightweight microcontroller execution.

(4) Agronomic emission mitigation: Empirical evaluation across aeration treatments demonstrated that a 20-min aeration schedule curtailed total $NH_3$ volatilization relative to 10-min and 30-min regimes while maintaining sufficient microbial respiration and compliant compost maturity, providing actionable guidance for low-emission composting operations.

## 2. Materials and methods

### 2.1. Experimental setup

#### 2.1.1. Bioreactor operation

The experimental data collection apparatus comprised an array of six sealed 20-L stainless-steel bioreactors (inner diameter 260 mm, height 380 mm), each loaded with an initial substrate volume of 16 L (moist substrate mass $M_{mat}$ = 10.5 kg). The reactors were partitioned into three aeration treatment groups with duplicate parallel runs (n = 2): Group 1 (Tanks 1–2), Group 2 (Tanks 3–4), and Group 3 (Tanks 5–6), subjected to 10-, 20-, and 30-min forced aeration durations, respectively, administered twice daily at 10:00 and 16:00. The supervisory control architecture integrated a custom-engineered master control cabinet, six wireless node acquisition enclosures, and six co-located multi-sensor arrays monitoring core matrix temperature, relative humidity, headspace carbon dioxide ($CO_2$), and gaseous ammonia ($NH_3$). Airflow was injected through the reactor base plenum at 12.0 $L \cdot min^{-1}$, traversing upward through the porous composting matrix. Effluent headspace gas discharged through a top vent, passed through a dedicated conditioning cartridge packed with 4–6 mm desiccant silica gel and molecular sieves to scrub moisture, and was conveyed by an intermittent suction sampling pump (EDLP600-D24B, 0.5 $L \cdot min^{-1}$ on a 10-min duty cycle) into an infrared $CO_2$ sensing chamber. The schematic configuration of the bioreactor system is illustrated in Fig. 1.

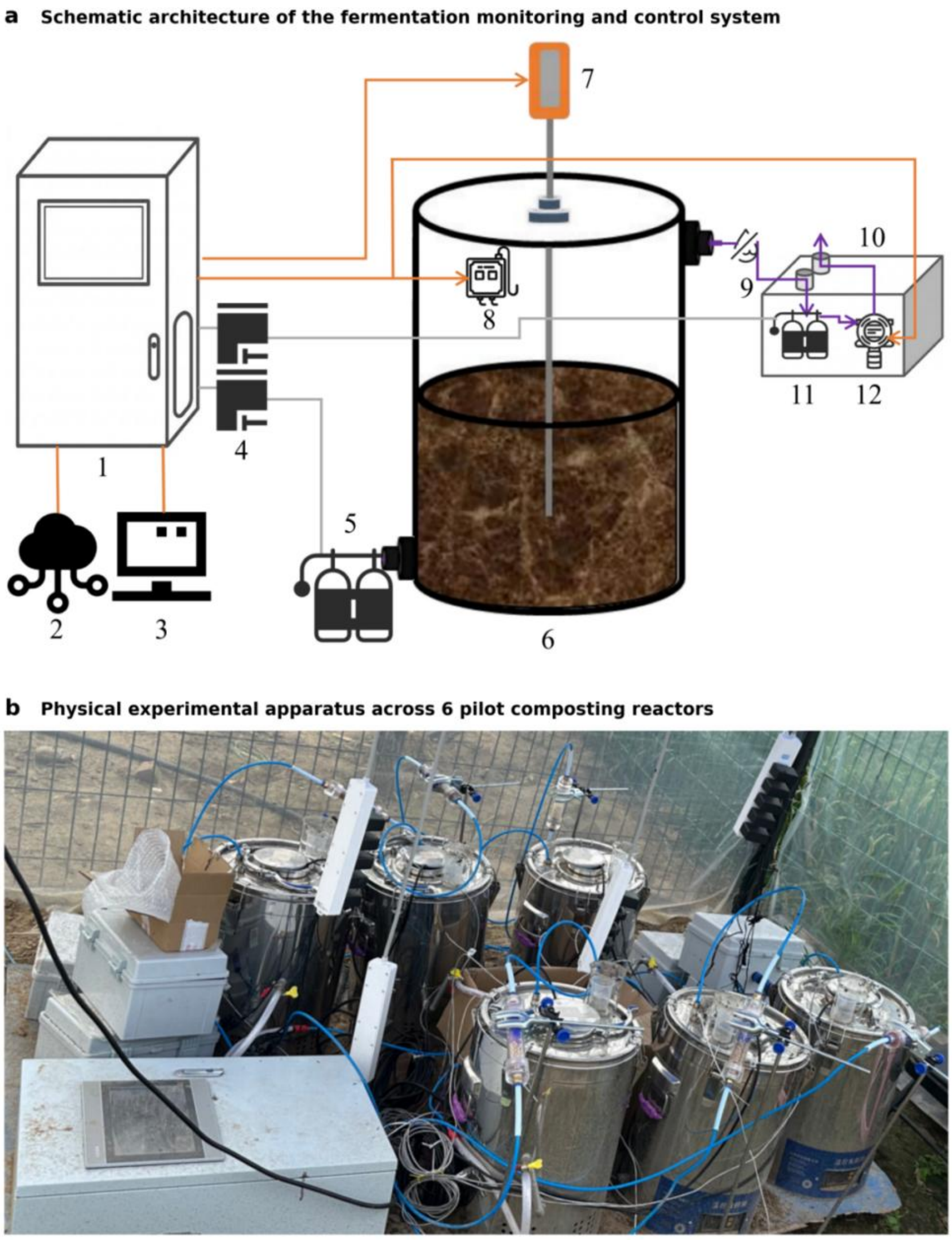

Fig. 1. Architectural schematic of the aerobic composting bioreactor experimental platform: (1) master control cabinet; (2) cloud platform; (3) touchscreen HMI display; (4) solid-state relay array; (5) forced aeration diaphragm pump; (6) sealed 20-L fermentation vessel; (7) temperature and relative humidity probes; (8) electrochemical $NH_3$ sensor; (9) desiccant and molecular sieve conditioning unit; (10) IP68 waterproof enclosure; (11) suction sampling pump; (12) non-dispersive infrared $CO_2$ sensor.

The master control cabinet was driven by an STM32F407 ARM Cortex-M4 microcontroller running FreeRTOS, which orchestrated multi-task microsecond preemptive scheduling for sensor acquisition, telemetry parsing, and relay actuation. The controller actuated 12 relay channels regulating six positive-displacement diaphragm aeration pumps (KZP12-B24, delivery rate 12.0 $L \cdot min^{-1}$) and six micro-diaphragm suction sampling pumps (EDLP600-D24B, sampling rate 0.5 $L \cdot min^{-1}$). Distributed wireless node enclosures communicated with the central cabinet via an RS-485 industrial fieldbus. A touch human-machine interface (Kunlun Tongtai TPC7032Nt) interfaced with the MCU over RS-232 to facilitate real-time edge monitoring and local data export, while an RS-485 IoT gateway streamed telemetric records to an industrial cloud repository. Process monitoring within each bioreactor employed a PTFE-sheathed Class-A PT100 thermal resistance probe (accuracy ±0.1 °C) and a capacitive sensor (SN-3000SD-TR-N01, accuracy ±3% RH) inserted into the geometric core of the substrate bed. Gaseous concentrations were acquired via a diffusion electrochemical $NH_3$ sensor (SN-3002-$NH_3$-N01-1000P-2, range 0–1000 ppm, accuracy ±3%) suspended in the reactor headspace, alongside a pump-suction non-dispersive infrared $CO_2$ sensor (JX-$CO_2$-103-100W, range 0–100,000 ppm, accuracy ±5% F.S. at 25 °C) enclosed within an IP68-rated node module, thereby guaranteeing operational stability under severe thermophilic, humid, and corrosive conditions.

Raw layer chicken manure (20.0 kg) was procured from the poultry breeding facility at the National Precision Agriculture Demonstration Base in Xiaotangshan, Beijing. After removing feathers and feed debris, the raw manure was ground and homogenized to pass a 4-mesh sieve. Rice husks (Xuzhou cultivar U8259) served as the carbonaceous bulking agent. The substrate was prepared by blending chicken manure, dry rice husks, fresh rice husks, and deionized water at a wet mass ratio of 5:1:1:1, establishing an initial moisture content of 58.2% and a C/N ratio of approximately 18.5. During initial loading and subsequent manual turning operations, 30 vertical aeration chimneys (diameter 2.0 cm) were formed using perforated rods throughout the substrate depth to enhance air permeability and convective exchange. The vessel base heating mantles were regulated at a constant temperature of 55 °C to simulate adiabatic large-scale pile thermal conditions. All multi-sensor telemetry channels were synchronously polled at a 10-min sampling interval, generating 4,033 continuous observation vectors per reactor throughout the 28-day composting campaign.

#### 2.1.2. Analytical methods

Experimental analytical protocols adhered strictly to Chinese agricultural industry standards for Organic Fertilizers (NY/T 525-2026) and Hygienic Requirements for Harmless Disposal of Night Soil (GB 7959-2012). Destructive core samples (250 g fresh weight) were collected on days 0, 1, 7, 14, 21, and 28 of the composting cycle. On days 7, 14, and 21, manual turning was executed alongside the supplementation of a commercial composite microbial inoculant (Junde bio-agent, 0.05% w/w wet basis) to promote humification. Subsamples were partitioned into dried and fresh fractions: a 100-g aliquot was air-dried at ambient room temperature, pulverized through a 110-mesh sieve, and stored in sealed desiccators for elemental total carbon (TC) and total nitrogen (TN) determination; a 125-g fresh aliquot was sealed and refrigerated at 4 °C for ammonium nitrogen ($NH_4^+$-N) and nitrate nitrogen ($NO_3^-$-N) analyses to quantify nitrogen speciation trajectories. Compost phytotoxicity was evaluated via the seed germination index (GI) using cucumber seeds (Cucumis sativus L., Jingyan Chunqiu Green No. 2): 5.0 g of fresh substrate was extracted with 50 mL of ultrapure water (1:10 w/v), sealed, agitated on an orbital shaker at 200 rpm for 120 min, and centrifuged at 3000 rpm for 15 min. A 20-mL aliquot of supernatant was collected to determine extract pH and seed GI.

#### 2.1.3. Data preprocessing

To ensure high numerical precision for neural optimization and subsequent symbolic distillation via PySINDy and PySR, sensor telemetry streams underwent outlier screening, Gaussian smoothing, and physical scaling. Substrate mass loss from destructive sampling was mathematically compensated using a mass-decay correction coefficient. Each bioreactor time series was chronologically partitioned into training, validation, conformal calibration, and test subsets according to a strict 6:1:1:2 chronological ratio (60% training, 10% validation, 10% calibration, and 20% test), yielding 144,360 sliding window sample instances across the six vessels. To evaluate cross-reactor spatial generalization, a two-fold leave-reactors-out evaluation was structured: Fold A trained on the initial 80% of Tanks 1, 3, and 5 and tested on the final 20% of Tanks 2, 4, and 6; Fold B swapped these reactor subsets symmetrically. Because manual pile turning occurred exactly three times (days 7, 14, and 21) and perturbed sensor readings for only a single 10-min time step per event, these localized turning artifacts were compensated via cubic Hermite spline interpolation.

#### 2.1.4. Computational environment

All deep learning models were trained on a high-performance computing workstation operating Ubuntu 20.04.6 LTS, equipped with an Intel Xeon Silver 4210R CPU (@ 2.40 GHz) and four NVIDIA GeForce RTX 4090 GPUs (24 GB VRAM each). The host environment ran NVIDIA display driver 570.133.07 supporting CUDA 12.8. The execution environment (Conda environment tsai_env) utilized Python 3.10.20, PyTorch 2.1.0+cu118, CUDA Runtime 11.8, and cuDNN 8.7.0. Model optimization employed the Adam optimizer with a batch size of 32, an initial learning rate of 1e-4, a maximum training duration of 30 epochs, and an early-stopping patience of 6 epochs on validation loss. To ensure statistical reliability and reproducibility, all benchmark models were independently executed across five distinct random initialization seeds (2021–2025), with individual training runs pinned to dedicated GPUs.

Under this standardized computational infrastructure, baseline volatilization and aeration event forecasting configurations were executed across identical data preprocessing pipelines, input lookback horizons (L = 144, 24 h), multi-step forecast horizons (H = 36, 6 h), batch dimensions, and random seeds. Methodological differences were confined strictly to backbone neural architectures, event specialist modules, multi-step rollout strategies, or physical constraint ablations. Comprehensive algorithmic specifications and corresponding evaluation protocols are detailed in Tables 1–6.

### 2.2. Biochemical reaction modeling

To establish rigorous physical boundaries for the subsequent data-driven time-series architecture, we modeled the biochemical reaction mechanisms governing the gas phase within the aerobic composting bioreactor headspace based on conservation of mass (Ni, 1999; Pagans et al., 2006). The baseline headspace free volume was specified as $V_{\text{free}} = 4.0$ L. Grounded in the mass balance principles of a generalized Continuous Stirred Tank Reactor (CSTR), the continuous temporal evolution of the transient ammonia concentration C(t) (in ppm) within the reactor headspace is governed by the following first-order ordinary differential equation (ODE) (Pagans et al., 2006):

$$V_{free}\frac{dC(t)}{dt} = G_{bio}(T, CO_2) - Q_{air}(t) \cdot C(t)$$

where $G_{bio}$(T, $CO_2$) denotes the biological $NH_3$ generation rate (mg·min$^{-1}$); $Q_{air}$(t) represents the volumetric aeration flow rate (L·min$^{-1}$), configured as a piecewise step variable assigned to 12.0 L·min$^{-1}$ during forced aeration and 0 L·min$^{-1}$ during the non-aerated baseline phase (Pagans et al., 2006). Over the 28-day composting cycle, the decomposition of solid organic matter and physical compaction induce matrix settlement, causing a gradual physical expansion of the headspace free volume (Ni, 1999). In our mass balance formulation, gas transport and interphase equilibrium are parameterized via concentration and ventilation flow rate. Because dynamic temporal fluctuations in headspace volume were not continuously measured in situ, $V_{free}$ was maintained at the nominal constant of 4.0 L, leaving subtle short-term variations to be captured implicitly by the time-series module from sensor records. This formulation circumvents the introduction of unobservable geometric state variables while ensuring numerical robustness.

Within the porous chicken manure matrix during high-temperature aerobic composting, $G_{bio}$ is jointly governed by the core substrate temperature T and the respiratory metabolic intensity of heterotrophic microorganisms (Pagans et al., 2006). Coupling the thermal sensitivity of biochemical reaction kinetics via a modified Arrhenius equation with $CO_2$ concentration as a proxy for microbial respiration, the biological $NH_3$ generation rate is formulated as a nonlinear function of temperature and $CO_2$ concentration (Ni, 1999):

$$G_{bio} = \alpha \cdot exp(k_1(T - T_{base})) \cdot \ln(CO_2)$$

where α = $M_{mat}$ · $\alpha_{specific}$, in which $M_{mat}$ = 10.5 kg represents the total wet mass of the initial substrate loading, and $\alpha_{specific}$ is the specific ammonia production coefficient per unit substrate mass; $T_{base}$ = 31.0 °C is the reference baseline temperature for thermal kinetic scaling; $k_1$ denotes the temperature sensitivity coefficient governing the ammonification reaction rate; and ln($CO_2$) quantifies the promotional effect of elevated microbial respiration on ammonia generation.

When subjected to intense operational disturbances, unconstrained deep neural networks are notoriously susceptible to generating negative concentration forecasts or unphysical artifacts that violate fundamental mass balances (Karniadakis et al., 2021). Grounded in the steady-state asymptotic behavior of the bioreactor and the multi-phase gas–liquid–solid partitioning equilibrium of $NH_3$, we derived dynamic concentration boundaries [$C_{min}$(t), $C_{max}$(t)] to constrain allowable model outputs (Raissi et al., 2019; Sander, 2015). The physical lower boundary $C_{min}$(t) is strictly bounded by non-negativity: $C_{min}(t) \equiv 0$ ppm.

The dynamic upper boundary $C_{max}$(t) is decoupled into two mutually exclusive regimes dictated by the operational aeration state $Q_{air}$(t) (Cranmer et al., 2020; Levenspiel, 1999; Sander, 2015):

$$C_{max,aerate}(t) = \frac{G_{bio}(T,CO_2)}{Q_{air}(t)}, Q_{air}(t)>0$$

$$C_{max,static}(t) = \beta \cdot exp(k_2(T - 60.0)), Q_{air}(t)=0$$

During forced aeration, continuous airflow rapidly flushes $NH_3$ from the reactor headspace, driving the gas concentration toward the convective dilution equilibrium $C_{max,aerate}$(t), where the transient accumulation rate can be approximated as $dC(t)/dt \approx 0$ (Levenspiel, 1999). Conversely, during non-aerated baseline periods, gas exchange is negligible, and headspace concentration is predominantly governed by the thermodynamic gas–liquid partitioning across the moist compost matrix, described by Henry's law (Sander, 2015). Using a thermophilic reference temperature of 60.0 °C, empirical saturation coefficients were calibrated for each duplicate bioreactor group: $\beta \in$ [521.0, 274.0, 367.0] ppm for Group 1 (Tanks 1–2), Group 2 (Tanks 3–4), and Group 3 (Tanks 5–6), respectively.

### 2.3. Network architecture

#### 2.3.1. Patch-based representation learning

At each 10-minute sampling step, we constructed an 8-dimensional process input vector rather than relying solely on the $M_{raw}$ = 4 raw sensory channels. The raw observations comprised temperature ($T$), relative humidity ($RH$), carbon dioxide ($CO_2$), and ammonia ($NH_3$). In synchronization with the diurnal scheduled aeration events, the data loading pipeline generated four auxiliary engineering variables: the binary event gate $m_t$, the empirical event ratio $R_{emp,t}$, the normalized intra-event relative elapsed time $\tau_t$, and the reactor categorical identifier Group_ID$_t$. These auxiliary variables were strictly derived from deterministic prior schedule information, vessel identity, and timestamps accessible at the forecast origin. Designating the lookback input window of length $L$ as $Z \in$ 鈫$^{\mathrm{L \times Cin}}$, the $t$-th input vector is formally defined as:

$$\mathbf{z}_t = \left[T_t,\ \mathrm{Humid}_t,\ \mathrm{CO}_{2,t},\ \mathrm{m}_t,\ R_{\mathrm{emp},t},\ \tau_t,\ \mathrm{Group_ID}_t,\ \mathrm{NH}_{3,t}\right]^{\mathrm{T}} \in \mathbb{R}^{C_{\mathrm{in}}}, C_{\mathrm{in}} = 8$$

The transition into the thermophilic phase within three days after inoculation served as a hallmark of successful composting initiation. Accordingly, a continuous historical lookback horizon of L = 432 time steps (72 h) was adopted, encompassing complete diurnal fluctuations and six consecutive aeration cycles. The network outputs multi-step predictions for the subsequent H = 6 time steps, projecting the $NH_3$ concentration trajectory from 10 to 60 min ahead. This horizon aligns directly with the actionable feedforward lead time required for intelligent aeration scheduling while preventing error compounding from unobserved operational disturbances over excessively extended forecast horizons.

We configured a patch length of P = 16 time steps (160 min), a temporal receptive field sufficient to capture the transient kinetic response preceding and following forced aeration. The stride was set to S = 8 steps (80 min), generating a 50% overlapping segmentation that enables adjacent patches to share boundary context, thereby preventing artificial boundary truncation when sharp emission peaks coincide with patch edges. Consequently, the 432-step historical sequence was partitioned into N = 53 patch tokens per univariate channel, as governed by N =(L - P) / S+ 1 = (432 - 16) / 8 + 1 = 53.

For each univariate channel, patches are projected into a $d_{\mathrm{model}}$ = 64-dimensional latent space via a shared linear embedding, augmented with learnable 1D positional encodings, and fed into the Transformer encoder. The 64-dimensional latent capacity provides sufficient representational width to encode localized temporal features across temperature, humidity, gas concentrations, and event variables, while curbing parameter inflation on the dataset derived from six pilot-scale bioreactors. Positional encodings preserve chronological ordering across the patched sequence, serving as the canonical approach for injecting temporal inductive biases into permutation-invariant self-attention architectures (Vaswani et al., 2017).

The encoder comprises $e_{\text{layers}} = 3$ stacked Transformer layers, hierarchically abstracting short-term transients to extended multiscale dependencies without inducing over-parameterization. Multi-head self-attention employs $n_{\text{heads}} = 4$ heads ($d_k = 16$ per head), enabling distinct attention subspaces to focus on overarching humification trends, thermal lags, and localized aeration bursts. The feed-forward network (FFN) dimension is set to $d_{\text{ff}} = 256$ ($4 \times d_{\text{model}}$) to supply nonlinear capacity, with a dropout rate of 0.1 applied throughout to mitigate overfitting under finite sample regimes.

The architecture incorporates a channel-independent (CI) PatchTST design: each univariate channel is patched and embedded independently, while all $C_{\text{in}} = 8$ channels share the identical linear projection, Transformer encoder backbone, and flattened projection head. Parameter sharing not only drastically suppresses model parameter count compared to channel-mixing architectures, but also regularizes representation learning by subjecting all variables to uniform temporal abstraction rules. Reversible Instance Normalization (RevIN) is integrated across each lookback window without learnable affine parameters, normalizing inputs by window instance statistics and de-normalizing output forecasts to counteract non-stationary distribution drift across composting stages and heterogeneous bioreactors. Finally, a linear flattening projection head aggregates the latent representations across channels to output the multi-step baseline $NH_3$ trajectory over the next 6 horizons.

### 2.3.2. Physics-informed residual learning

Pure data-driven empirical loss functions minimize statistical discrepancies between predictions and observations but cannot penalize physically unfeasible artifacts, such as negative gas concentrations or explosive violations of conservation of mass. To embed domain-specific physical principles into neural training, we formulated the thermodynamic concentration boundaries and gas-phase mass balance derived in Section 2.2 as differentiable soft penalty constraints. Physics-informed machine learning incorporates governing differential equations and inequality boundaries via penalty residuals into the loss formulation, optimizing data fidelity subject to physical consistency (Cuomo et al., 2022; Karniadakis et al., 2021; Raissi et al., 2019). We unified five unilateral boundary penalties into a composite thermodynamic barrier loss:

$$\mathcal{L}_{\text{boundary}} = \lambda_{\text{boundary}}\left(\mathcal{L}_{\text{lower}} + \mathcal{L}_{\text{upper,static}} + \mathcal{L}_{\text{upper,aerate}} + \mathcal{L}_{\text{upper,event}} + \mathcal{L}_{\text{upper,ratio}}\right)$$

where $\mathcal{L}_{\text{lower}}$ penalizes negative concentration predictions throughout all operational periods; $\mathcal{L}_{\text{upper,static}}$ enforces the Henry equilibrium upper bound exclusively during unperturbed baseline periods (mt = 0) away from transition boundaries; $\mathcal{L}_{\text{upper,aerate}}$ is activated during active ventilation ($Q_{\text{air}} > 0$) to constrain convective dilution; and $\mathcal{L}_{\text{upper,event}}$ alongside $\mathcal{L}_{\text{upper,ratio}}$ penalizes excessive $NH_3$ concentration spikes and response ratio escalations during forced aeration events (mt ≥ 0.5). The dynamic empirical event multiplier upper bound is specified as $R_{\text{emp}} + 0.35$.

To incorporate biochemical kinetics involving temporal derivatives, physics-informed neural networks (PINNs) typically minimize the mean squared residual of governing differential equations within the loss objective, encouraging network outputs to satisfy physical laws across collocation points (Raissi et al., 2019) via backward Euler or finite-difference approximations (Lim et al., 2022; Thakur et al., 2024). Building upon this paradigm, we discretized the gas-phase mass balance from Section 2.2 across the uniform sampling grid ($\Delta t$ = 10 min). Letting t denote the current prediction horizon and t-1 its preceding step, $C_{\text{t}}$ and $C_{\text{t-1}}$ represent the composite, de-normalized $NH_3$ concentration predictions. Approximating the temporal accumulation rate with a first-order backward finite difference, the discrete mass balance residual is defined as:

$$\mathcal{R}_{\text{ode},t} = V_{\text{free}}\frac{C_t - C_{t-1}}{\Delta t} - G_{\text{bio}}(T_t, CO_{2,t}) + Q_{\text{air},t} \cdot C_t$$

where $V_{\text{free}}$ $(C_t - C_{\text{t}-1}) / \Delta t$ represents the net accumulation (storage) flux of $NH_3$ within the gaseous headspace. $T_t$ is the core substrate temperature at time $t$, restored from normalized inputs to degrees Celsius; $CO_{2,\text{t}}$ is the concurrent carbon dioxide sensor measurement mapped to physical units for computing $G_{\text{bio}}$. The biological generation flux $G_{\text{bio}}$ ($T_t$, $CO_{2,\text{t}}$) follows the modified Arrhenius and respiration kinetics defined in Section 2.2. $Q_{\text{air,t}} \cdot C_t$ represents the advective removal flux induced by forced aeration. When the storage flux exactly balances biological generation minus convective displacement, $\mathcal{R}_{\text{ode,t}} = 0$;

non-zero residuals quantify the discrepancy between the forecast trajectory and theoretical conservation of mass.

Following standard PINN formulations that minimize the mean squared residual of governing equations (Raissi et al., 2019), the normalized ODE loss across all valid consecutive difference pairs is formulated as:

$$\mathcal{L}_{\text{ode}} = \frac{1}{|\mathcal{M}_{\text{ode}}|} \sum\nolimits_{t \in \mathcal{M}_{\text{ode}}} \left( \frac{\mathcal{R}_{\text{ode},t}}{Scale_{\text{ode}}} \right)^2$$

where $\mathcal{M}_{\text{ode}}$ denotes the index set of valid finite-difference pairs, and $|\mathcal{M}_{\text{ode}}|$ denotes the cardinality of this set. To prevent spurious gradient spikes induced by discontinuous operational switching, physical residuals are evaluated strictly when two adjacent time steps share identical aeration and event gating states; boundary transitions caused by blower activation/deactivation or event gating switches are masked out. $\text{Scale}_{\text{ode}}$ is dynamically computed as the stop-gradient mean absolute magnitude of the storage, biological generation, and convective removal fluxes, ensuring that the dimensionless normalized residual squares remain balanced across diverse composting phases and operational regimes. The denominator is stabilized by adding a numerical epsilon ($10^{-5}$) to prevent division by zero when $|\mathcal{M}_{\text{ode}}| = 0$. If a forecast trajectory contains fewer than two valid consecutive steps or no valid pairs exist within a minibatch, $\mathcal{L}_{\text{ode}}$ is set to 0, smoothly bypassing the differential constraint without interrupting optimization.

Building upon the joint optimization framework of empirical data fidelity and physical residuals, we tailored a multi-task global training objective designed to align with the core evaluation targets: baseline tracking, forced-aeration event responsiveness, and peak amplitude accuracy ($E_{\text{peak}}$):

$$\begin{aligned}\mathcal{L}_{\text{total}} = \; & \lambda_{\text{raw}}\mathcal{L}_{\text{raw}} + \mathcal{L}_{\text{normal}} + \lambda_{\text{event,raw}}\mathcal{L}_{\text{event,raw}} + \lambda_{\text{event,ratio}}\mathcal{L}_{\text{event,ratio}} \\ & + \lambda_{\text{event,peak}}\mathcal{L}_{\text{event,peak}} + \lambda_{\text{trans}}\mathcal{L}_{\text{trans}} + \gamma(e) \cdot \left(\mathcal{L}_{\text{boundary}} + \lambda_{\text{ode}}\mathcal{L}_{\text{ode}}\right) + \mathcal{L}_{\text{param}}\end{aligned}$$

The multi-task empirical data loss components encompass global numerical fidelity, normal-regime baseline accuracy, aeration-event numerical values, event multiplication ratios, peak emission amplitudes, and transient pre-/post-aeration response dynamics. In the primary experiments, the penalty weighting coefficients were set to $\lambda_{\text{boundary}} = \lambda_{\text{ode}} = 0.01$. The progressive physics warming schedule $\gamma(e)$ was configured to remain at 0 for the initial 5 epochs, ramp linearly from 0 to 1 over epochs 6 to 10, and remain constant at 1 thereafter. The parameter regularizer $\mathcal{L}_{\text{param}}$ enforces physical plausibility and non-negativity across all learned kinetic coefficients. These thermodynamic and kinetic constraints prevent predicted trajectories from violating boundary limits, governing ODE mass-balance dynamics, and kinetic parameter feasibility. By dynamically balancing empirical accuracy and physical consistency throughout training, the model achieves high forecasting precision while faithfully capturing baseline volatilization, forced-aeration release dynamics, peak concentrations, and transient regime transitions.

#### 2.3.3. Block-aware split conformal prediction

The temporal block aggregation scheme was designed to mitigate the invalid assumption of independent and identically distributed (i.i.d.) observations caused by pronounced autocorrelation among consecutive 10-minute sampling steps. Stratified by random seed and bioreactor identifier, calibration set observations were partitioned into contiguous, non-overlapping 1-hour temporal blocks. Across five random seeds, individual calibrators were constructed across six forecast horizons ($h \in \{1, \ldots, 6\}$) and two operational regimes (normal vs. aeration event), yielding a total of 60 calibrators. Each calibrator operated strictly on its stratum-specific data subset: for each random seed, the six event-horizon calibrators comprised 66 temporal blocks each, whereas the normal-regime and aeration-regime calibrators encompassed 402 and 408 temporal blocks, respectively. A nominal coverage level of $1 - \alpha_{\text{CP}} = 90\%$ ($\text{PICP}_{90}$) was prescribed, ensuring that the constructed intervals cover at least 90% of future realizations with finite-sample distribution-free validity (Angelopoulos and Bates, 2023). Under the split conformal inference framework, the maximum absolute prediction error across each 1-hour temporal block was computed on the calibration set. Sorting these block-maximum non-conformity scores in ascending order yields the empirical quantile threshold that defines the symmetric prediction interval. Specifically, within each forecast horizon h and operational regime r, the conformal error radius $q_{h,r}$ was evaluated via the finite-sample quantile ranking rule:

$$q_{h,r} = s\left(\min\left\{n_{h,r}, \lceil (n_{h,r}+1)(1-\alpha) \rceil\right\}\right)$$

where $n_{h,r}$ denotes the effective number of calibration blocks within the stratum (h, r), s(k) represents the k-th order statistic of the sorted block-maximum non-conformity scores, $\alpha_{CP}$ = 0.10 denotes the nominal miscoverage rate, and r designates the operational regime. The quantile rank is bounded within [1, $n_{h,r}$] to prevent out-of-bounds indexing in small-sample strata. Consequently, the (1 - $\alpha_{CP}$) conformal prediction interval for horizon h at time step t is formulated as:

$$C_{t,h}^{(1-\alpha_{CP})} = \left[ y_{t,h} - q_{h,r}, y_{t,h} + q_{h,r} \right]$$

where $y_{t,h}$ represents the point prediction of headspace $NH_3$ concentration generated by the model at time step t for forecast horizon h.

#### 2.3.4. Symbolic regression

Symbolic regression was conducted as a post-hoc interpretability analysis to distill governing relationships from the trained neural network. The kinetic evolution of headspace $NH_3$ exhibits fundamentally distinct regimes: the normal stationary phase is characterized by slow, quasi-equilibrium volatilization, whereas forced aeration triggers abrupt transient surge-and-decay dynamics within short timeframes. Consequently, a monolithic symbolic expression cannot adequately capture both operational phases. To maintain physical fidelity, the baseline point forecasting architecture was decoupled, utilizing PINN-PatchTST for normal-regime dynamics and the GPR event specialist for aeration-pulse transients, which were concatenated across operational switching boundaries.

For the normal regime, Sparse Identification of Nonlinear Dynamics (PySINDy) was applied using the single-step outputs of the frozen PINN-PatchTST as regression targets. The candidate feature library incorporated historical sensor observations across a continuous lookback window of 432 steps (72 h). Multiple hyperparameter configurations were benchmarked on the validation set to select the optimal model that maximized fidelity to the PINN-PatchTST teacher while preserving algebraic parsimony and numerical stability. The discrete delay-map formulation implemented via PySINDy belongs to the broader class of sparse identification techniques for nonlinear dynamical systems (Brunton et al., 2016), possessing the general form:

$$\dot{x}(t)\Theta(x(t))\Xi$$

where x(t) represents the system state vector, $\dot{x}(t)$ denotes the predicted rate of state evolution, $\Theta(x)$ is a candidate function library comprising linear, polynomial, and nonlinear candidate basis functions, and $\Xi$ is the sparse coefficient matrix. By enforcing sparsity via sequential thresholded least squares, PySINDy isolates a minimal subset of active terms, yielding an interpretable state-transition mapping. In this work, the method was employed to approximate the discrete delay map of the frozen PINN-PatchTST; hence, the resulting formulations represent discrete-time candidate recurrence relations. Because PySINDy evaluates the single-step reproduction of the frozen teacher model's outputs, its objective is to render the black-box neural predictions into transparent, algebraically verifiable candidate equations rather than postulating a universal biochemical kinetic law. For the forced aeration regime, the input feature vector for symbolic regression via genetic programming (PySR) was defined as:

$$\mathbf{x}_i = \left[g_{1,i},\ g_{2,i},\ g_{3,i},\ \tau_i,\ T_i,\ \mathrm{CO}_{2,i},\ R_{\mathrm{emp},i},\ \frac{h_i}{6}\right]^{\top} \in \mathbb{R}^8$$

where $g_{1,i}$, $g_{2,i}$, $g_{3,i} \in \{0, 1\}$ are binary one-hot indicators denoting reactor aeration group membership ($g_{1,i} + g_{2,i} + g_{3,i} = 1$), $\tau_i$ is elapsed composting time, $T_i$ and CO2,i are current pile temperature and headspace carbon dioxide concentration, $R_{\mathrm{emp},i}$ represents the empirical event ratio, and $h_i \in \{1, \ldots, 6\}$ denotes the multi-step forecasting horizon normalized by $h_i / 6$. PySR treats symbolic expression discovery as a multi-objective Pareto optimization that balances predictive accuracy against expression complexity (Cranmer, 2023). Guided by these criteria, candidate expressions on the Pareto frontier were selected based on validation set performance, favoring formulas that exhibited high fidelity to the teacher model, parsimonious algebraic structure, and bounded numerical stability. The symbolic expressions distilled by PySR serve strictly to approximate the GPR event response and provide auditable symbolic transparency; final quantitative deployment forecasts during aeration pulses were generated by the calibrated GPR specialist.

The distilled formulations delineate discrete-time delay mappings and algebraic approximations for the stationary and forced-aeration regimes. During multi-step rollout, raw symbolic outputs were rescaled to physical sensor units, and negative concentrations were projected to zero ($y_{clipped} = \max(0, y)$) as standard numerical post-processing. Symbolic expressions and their fitted numerical coefficients were recorded across five random seeds to quantify stochastic optimization variability. Because varying initialization seeds can yield diverse algebraic topologies on the Pareto frontier, these distilled models are explicitly reported as candidate symbolic relationships. Crucially, evaluation metrics of the symbolic surrogates assess distillation fidelity to the neural teacher and should not be equated directly with the end-to-end forecasting metrics of the complete pipeline.

### 2.4. Evaluation metrics

All evaluation metrics were calculated under a frozen benchmarking protocol wherein ground-truth observations and temporal train-validation-test partitions were fixed prior to model assessment. Forecasts generated for each bioreactor at corresponding target timestamps were strictly paired with observed values, and performance was evaluated separately across the full test set, the stationary normal regime, and the forced-aeration event regime. Across all tables, results are reported as the mean ± standard deviation over five independent random seeds to quantify computational and initialization variance.

Root Mean Squared Error (RMSE) serves as the primary evaluation metric, penalizing substantial prediction outliers more heavily. The coefficient of determination ($R^2$) provides a complementary assessment of the proportion of variance explained by the model across the test series. Mean Absolute Error (MAE) provides insufficient penalty for transient peak errors ($E_{peak}$), whereas relative metrics such as Mean Absolute Percentage Error (MAPE), symmetric MAPE (sMAPE), and Weighted Absolute Percentage Error (WAPE) suffer from numerical instability and extreme divergence when near-zero $NH_3$ baseline concentrations appear in the denominator (Hyndman and Koehler, 2006). Mean Absolute Scaled Error (MASE) was employed to benchmark forecast accuracy across diverse reactors and scale ranges, using an in-sample 24-h seasonal naive baseline ($m = 144$ steps); a value of MASE $<$ 1 signifies that the model outperforms the naive seasonal persistence benchmark (Lim et al., 2022). Because stationary normal periods constitute over 93% of the complete sampling timeline, aggregate metrics alone tend to obscure severe local inaccuracies during brief forced-aeration episodes. To isolate predictive skill during forced aeration, the event-increment coefficient of determination ($R^2_{\Delta,\mathrm{event}}$) is formulated as follows:

$$R^2_{\Delta,\mathrm{event}} = 1 - \frac{\sum_{i \in E} \left[ \left( \hat{y}_i^{\mathrm{event}} - \hat{y}_i^{\mathrm{base}} \right) - \left( y_i^{\mathrm{raw}} - y_i^{\mathrm{clean}} \right) \right]^2}{\sum_{i \in E} \left[ \left( y_i^{\mathrm{raw}} - y_i^{\mathrm{clean}} \right) - \overline{\Delta y}_E \right]^2}$$

Within the set of aeration event samples 洞 the predicted event increment for the i-th sample is defined as the difference between the full event model prediction $\hat{y}_i^{\mathrm{event}}$ and the baseline normal prediction $\hat{y}_i^{\mathrm{base}}$. Similarly, the observed event increment is defined as the difference between the raw measured concentration $y_i^{\mathrm{raw}}$ and the baseline reference concentration $y_i^{\mathrm{clean}}$ obtained via nonlinear spline interpolation. The metric $R^2_{\Delta,\mathrm{event}}$ evaluates the standard coefficient of determination over these paired increments against the mean observed increment $\Delta\bar{y}$洞across 洞 explicitly quantifying the model's capacity to explain transient emission surges beyond baseline drift.

Peak concentration error ($E_{\mathrm{peak}}$) and peak timing offset ($E_{\mathrm{time}}$) evaluate pulse dynamics from complementary perspectives: $E_{\mathrm{peak}} = |\hat{y}_{\max} - y_{\max}|$ quantifies the absolute discrepancy between predicted and observed maximum $NH_3$ concentrations (ppm), whereas $E_{\mathrm{time}} = |\hat{t}_{\max} - t_{\max}|$ measures the temporal phase displacement of the peak in minutes. Thus, $E_{\mathrm{peak}}$ measures amplitude distortion, whereas $E_{\mathrm{time}}$ captures temporal latency or premature triggering. A model might accurately estimate pulse amplitude while exhibiting lead or lag, or alternatively synchronize perfectly in time while severely under- or over-estimating peak height. Evaluating both metrics decouples amplitude fidelity from phase synchronization.

A comprehensive review by Wheatcroft (2019) detailed the general formulation of skill scores based on model-versus-reference error ratios; Mittermaier (2008) thoroughly examined the utilization of persistence as a baseline reference forecast; and Mayer and Yang (2024) formalized the corresponding RMSE Skill Score structure. The RMSE Skill Score relative to naive persistence adopts the standard relative improvement formulation widely utilized in meteorological and time-series verification, reported as a percentage:

$$\text{Skill} = 100\left(1 - \frac{\text{RMSE}model}{\text{RMSE}Persistence}\right)$$

Drawing upon physical feasibility criteria in scientific machine learning (Karniadakis et al., 2021; Raissi et al., 2019), the Physical Violation Rate (PVR) audits whether raw network outputs satisfy thermodynamic boundaries prior to any post-hoc output clipping or projection:

$$\text{Raw PVR} = 100N^{-1}\sum\nolimits_{t} I\,(y_t < 0)$$

$$\text{Total PVR} = 100N^{-1}\sum\nolimits_{t} I\,(y_t < Cmin,t\ \vee\ y_t > Cmax,t)$$

where $y_t$ denotes the raw model forecast at time step t, N is the total number of forecast instances, and $I(\cdot)$ is the indicator function. Raw PVR audits the frequency of negative concentration violations ($y_t < 0$), whereas Total PVR encompasses violations of both the physical non-negativity lower bound and the regime-dependent thermodynamic upper bounds ($C_{max,static}$ or $C_{max,event}$), with dual violations at any single step counted only once. For conformal calibration, the non-conformity scores were computed using 60-minute non-overlapping block-maximum absolute residuals to safeguard finite-sample coverage guarantees under serial correlation (Vovk et al., 2005). Because reporting $PICP_{90}$ alone incentivizes excessively wide intervals and reporting MPIW alone ignores under-coverage, the Winkler Interval Score ($IS_{0.1}$) was included as a strictly proper scoring rule that jointly penalizes interval width and boundary exceedances (Gneiting and Raftery, 2007; Kuleshov et al., 2018). Accordingly, $PICP_{90}$, MPIW, and $IS_{0.1}$ are reported in tandem to provide a balanced evaluation of uncertainty quantification.

## 3. Results and discussion

### 3.1. Predictive performance and ablation analysis

To evaluate whether the proposed framework effectively balances forecasting accuracy across both stationary periods and forced-aeration transients, ten competitive model architectures were benchmarked across all six pilot bioreactors. Table 1 summarizes the overall predictive accuracy, regime-specific performance (normal vs. aeration event), peak amplitude and timing errors during aeration pulses, and unconstrained physical violation rates (PVR).

Table 1. Comprehensive performance comparison of ten forecasting models for ammonia emissions evaluated on the strictly chronological test set (mean ± standard deviation across five random seeds). Bold text indicates optimal results; underline indicates second-best results.

| **Model** | **Overall RMSE** | **Overall $R^2$** | $R^2_{normal}$ | $R^2_{event}$ | $E_{peak}$ | $E_{time}$ | **MASE** | **Total PVR** | **Raw PVR** |
|---|---|---|---|---|---|---|---|---|---|
| PINN-PatchTST-GPR(Ours) | 0.8256 ± 0.0132 | 0.9817 ± 0.0006 | 0.9744 ± 0.0011 | 0.9711 ± 0.0020 | 1.6924 ± 0.0837 | 0.8 ± 0.2 | 0.0124 ± 0.0001 | 0.0000 ± 0.0000 | 0.0000 ± 0.0000 |
| NLinear | 1.0217 ± 0.0128 | 0.9720 ± 0.0007 | 0.9622 ± 0.0011 | 0.9663 ± 0.0034 | 2.0224 ± 0.0961 | 0.7 ± 0.4 | 0.0155 ± 0.0002 | 0.0000 ± 0.0000 | 0.0000 ± 0.0000 |
| DLinear | 1.2613 ± 0.1565 | 0.9568 ± 0.0106 | 0.9355 ± 0.0195 | 0.9618 ± 0.0039 | 2.2889 ± 0.3536 | 0.9 ± 0.6 | 0.0205 ± 0.0037 | 0.0000 ± 0.0000 | 0.0000 ± 0.0000 |
| Linear | 1.5886 ± 0.3895 | 0.9290 ± 0.0314 | 0.8841 ± 0.0568 | 0.9579 ± 0.0087 | 2.3214 ± 0.4779 | 0.6 ± 0.0 | 0.0283 ± 0.0091 | 0.0007 ± 0.0015 | 0.0007 ± 0.0015 |
| PatchTST | 2.3797 ± 0.6682 | 0.8384 ± 0.0907 | 0.7641 ± 0.1497 | 0.8469 ± 0.0680 | 4.6632 ± 0.9769 | 0.6 ± 0.5 | 0.0246 ± 0.0033 | 0.1462 ± 0.1258 | 0.1462 ± 0.1258 |
| Transformer | 2.1742 ± 0.3931 | 0.8698 ± 0.0464 | 0.9020 ± 0.0439 | 0.6807 ± 0.1146 | 8.3061 ± 1.8652 | 5.6 ± 3.8 | 0.0304 ± 0.0075 | 0.0000 ± 0.0000 | 0.0000 ± 0.0000 |
| Persistence | 4.1089 ± 0.0000 | 0.5469 ± 0.0000 | 0.8672 ± 0.0000 | -0.2310 ± 0.0000 | 11.7500 ± 0.0000 | 10.4 ± 0.0 | 0.0352 ± 0.0000 | 0.0000 ± 0.0000 | 0.0000 ± 0.0000 |
| iTransformer | 12.3561 ± 1.2078 | -3.1283 ± 0.8150 | -4.1959 ± 1.1169 | -4.9720 ± 1.1122 | 40.1687 ± 3.3201 | 1.0 ± 1.2 | 0.2271 ± 0.0262 | 0.0887 ± 0.0933 | 0.0887 ± 0.0933 |
| Autoformer | 37.4413 ± 6.7746 | -37.6042 ± 13.3197 | -68.3513 ± 25.0614 | -10.8965 ± 1.8076 | 41.8262 ± 5.9536 | 13.1 ± 2.0 | 0.6766 ± 0.1365 | 7.5222 ± 3.3906 | 7.4072 ± 3.4127 |
| Informer | 56.9387 ± 8.7926 | -87.6596 ± 26.4911 | -160.7484 ± 48.6345 | -16.0002 ± 5.1563 | 42.5578 ± 7.5668 | 8.0 ± 2.4 | 1.2047 ± 0.1863 | 0.0048 ± 0.0072 | 0.0000 ± 0.0000 |

As evidenced in Table 1, the proposed model (Ours) attained an overall RMSE of 0.8256 ± 0.0132 ppm, an $R^2$ of 0.9817 ± 0.0006, and a MASE of 0.0124 ± 0.0001, consistently achieving optimal performance across all global metrics. Compared to the second-best model (NLinear), Ours reduced the mean RMSE by 19.2%, whereas the baseline

Persistence model yielded a substantially higher RMSE of 4.1089 ppm. During forced aeration episodes, Ours achieved an $E_{\text{peak}}$ of 1.6924 ± 0.0837 ppm and an $E_{\text{time}}$ of 0.8 ± 0.2 min, ranking best and second-best among all evaluated models, respectively. Although bare PatchTST achieved a marginally lower $E_{\text{time}}$, it exhibited severe physical constraint violations (Raw PVR = 3.42%) and markedly inferior overall predictive accuracy. This multi-objective superiority is further highlighted by the joint distribution of global versus event-specific $R^2$ and the trade-off between peak concentration and timing errors illustrated in Fig. 2.

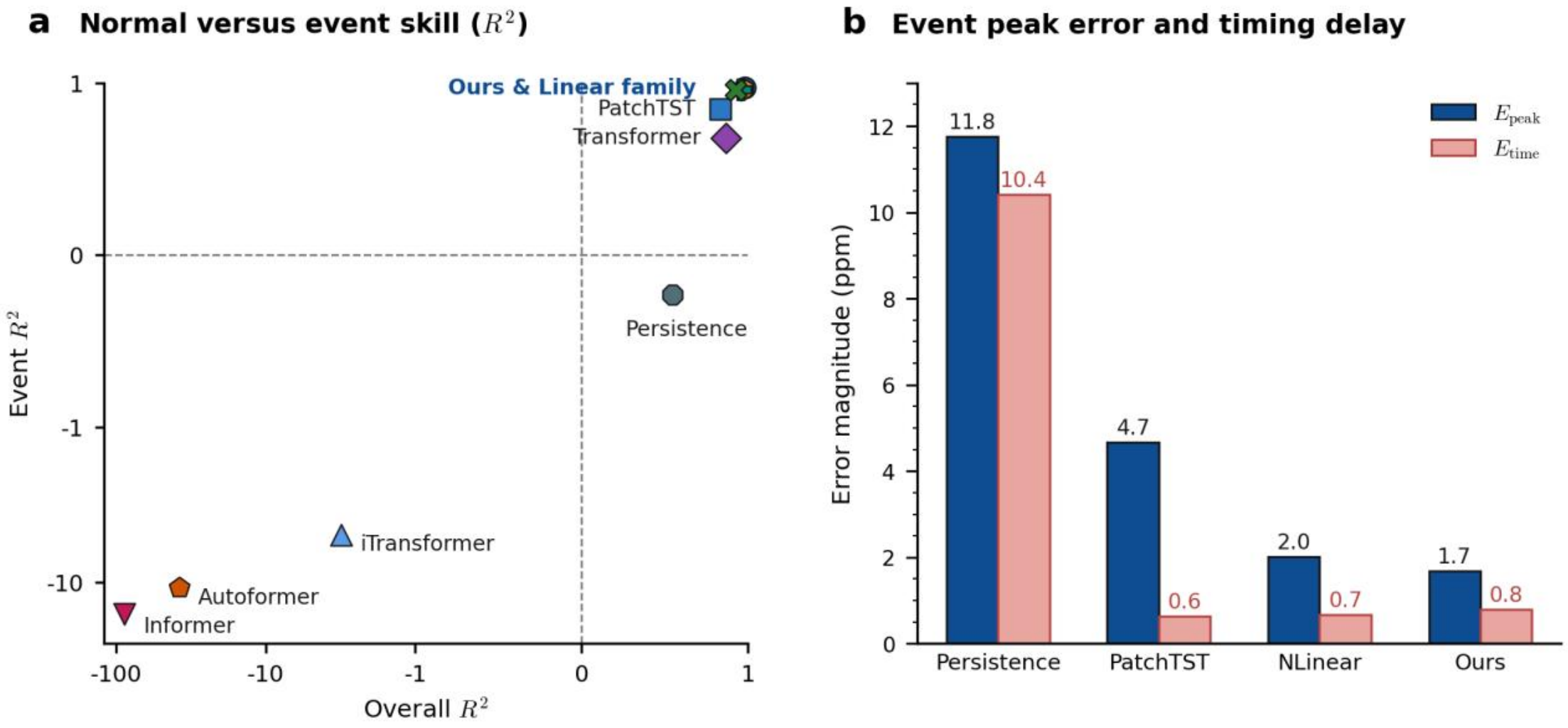

Fig. 2. Comparative evaluation of predictive accuracy and event dynamics across benchmark models. (a) Joint distribution of global $R^2$ versus event-regime $R^2$, demonstrating that Ours and linear baselines reside within the positive quadrant, whereas Persistence and unconstrained Transformer baselines experience severe degradation during aeration. (b) Scatter plot of peak concentration error ($E_{\text{peak}}$) versus peak timing offset ($E_{\text{time}}$); Ours achieves near-optimal concentration tracking ($E_{\text{peak}} \approx 1.7$ ppm, $E_{\text{time}} \approx 0.8$ min), sharply contrasting with Persistence (11.8 ppm and 10.4 min, respectively).

These comparisons reveal that aggregate forecasting discrepancies are predominantly governed by forced-aeration pulse dynamics. Under the pre-event single-shot protocol, models generate a 6-step horizon forecast immediately prior to aeration onset, evaluated strictly on the $h = 1$ step. Under the rolling forecast protocol, models dynamically re-forecast the 6-step horizon at each successive timestamp using newly incoming observations. Comparative results for both protocols are summarized in Table 2.

Table 2. Performance comparison of benchmark models under abrupt aeration conditions for single pre-aeration origin forecasting and rolling h = 1 step forecasting (mean ± standard

deviation across 5 random seeds). Bold indicates the best result; underline indicates the second-best result.

| Model | Pre-aeration RMSE | Pre-aeration $R^2$ | Rolling RMSE | Rolling $R^2$ |
|---|---|---|---|---|
| PINN-PatchTST-GPR(Ours) | 2.1716 ± 0.0813 | 0.9697 ± 0.0023 | 2.1609 ± 0.0812 | 0.9700 ± 0.0022 |
| NLinear | 2.2973 ± 0.0514 | 0.9661 ± 0.0015 | 2.6236 ± 0.3261 | 0.9552 ± 0.0118 |
| DLinear | 3.0155 ± 0.2504 | 0.9412 ± 0.0098 | 3.0514 ± 0.5245 | 0.9387 ± 0.0221 |
| Linear | 4.0388 ± 0.8624 | 0.8913 ± 0.0463 | 5.0439 ± 1.7719 | 0.8204 ± 0.1116 |
| PatchTST | 4.3828 ± 1.0367 | 0.8710 ± 0.0613 | 5.7495 ± 2.3468 | 0.7592 ± 0.2130 |
| Transformer | 7.6547 ± 3.0101 | 0.5769 ± 0.3560 | 7.9368 ± 2.9748 | 0.5497 ± 0.3318 |
| Persistence | 16.1858 ± 0.0000 | -0.6836 ± 0.0000 | 12.6006 ± 0.0000 | -0.0204 ± 0.0000 |
| iTransformer | 31.3047 ± 4.6277 | -5.4081 ± 1.9004 | 30.1972 ± 4.9724 | -4.9873 ± 1.8844 |
| Autoformer | 45.4688 ± 9.4181 | -12.7424 ± 5.7017 | 39.7601 ± 3.8670 | -9.2365 ± 1.9018 |
| Informer | 52.5096 ± 7.7304 | -17.0270 ± 5.1860 | 50.2274 ± 7.9323 | -15.5365 ± 4.9948 |

As presented in Table 2, Ours achieved an RMSE of 2.1716 ± 0.0813 ppm in single pre-aeration origin forecasting and 2.1609 ± 0.0812 ppm in rolling h = 1 step forecasting, whereas the corresponding persistence baseline yielded substantially higher errors of 16.1858 ± 0.0000 ppm and 12.6006 ± 0.0000 ppm, respectively. Although NLinear demonstrated competitive accuracy in single pre-aeration forecasting with an RMSE of 2.2973 ± 0.0514 ppm, its rolling prediction error degraded to 2.6236 ± 0.3261 ppm. To decouple the effects of physical loss components from underlying architectural capabilities, we fixed the PatchTST backbone, the GPR event specialist, and all non-physical losses while systematically varying only the weights of the mass-balance ODE residual and dynamic boundary terms, with the resulting ablation summarized in Table 3.

Table 3. Ablation comparison of physics-informed and non-negativity constraint configurations under a fixed event specialist (mean ± standard deviation across 5 random seeds). Bold indicates the best result; underline indicates the second-best result.

| Physics Configuration | Overall RMSE | $R^2_{\text{normal}}$ | Pre-event RMSE | $R^2_{\Delta,\text{event}}$ |
|---|---|---|---|---|
| Soft bounds only | 0.8239 ± 0.0120 | 0.9785 ± 0.0006 | 2.1625 ± 0.0766 | 0.9602 ± 0.0033 |
| Full PINN (Ours) | 0.8256 ± 0.0132 | 0.9744 ± 0.0011 | 2.1716 ± 0.0813 | 0.9600 ± 0.0037 |
| No physics | 0.8276 ± 0.0134 | 0.9783 ± 0.0005 | 2.1840 ± 0.0822 | 0.9598 ± 0.0038 |
| ODE only | 0.8277 ± 0.0134 | 0.9783 ± 0.0005 | 2.1846 ± 0.0829 | 0.9598 ± 0.0038 |

As demonstrated in Table 3, the overall RMSE values for Soft bounds only, Full PINN, No physics, and ODE only were 0.8239 ± 0.0120 ppm, 0.8256 ± 0.0132 ppm, 0.8276 ± 0.0134 ppm, and 0.8277 ± 0.0134 ppm, respectively. The maximum pairwise difference in mean RMSE across the four configurations was only 0.0038 ppm, which falls within the standard deviation observed across the five random seeds. Thus, under the prevailing operating regimes and loss weightings, the physical constraints do not confer an independently attributable gain in point-forecasting RMSE, nor does the full configuration induce any perceptible accuracy penalty. Although Full PINN did not attain the absolute numerical minimum RMSE, we adopted it as the primary architecture because it simultaneously enforces ODE kinetic dynamics and dynamic concentration boundaries, thereby providing a complete mechanistic interface under essentially identical predictive accuracy; this choice represents a deliberate design decision rather than an empirical claim of superior point accuracy. Physical violation rates (PVR) were not employed to rank configurations in this table and remain documented in Table 1. To examine whether representations learned on a subset of bioreactors generalize effectively to unseen vessels, we established a two-fold cross-reactor evaluation protocol (Fold A/B), with the results reported in Table 4.

Table 4. Model performance comparison under Fold A/B cross-reactor partitioning (mean ± standard deviation across 5 random seeds). Bold indicates the best result; underline indicates the second-best result.

| Model | Fold A Overall RMSE | Fold B Overall RMSE | Equal Fold Mean RMSE | Fold A Overall $R^2$ | Fold B Overall $R^2$ | Equal Fold Mean $R^2$ |
|---|---|---|---|---|---|---|
| PINN-PatchTST-GPR(Ours) | 0.9671 ± 0.0403 | 0.8318 ± 0.0164 | 0.8994 | 0.9753 ± 0.0021 | 0.9811 ± 0.0007 | 0.9782 |
| NLinear | 1.0201 ± 0.0131 | 0.9974 ± 0.0051 | 1.0088 | 0.9725 ± 0.0007 | 0.9728 ± 0.0003 | 0.9727 |
| PatchTST | 1.8258 ± 0.2252 | 1.0623 ± 0.0347 | 1.4441 | 0.9108 ± 0.0230 | 0.9692 ± 0.0020 | 0.9400 |
| Linear | 1.6991 ± 0.4068 | 1.4470 ± 0.2391 | 1.5730 | 0.9202 ± 0.0352 | 0.9416 ± 0.0182 | 0.9309 |
| DLinear | 1.9151 ± 0.5818 | 1.3819 ± 0.3630 | 1.6485 | 0.8959 ± 0.0606 | 0.9450 ± 0.0317 | 0.9204 |
| Transformer | 3.1986 ± 1.1016 | 2.9059 ± 1.0236 | 3.0522 | 0.7040 ± 0.1764 | 0.7465 ± 0.1660 | 0.7253 |
| Persistence | 4.1377 ± 0.0000 | 4.0799 ± 0.0000 | 4.1088 | 0.5476 ± 0.0000 | 0.5454 ± 0.0000 | 0.5465 |
| iTransformer | 11.9115 ± 1.3486 | 11.2142 ± 2.5261 | 11.5628 | -2.7876 ± 0.8484 | -2.5737 ± 1.5589 | -2.6807 |
| Autoformer | 30.2303 ± 6.9972 | 45.4855 ± 13.7168 | 37.8579 | -24.1836 ± 11.6013 | -59.6110 ± 33.8273 | -41.8973 |
| Informer | 43.9853 ± 12.9064 | 49.5995 ± 15.2669 | 46.7924 | -53.6450 ± 31.2373 | -71.2750 ± 41.6181 | -62.4600 |

Table 4 indicates that Ours achieved the lowest overall RMSE across both folds, yielding 0.9671 ± 0.0403 ppm in Fold A and 0.8318 ± 0.0164 ppm in Fold B, with corresponding overall $R^2$ values of 0.9753 ± 0.0021 and 0.9811 ± 0.0007, respectively. When equally weighted across both folds, Ours attained an average RMSE of 0.8994 ppm and an $R^2$ of 0.9782, outperforming NLinear (1.0088 ppm and 0.9727). PatchTST, Linear, and DLinear exhibited intermediate average RMSEs of 1.4441–1.6485 ppm and $R^2$ values of 0.9204–0.9400, while vanilla Transformer and Persistence accumulated substantially larger errors (mean RMSEs of 3.0522 ppm and 4.1088 ppm, respectively). In contrast, iTransformer, Autoformer, and Informer experienced severe generalization failure, with mean $R^2$ collapsing to -2.6807, -41.8973, and -62.4600, indicating that unconstrained global-attention models struggle to generalize across distinct bioreactor units. These findings demonstrate that Ours maintains strong predictive consistency under cross-reactor evaluation. The seed-to-seed variations across physical ablation configurations and the two-fold cross-reactor transfer dynamics are further illustrated in Figure 3.

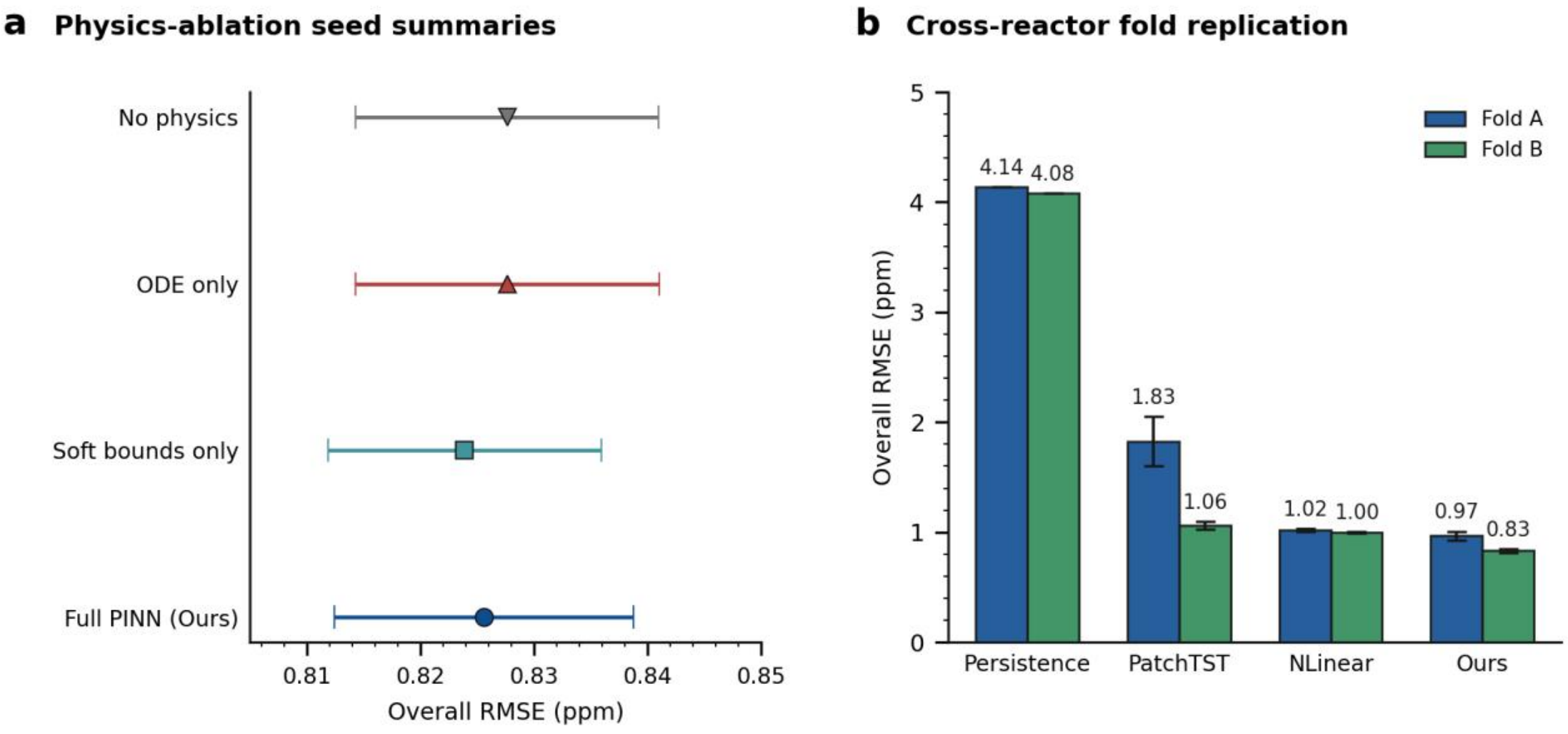

Fig. 3. Physical-constraint effects and cross-reactor replication. (a) Comparison of RMSE distributions across four physical-constraint ablation configurations, where the substantial interval overlap is consistent with the minimal point-accuracy differences reported in Table 3; error bars represent sample standard deviations across 5 random seeds. (b) Overall RMSE of Ours in comparison with baseline models under Fold A and Fold B cross-reactor partitions, demonstrating consistently lower error than competing baselines despite inter-fold variability. Overall, the empirical evidence supports the predictive robustness of the proposed framework under varying bioreactor and aeration conditions.

### 3.2. Uncertainty quantification

To objectively evaluate the robustness of alternative rolling forecasting strategies during transient $NH_3$ concentration extrapolation, we designed a multi-stage temporal testing protocol. Because forced aeration generates prominent, rapid surges in $NH_3$ concentration, a single aggregated metric computed over the full fermentation cycle can easily obscure localized performance degradation during high-emission episodes. Consequently, the evaluation was partitioned into three distinct metric intervals: full-cycle global RMSE and $R^2$, a peak phase defined within ±10 min of the concentration extremum, and a recovery phase tracking the decline back to baseline levels. The comparative results across rolling strategies and operational phases are presented in Table 5.

Table 5. Predictive performance and relative skill score comparison of rolling forecasting strategies across peak and recovery phases of aeration events (mean ± standard deviation

across 5 random seeds). Bold indicates the best result; underline indicates the second-best result.

| Rolling Strategy | Overall RMSE | Overall $R^2$ | Skill vs Persistence | Peak Phase RMSE | Recovery Phase RMSE |
|---|---|---|---|---|---|
| Frozen PINN (Ours) | 1.8547 ± 0.0660 | 0.9771 ± 0.0016 | 86.37 ± 0.49 % | 2.2450 ± 0.1754 | 0.7204 ± 0.0566 |
| Re-anchored PINN | 1.8622 ± 0.0643 | 0.9769 ± 0.0016 | 86.31 ± 0.47 % | 2.2584 ± 0.1691 | 0.7228 ± 0.0486 |
| Persistence | 13.6039 ± 0.0000 | -0.2328 ± 0.0000 | 0.00 ± 0.00 % | 22.3150 ± 0.0000 | 1.0897 ± 0.0000 |
| ODE Rollout | 50.3355 ± 0.2969 | -15.8782 ± 0.1991 | -270.01 ± 2.18 % | 9.8661 ± 0.0380 | 69.2380 ± 0.4616 |

As summarized in Table 5 across the multi-stage rolling evaluation, the proposed Frozen PINN demonstrated superior performance across all evaluation dimensions. Frozen PINN attained an overall RMSE of 1.8547 ± 0.0660 ppm, compared to 1.8622 ± 0.0643 ppm for Re-anchored PINN; the corresponding peak-phase RMSEs were 2.2450 ± 0.1754 ppm and 2.2584 ± 0.1691 ppm, while the recovery-phase RMSEs were 0.7204 ± 0.0566 ppm and 0.7228 ± 0.0486 ppm, respectively. This indicates that under prevailing measurement noise and physical constraints, residual perturbations introduced by the re-anchoring mechanism slightly outweighed its calibration benefit. Relative to the persistence baseline, Frozen PINN achieved a remarkable forecast skill score of 86.37 ± 0.49%. In sharp contrast, the purely mechanistic ODE Rollout suffered catastrophic open-loop integration divergence, drifting to an RMSE of 69.2380 ppm during the recovery phase, which confirms that uncorrected differential equations cannot maintain numerical stability over extended rollouts. Concurrently, the persistence baseline collapsed during the aeration peak (RMSE = 22.3150 ppm), highlighting the critical role of physics-informed inductive biases in capturing rapid concentration surges.

To determine whether the prediction intervals reliably envelop ground-truth observations while avoiding excessive conservatism—given that wider intervals trivially achieve higher coverage—we simultaneously report the 90% prediction interval coverage probability ($PICP_{90}$), the mean prediction interval width (MPIW), and the Winkler interval score ($IS_{0.1}$), which jointly balances interval sharpness with penalties for coverage misses (Gneiting and Raftery, 2007; Khosravi et al., 2011). Figure 4 illustrates the empirical findings across all six bioreactor tanks from three complementary perspectives: forecast trajectories during aeration events, tank- and horizon-stratified $PICP_{90}$, and tank-wise comparisons of MPIW alongside composite $IS_{0.1}$ scores.

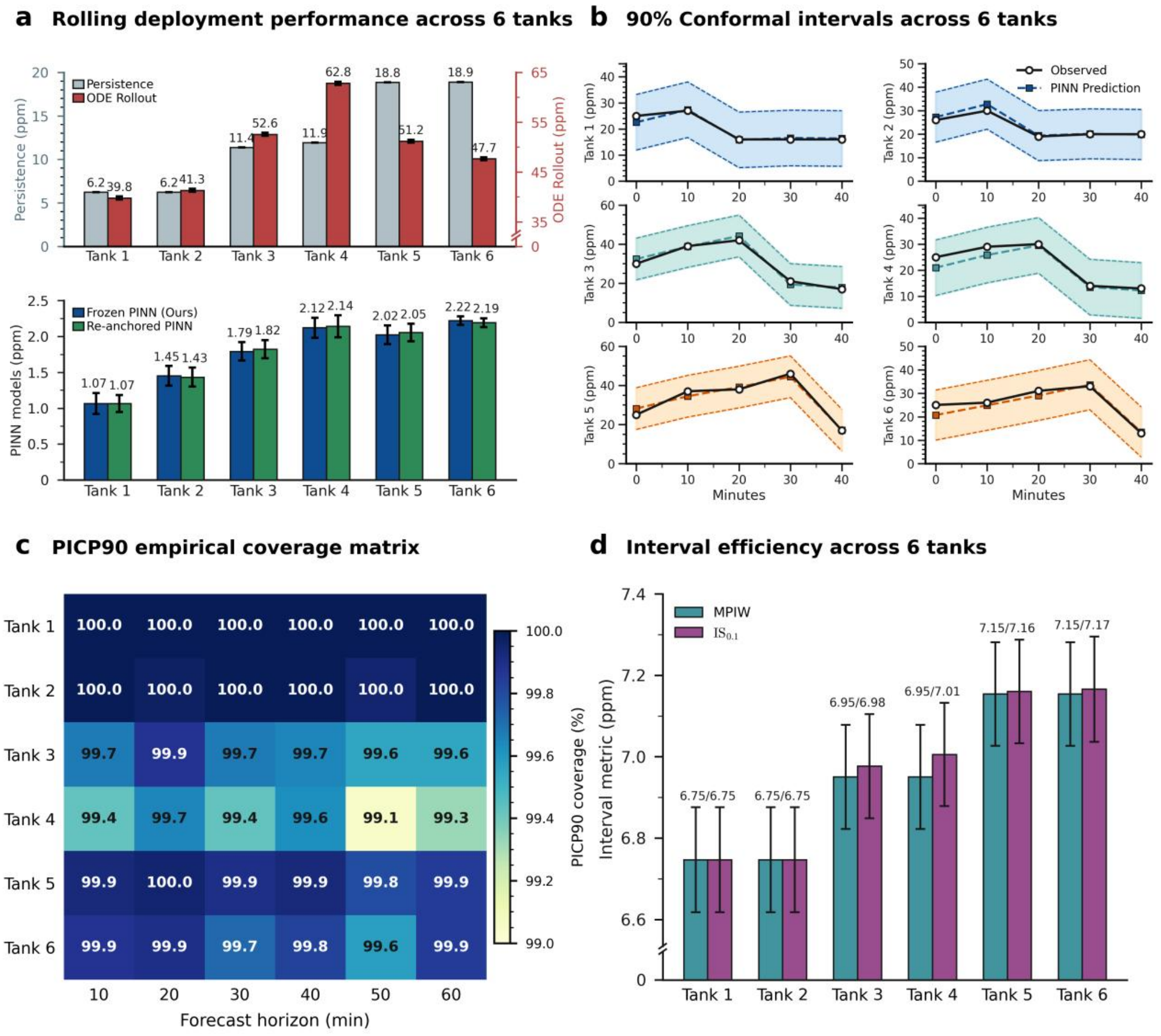

Within the calibration set, continuous 60-min observation sequences were segmented into non-overlapping temporal blocks. The maximum absolute forecasting residual within each block served as the non-conformity metric to calibrate regime- and horizon-stratified prediction intervals, adhering to the Block-Aware Split Conformal Prediction (BASCP) protocol (Angelopoulos and Bates, 2023). Averaged across all six bioreactors and all forecast horizons over five random seeds (comprising 28,872 evaluation points per seed), the overall empirical metrics were: $PICP_{90} = 99.792 \pm 0.047\%$, $MPIW = 6.951 \pm 0.128$ ppm, and $IS_{0.1} = 6.968 \pm 0.128$ ppm, corresponding to an average interval width of approximately 7 ppm. To verify that macroscopic averages do not mask localized coverage deficiencies within specific reactors or lookahead steps, Figure 4c details the reactor-by-horizon $PICP_{90}$ matrix, while Figure 4d compares reactor-specific MPIW and composite Winkler scores.

During forced aeration episodes, the calibrated prediction intervals dynamically expanded, ensuring robust encapsulation of volatile observations. Specifically, the aeration regime exhibited an empirical $PICP_{90}$ of 99.961 ± 0.086% with an MPIW of 19.349 ± 0.630 ppm, whereas the normal stationary regime yielded 99.775 ± 0.046% and 5.728 ± 0.147 ppm, respectively, representing a 3.38-fold expansion in interval width during active aeration. While wider bands successfully enveloped nearly all transient $NH_3$ spikes and demonstrated resilience against strong process disturbances, excessively broad intervals can diminish actionable granularity for feedforward ventilation control. Consequently, we jointly evaluated MPIW and the proper scoring rule $IS_{0.1}$ to simultaneously reward sharpness and penalize coverage breaches, rather than imposing an arbitrary, ungrounded width heuristic (such as a 3 ppm cap).

Stratified evaluations across forecasting horizons and individual bioreactors revealed that the high empirical coverage was uniformly distributed rather than driven by localized conditions. Across the six multi-step lookahead horizons, $PICP_{90}$ ranged from 99.655% to 99.892% with corresponding MPIWs of 6.711 to 7.339 ppm; across the six bioreactors, $PICP_{90}$ spanned 99.418% to 100.000% with MPIWs between 6.747 and 7.155 ppm. This uniformity demonstrates that valid finite-sample coverage is preserved across varying lead times and reactor units rather than supported by anomalous localized subsets (Angelopoulos and Bates, 2023). In summary, while prediction intervals expanded noticeably under strong forced-aeration disturbances, they enveloped virtually all empirical trajectories while preserving actionable resolution for operational decision-making.

### 3.3. Symbolic extraction of kinetics

Following the Sparse Identification of Nonlinear Dynamics (SINDy) methodology for discovering parsimonious governing relationships through candidate function libraries and sparse regression (Brunton et al., 2016), the representative candidate equation distilled under the frozen benchmark artifact with random seed = 2023 is formulated as:

$$\widehat{\tilde{C}}_{t+1}^{(2023)} = 0.016015\tilde{C}_{t-3} + 0.024875\tilde{C}_{t-11} + 0.053337\tilde{C}_{t-17} + 0.030142\tilde{C}_{t-35} + 0.021957\tilde{C}_{t-71} - 0.039474\tilde{C}_{t-215} + 0.192784\tilde{M}_{t,0:5} + 0.549972\tilde{M}_{t,0:17} + 0.331500\tilde{M}_{t,0:71} - 0.195308\tilde{M}_{t,0:143}$$

where $\hat{C}_{t+1}$ denotes the standardized $NH_3$ concentration forecast for the next sampling step; $\hat{C}_{t-k}$ represents the standardized $NH_3$ concentration observed at k antecedent sampling points, corresponding to historical lookback lags of 3, 11, 17, 35, 71, and 215 time steps (spanning 30 min to 35.8 h), respectively; and $\hat{M}_{t,0:k}$ denotes the standardized rolling-average concentration over the preceding k steps up to time t. Circumflexes denote variables standardized to zero mean and unit variance. Sparse regression under this seed pruned temperature, relative humidity, $CO_2$, and constant bias terms. During inference, model predictions are inverse-transformed back to physical units (ppm) and subjected to strict non-negativity enforcement ($C \geq 0$). These fidelity metrics quantify the reconstruction accuracy of the distilled symbolic surrogate relative to the teacher model (PINN-PatchTST), rather than independent forecasting errors against ground-truth $NH_3$ measurements. Because the selected candidate terms vary slightly across different random seeds, this distilled equation should be interpreted as an empirical surrogate capturing dominant delay dynamics within the operational manifold, rather than a universal kinetic conservation law. Across the five evaluation seeds, PINN-PatchTST → PySINDy achieved an RMSE of 0.4624 ± 0.0758 ppm and an $R^2$ of 0.9896 ± 0.0035 relative to the teacher backbone. Table 6 summarizes the mean performance and variability across five random seeds, benchmarking alternative event-response strategies during aeration episodes:

Table 6. Ablation of aeration event specialist branches and fidelity of neural-symbolic distillation (mean ± standard deviation across 5 random seeds). Bold indicates optimal performance; underline indicates second-best performance.

| **Model** | **Event RMSE** | $R^2_{\Delta,event}$ | $E_{peak}$ | $E_{time}$ |
|---|---|---|---|---|
| No Specialist | 11.5943 ± 0.2294 | -1.5534 ± 0.0000 | 13.2902 ± 0.1843 | 10.44 ± 1.19 |
| GPR → PySR | 2.4084 ± 0.0864 | 0.9523 ± 0.0044 | 1.5554 ± 0.2592 | 2.11 ± 1.17 |

| Model | Event RMSE | $R^2_{\Delta,\text{event}}$ | $E_{\text{peak}}$ | $E_{\text{time}}$ |
|---|---|---|---|---|
| Lookup | 2.2735 ± 0.0351 | 0.9607 ± 0.0009 | 1.6468 ± 0.0499 | 0.94 ± 0.15 |
| GAMM | 2.2559 ± 0.0890 | 0.9608 ± 0.0011 | 1.7053 ± 0.1126 | 0.92 ± 0.16 |
| PySR | 3.1794 ± 1.1887 | 0.9357 ± 0.0194 | 2.3002 ± 0.8536 | 1.08 ± 0.21 |
| GPR | 2.1716 ± 0.0813 | 0.9600 ± 0.0037 | 1.6924 ± 0.0837 | 0.81 ± 0.15 |

In the absence of an event specialist, the baseline model exhibited an aeration event RMSE of 11.5943 ± 0.2294 ppm, an event-increment coefficient of determination ($R^2_{\Delta,\text{event}}$) of -1.5534, a peak concentration error ($E_{\text{peak}}$) of 13.2902 ± 0.1843 ppm, and a peak timing offset ($E_{\text{time}}$) of 10.44 ± 1.19 min. Incorporating the GPR event specialist dramatically reduced the event RMSE to 2.1716 ± 0.0813 ppm and curtailed $E_{\text{time}}$ to 0.81 ± 0.15 min. This marked improvement confirms that dedicated machine learning modules tailored to forced-aeration transients provide substantially superior tracking of emission peak magnitudes and their temporal onset.

To determine whether the non-parametric GPR event response could be compressed into a concise, human-auditable mathematical expression, symbolic regression via genetic programming was conducted using PySR by optimizing the Pareto trade-off between predictive error and formula complexity (Cranmer, 2023). Under random seed = 2023, the extracted Pareto-optimal expression is formulated as:

$$f_{\text{PySR}}^{(2023)}(\mathbf{x}_t) = g_{2,t} \cdot R_{\text{emp},t}\left(\frac{h_t}{6} - 0.53659093\right)\left(R_{\text{emp},t} - 1.7477332\right)$$

where 0.53659093 and 1.7477332 are dimensionless empirical constants obtained from training data regression, possessing no direct biological meaning. The output of this equation represents a candidate correction in logarithmic event space, which is converted to a multiplicative event response during model evaluation. This expression is intended solely as an interpretable surrogate approximating the GPR event response rather than as a complete substitute for the full non-parametric GPR specialist.

Under the frozen test benchmark, the distilled PySR surrogate approximating the GPR response attained an event RMSE of 2.4084 ± 0.0864 ppm, an $R^2_{\Delta,\text{event}}$ of 0.9523 ± 0.0044, an $E_{\text{peak}}$ of 1.5554 ± 0.2592 ppm, and an $E_{\text{time}}$ of 2.11 ± 1.17 min. Although the exact algebraic structures differed slightly across the five random seeds, all discovered formulas consistently fulfilled the low-complexity, human-auditable design criterion for approximating GPR event dynamics. To rigorously compare the ability of alternative event specialist architectures to capture abrupt post-aeration $NH_3$ concentration surges while maintaining the normal-regime forecasting branch, data partitioning, and evaluation protocols strictly unchanged, six aeration event configurations were evaluated, as shown in Figure 5:

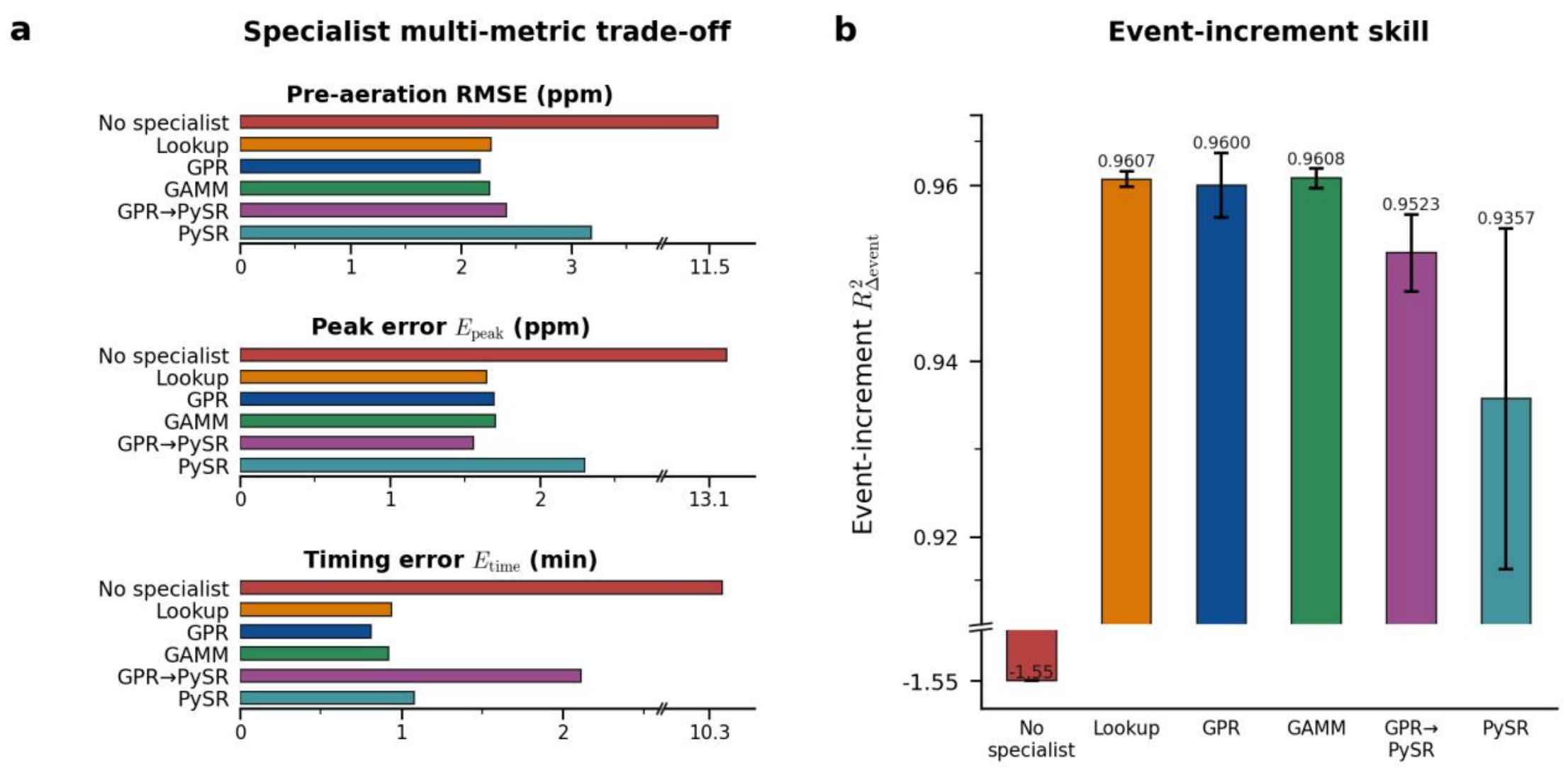

As shown in Figure 5a, without an event specialist ('No specialist'), the pre-aeration event RMSE, $E_{\text{peak}}$, and $E_{\text{time}}$ reached 11.5943 ± 0.2294 ppm, 13.2902 ± 0.1843 ppm, and 10.44 ± 1.19 min, respectively, which were substantially higher than those of configurations with event specialists. Integrating event specialists reduced these three error metrics to approximately 2.17–3.18 ppm, 1.56–2.30 ppm, and 0.81–2.11 min. GPR yielded the lowest event RMSE and $E_{\text{time}}$ (2.1716 ± 0.0813 ppm and 0.81 ± 0.15 min, respectively); the distilled GPR → PySR surrogate delivered the lowest $E_{\text{peak}}$ (1.5554 ± 0.2592 ppm), although its $E_{\text{time}}$ increased to 2.11 ± 1.17 min. Figure 5b indicates that while No specialist produced an unacceptable $R^2_{\Delta,\text{event}}$ of -1.55, the empirical Lookup table, GPR, and GAMM achieved high values of 0.9607, 0.9600, and 0.9608, respectively, while GPR → PySR and standalone PySR achieved 0.9523 and 0.9357. These results demonstrate that event specialists substantially improve predictive precision during aeration episodes, with GPR representing the optimal overall event specialist.

Both distilled symbolic branches consist entirely of explicit finite arithmetic operators. In an embedded implementation incorporating a circular sliding window, offline parameter fusion for standardization coefficients, and pre-normalized event step indices, computing a single forward forecast using the PySINDy formula requires merely 14 multiplications, 12 additions, and 6 subtractions; the PySR formula requires 5 multiplications, 2 subtractions, and 1 exponential operation. This lightweight architecture requires buffering at most 216 historical $NH_3$ observations (36 h of antecedent context at 10-min resolution), occupying approximately 0.8–1.8 KB of SRAM depending on whether Q15/Q31 fixed-point or IEEE 754 float32 representation is employed. Consequently, 8-bit microcontrollers (e.g., AVR or PIC) can run the fixed-point implementation, resource-constrained 32-bit MCUs (e.g., ARM Cortex-M0+/M3) can execute software floating-point calculations, and MCUs equipped with a hardware Floating Point Unit (FPU, e.g., ARM Cortex-M4F/M7) can execute full-precision floating-point inference with execution latencies on the order of ≈ 1.2 ms. This computational and memory overhead is orders of magnitude lower than the deep neural backbone (Ours), operating entirely without deep learning runtime frameworks or GPU accelerators, thereby offering unambiguous feasibility for real-time edge embedded deployment in resource-constrained composting facilities.

### 3.4. Aeration strategy optimization

Although individual reactors exhibited slight variations in peak thermal elevation and cooling rates, all six temperature profiles consistently progressed through the characteristic three-phase composting cycle: the mesophilic heating phase, the thermophilic phase, and the maturation cooling phase. Throughout the 28-day composting trial, all six bioreactors demonstrated this characteristic thermodynamic progression, accompanied by continuous reductions in moisture content, characteristic temporal trajectories of $NH_3$ and $CO_2$ off-gas concentrations, and phased biochemical transitions in substrate pH, seed germination index (GI), total carbon (TC), total nitrogen (TN), ammonium nitrogen ($NH_4^+$-N), and nitrate nitrogen ($NO_3^-$-N), as illustrated in Figure 6:

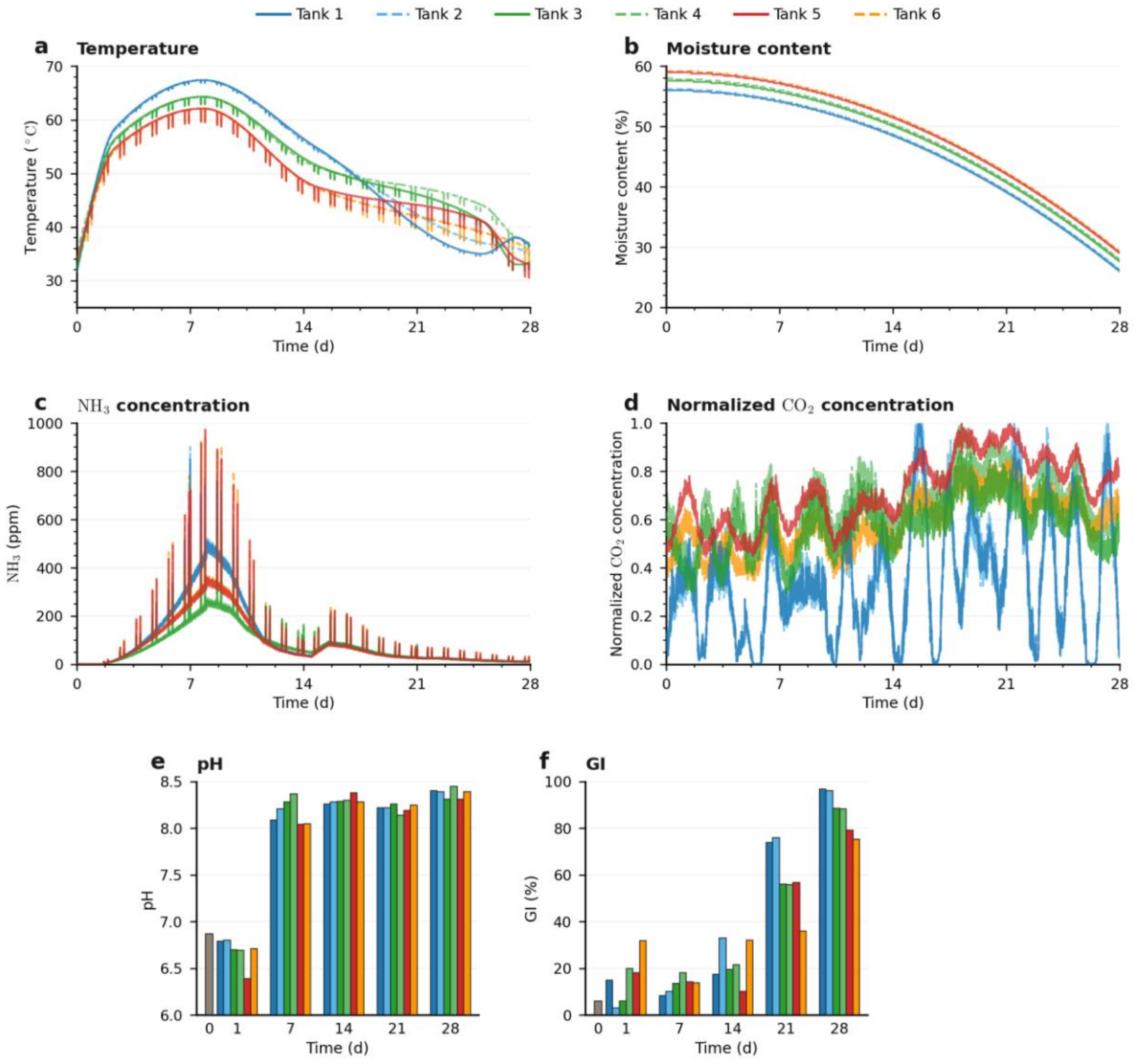

Core temperatures rose rapidly during days 0–3 and reached peak values on day 8. The peak temperatures recorded were 67.4 °C in the 10-min aeration group, 64.3 °C in the 20-min group, and 62.1 °C in the 30-min group; subsequently, temperatures in all reactors gradually declined, stabilizing at 33–37 °C by day 28. During forced aeration pulses, transient core temperature drops of 0.5–2.7 °C were observed, followed by gradual recovery to pre-aeration levels after aeration ceased. Substrate moisture content declined steadily from an initial 56%–59% to 26%–29% by day 28. Headspace $NH_3$ concentrations remained undetectable during the first two days, surged concurrently with core heating, and attained primary peaks on day 8. The 30-min aeration group registered the highest primary peak concentrations (938–973 ppm), followed by the 10-min group (849–901 ppm), whereas the 20-min group exhibited the lowest peak concentrations (563–643 ppm). Minor secondary emission fluctuations were observed around day 14 and subsequently attenuated. Overall, the 20-min aeration schedule produced the lowest total emission surge, whereas the 10-min group exhibited the highest cumulative emission response. Normalized $CO_2$ concentrations displayed periodic respiratory oscillations while maintaining elevated levels throughout the latter half of the composting cycle, with the 10-min group displaying more pronounced troughs and the 30-min group maintaining a higher baseline.

Substrate pH stood at 6.87 on day 0, decreased temporarily to 6.39–6.80 on day 1 due to the accumulation of short-chain volatile fatty acids, and then rose markedly to 8.04–8.37 by day 7 as microbial ammonification released base cations, remaining stable between 8.14 and 8.45 throughout days 14–28. Seed germination index (GI) remained below the phytotoxicity threshold (<50%) during the initial 14 days with intermittent fluctuations, increased sharply after day 21, and culminated at 75.17%–96.84% on day 28, confirming thorough phytotoxicity elimination (GI > 70%). Regarding terminal GI, the 10-min aeration group exhibited the highest maturity indices, whereas pH remained within a comparable alkaline range across all three treatment groups. In summary, Figure 6 holistically illustrates the coupled hydrothermal dynamics, gas emission profiles, maturity progression, and transient aeration responses across different forced aeration durations.

In assessing aeration performance and overall composting efficacy, operational optimization in this study prioritized real-time gas emission mitigation—specifically suppressing peak headspace $NH_3$ concentrations and cumulative gaseous nitrogen dissipation—complemented by compost stabilization rate and thermophilic sanitation duration. Product maturation and harmless sanitary criteria were evaluated with reference to prevailing agricultural and environmental standards, notably the Chinese Agricultural Industry Standard *Organic Fertilizer* (NY/T 525-2021) and the National Standard *Hygienic Requirements for Harmless Disposal of Night Soil* (GB 7959-2012), which establish core compliance thresholds for thermophilic temperature maintenance (>55 °C for ≥ 5 d) and terminal phytotoxicity elimination (GI ≥ 70%). While traditional destructive laboratory assays such as total carbon (TC), total nitrogen (TN), ammonium nitrogen ($NH_4^+$-N), and nitrate nitrogen ($NO_3^-$-N) provide retrospective biochemical verification, non-destructive, high-frequency continuous monitoring of thermodynamic state variables, $CO_2$ respiration, and headspace $NH_3$ off-gas dynamics served as the principal online control signals for predictive modeling and real-time aeration scheduling.

In accordance with institutional pilot research protocols and intellectual property frameworks governing collaborative bioreactor facilities, comprehensive destructive laboratory assays for solid-phase chemical fractions (total carbon, total nitrogen, ammonium nitrogen, and nitrate nitrogen) were scheduled for post-trial standardized verification under separate analytical protocols. The present study deliberately centered on the in-situ, continuous sensor-derived gaseous and thermodynamic observables (temperature, relative humidity, $CO_2$, and $NH_3$), which represent the essential, high-frequency physical state variables required for automated feedforward aeration control and edge-embedded predictive mitigation systems.

### 3.5. Implications and limitations

The engineering significance of the proposed framework (Ours) lies in furnishing predictive intelligence for supervisory aeration control, although it is not currently configured to execute direct actuation commands on behalf of the controller. Specifically, the aeration event branch provides anticipatory indications of impending variations in $NH_3$ concentration prior to scheduled aeration times, functioning as an actionable feedforward signal. Concurrently, the nominal 90% Block-Aware Split Conformal Prediction (BASCP) intervals empower supervisory control systems to select either conservatively bounded or operationally proactive aeration policies. Furthermore, the simplified symbolic relationships distilled via PySINDy and PySR offer transparent, rapidly verifiable mathematical approximations. As shown in Table 5, under continuous rolling forecasting, the Frozen PINN—which deploys the pre-trained model without subsequent parameter updates—exhibits overall performance comparable to the Re-anchored PINN, which recalibrates predictions upon receiving each successive batch of observations. Consequently, in the offline feedforward evaluations of this study, the Frozen PINN delivers sufficiently stable forecasts without requiring stepwise online model adjustments. Nevertheless, this empirical stability remains contingent upon the six bioreactors investigated under prescribed baseline conditions; when feedstock substrates, ambient conditions, or aeration agronomy vary, the Re-anchored PINN may still prove advantageous for correcting potential predictive drift. Moreover, the conformal prediction intervals delineate empirical uncertainty bounds under the present testing protocol rather than absolute operational safety envelopes, and they do not unconditionally guarantee identical coverage rates across unobserved operational batches (Angelopoulos and Bates, 2023; Gneiting and Raftery, 2007). Similarly, the symbolic formulas in Table 6 serve strictly as interpretable surrogate approximations of the event specialist and cannot supersede Gaussian Process Regression (GPR), nor do they constitute fully validated closed-loop aeration control equations (Cranmer et al., 2020). Although symbolic regression condenses complex nonlinear dynamics into closed-form expressions that can be deployed on resource-constrained edge computing devices for near-real-time inference, practical embedded deployment feasibility requires further validation through in-situ hardware benchmarking and closed-loop aeration trials.

The robustness and generalizability of these findings are also subject to constraints inherent in the experimental bioreactor layout. Because each of the three aeration strategies was evaluated across only two parallel replicate bioreactors ($n = 2$), observed inter-group differences represent indicative operational tendencies rather than definitive statistical proof of asymptotic emission stabilization. In Table 3, all evaluated models exhibited zero negative concentration predictions, demonstrating that no physical non-negativity violations occurred across the empirical test sequences; however, an observed zero pre-clipping physical violation rate (Raw PVR = 0%) does not independently prove that the governing material balance equations will be universally satisfied across unexamined operational agronomy. Furthermore, critical biochemical maturity endpoints—including carbon-to-nitrogen (C/N) ratio, total nitrogen, ammonium nitrogen ($NH_4^+$-N), and nitrate nitrogen ($NO_3^-$-N)—have not yet been integrated into a complete analytical evidence chain. Consequently, the present findings cannot definitively establish a multi-objective optimal trade-off schedule that simultaneously achieves minimal $NH_3$ emissions and full compliance with compost maturity and agronomic quality standards.

## 4. Conclusion

To resolve the dual modeling challenges posed by slow stationary background emissions and sharp forced-aeration release transients during aerobic chicken manure composting, this study developed the physics-informed neural network framework (Ours), coupling long-sequence multivariate sensor telemetry, headspace kinetic mass balance constraints, and a specialized Gaussian process aeration event head. Across continuous 28-day composting experiments in six parallel bioreactors, the proposed framework achieved an overall root mean squared error (RMSE) of 0.8256 ± 0.0132 ppm and a coefficient of determination ($R^2$) of 0.9817 ± 0.0006. During active forced-aeration episodes, the model achieved an event-regime $R^2$_{event} of 0.9711 ± 0.0020, a peak concentration error ($E_{\text{peak}}$) of 1.6924 ± 0.0837 ppm, and a peak timing offset ($E_{\text{time}}$) of 0.8 ± 0.2 min. In the rigorous frozen rolling forecasting benchmark, the Frozen PINN yielded an RMSE of 1.8547 ± 0.0660 ppm, achieving an 86.37 ± 0.49% error reduction compared with the standard persistence baseline.

Block-Aware Split Conformal Prediction (BASCP) provided distribution-free finite-sample uncertainty guarantees, achieving an overall empirical coverage probability ($PICP_{90}$) of 99.792 ± 0.047%, a mean prediction interval width (MPIW) of 6.951 ± 0.128 ppm, and a Winkler interval score ($IS_{0.1}$) of 6.968 ± 0.128 ppm at the nominal 90% confidence level. During forced aeration pulses, empirical coverage remained robust ($PICP_{90}$ = 99.961 ± 0.086%) while MPIW expanded to 19.349 ± 0.630 ppm, confirming that wider intervals are necessary to maintain valid coverage during steep transient fluctuations. Physical constraint ablation demonstrated that soft boundary regularization alone achieved marginally lower point-prediction error than the full PINN, corroborating that the kinetic ODE term acts primarily as an inductive regularizer against non-physical trajectory excursions. The explicit formulas distilled via PySINDy and PySR (Table 6) provide interpretable, lightweight surrogate approximations of dynamic event responses (Brunton et al., 2016; Cranmer et al., 2020). Under the investigated six-bioreactor agronomy, the proposed framework reliably delivers anticipatory forecasts ahead of scheduled aeration cycles, establishing a quantitative foundation for closed-loop supervisory control informed by predictive uncertainty bounds and symbolic edge surrogates. As illustrated in Figure 6, an intermittent aeration duration of 20 min achieved optimal overall $NH_3$ mitigation. Future investigations should incorporate full solid-phase chemical maturity profiles (C/N, total nitrogen, ammonium, and nitrate) across pilot-scale facilities to establish a Pareto-optimal control strategy that concurrently minimizes gaseous nitrogen emissions and maximizes compost organic humification.

CRediT authorship contribution statement

Lanbin Wei: Methodology, Software, Validation, Formal analysis, Investigation, Data curation, Visualization, Writing – original draft, Writing – review & editing.

Haiyong Weng:Funding acquisition, Resources, Project administration, Supervision, Writing – review & editing.

Declaration of competing interest

The authors declare that they have no known competing financial interests or personal relationships that could have appeared to influence the work reported in this paper.

Acknowledgements

This study was supported by the Guidance Project for the Construction of the Fujian Provincial Key Laboratory of Agricultural Information Perception Technology (Grant No. KJG22052A)..

Data availability

The data and code that support the findings of this study will be made available from the corresponding author upon reasonable request.

ORCID

Lanbin Wei https://orcid.org/0009-0006-3690-7343

Haiyong Weng https://orcid.org/0000-0002-4548-2109

**References**

[1] Angelopoulos, A.N., Bates, S., 2023. Conformal Prediction: A Gentle Introduction. Found. Trends Mach. Learn. 16 (4), 494–591. https://doi.org/10.1561/2200000101.

[2] Bari, Q.H., Koenig, A., Guihe, T., 2000. Kinetic analysis of forced aeration composting - I. Reaction rates and temperature. Waste Manage. Res. 18 (4), 303–312. https://doi.org/10.1177/0734242X0001800402.

[3] Bernal, M.P., Alburquerque, J.A., Moral, R., 2009. Composting of animal manures and chemical criteria for compost maturity assessment: A review. Bioresour. Technol. 100 (22), 5444–5453. https://doi.org/10.1016/j.biortech.2008.11.027.

[4] Brunton, S.L., Proctor, J.L., Kutz, J.N., 2016. Discovering governing equations from data by sparse identification of nonlinear dynamical systems (SINDy). Proc. Natl. Acad. Sci. U.S.A. 113 (15), 3932–3937. https://doi.org/10.1073/pnas.1517384113.

[5] Cranmer, M., Sanchez-Gonzalez, A., Battaglia, P., Xu, R., Cranmer, K., Spergel, D., Ho, S., 2020. Discovering symbolic models from deep learning with inductive biases. In: Adv. Neural Inf. Process. Syst. 33, 17429–17442.

[6] Cranmer, M., 2023. Interpretable Machine Learning for Science with PySR and SymbolicRegression.jl. arXiv:2305.01582. https://doi.org/10.48550/arXiv.2305.01582.

[7] Cuomo, S., Di Cola, V.S., Giampaolo, F., Rozza, G., Raissi, M., Piccialli, F., 2022. Scientific machine learning through physics-informed neural networks: Where we are and what's next. J. Sci. Comput. 92 (3), 88. https://doi.org/10.1007/s10915-022-01939-z.

[8] Ding, X., 2022. Study on heat, moisture and gas regulation mechanism of high-efficiency aerobic fermentation bioreactor. Trans. Chin. Soc. Agric. Mach. 53 (9), 310–318 (in Chinese with English abstract).

[9] FAO, 2023. The State of Food and Agriculture 2023: Revealing the true cost of food to transform agrifood systems. Food and Agriculture Organization of the United Nations, Rome. https://doi.org/10.4060/cc7724en.

[10] Galgani, P., van der Voet, E., Korevaar, G., 2014. Composting, anaerobic digestion and biochar production in Ghana: Environmental-economic assessment in the context of voluntary carbon markets. Waste Manage. 34 (12), 2454–2465.

https://doi.org/10.1016/j.wasman.2014.07.027.

[11] Girón-Rojas, C., Gil, E., Garcia-Ruiz, A., Iglesias, N., López, M., 2020. Assessment of biowaste composting process for industrial support tool development through macro data approach. Waste Manage. 105, 364–372. https://doi.org/10.1016/j.wasman.2020.02.019.

[12] Gneiting, T., Raftery, A.E., 2007. Strictly proper scoring rules, prediction, and estimation. J. Am. Stat. Assoc. 102 (477), 359–378. https://doi.org/10.1198/016214506000001437.

[13] Haug, R.T., 1993. The Practical Handbook of Compost Engineering. CRC Press, Boca Raton.

[14] Hyndman, R.J., Koehler, A.B., 2006. Another look at measures of forecast accuracy. Int. J. Forecast. 22 (4), 679–688. https://doi.org/10.1016/j.ijforecast.2006.03.001.

[15] Karniadakis, G.E., Kevrekidis, I.G., Lu, L., Perdikaris, P., Wang, S., Yang, L., 2021. Physics-informed machine learning. Nat. Rev. Phys. 3 (6), 422–440. https://doi.org/10.1038/s42254-021-00314-5.

[16] Khosravi, A., Nahavandi, S., Creighton, D., Atiya, A.F., 2011. Comprehensive review of neural network-based prediction intervals and new advances. IEEE Trans. Neural Networks 22 (9), 1341–1356. https://doi.org/10.1109/TNN.2011.2162110.

[17] Kuleshov, V., Fenner, N., Ermon, S., 2018. Accurate uncertainties for deep learning using calibrated regression. In: Dy, J., Krause, A. (Eds.), Proceedings of the 35th International Conference on Machine Learning. PMLR 80, pp. 2796–2804.

[18] Kupper, T., Häni, C., Neftel, A., Kincaid, C., Büchler, M., Amon, B., VanderZaag, A., 2020. Ammonia and greenhouse gas emissions from slurry storage - A review. Agric. Ecosyst. Environ. 300, 106963. https://doi.org/10.1016/j.agee.2020.106963.

[19] Levenspiel, O., 1999. Chemical Reaction Engineering, third ed. John Wiley & Sons, New York.

[20] Lim, K.L., Dutta, R., Rotaru, M., 2022. Physics Informed Neural Network using Finite Difference Method. In: 2022 IEEE International Conference on Systems, Man, and Cybernetics (SMC), pp. 1828–1833. https://doi.org/10.1109/SMC53654.2022.9945171.

[21] Liu, Z., Dong, H., Tao, X., et al., 2022. Policy evolution and emission reduction paths of livestock and poultry manure resource utilization in China. Trans. Chin. Soc. Agric. Eng. 38 (14), 215–224 (in Chinese with English abstract).

[22] Ma, R., Li, G., Yuan, J., et al., 2021. Effects of initial C/N ratio and bulking agents on

maturity and odor emissions during chicken manure hyperthermophilic composting. Environ. Sci. 42 (5), 2510–2519 (in Chinese with English abstract).

[23] Mason, I.G., 2006. Mathematical modelling of the composting process: A review. Waste Manage. 26 (1), 3–21. https://doi.org/10.1016/j.wasman.2005.01.021.

[24] Mayer, M.J., Yang, D., 2024. Potential root mean square error skill score. J. Renewable Sustainable Energy 16 (1), 016501. https://doi.org/10.1063/5.0187044.

[25] Mittermaier, M.P., 2008. The potential impact of using persistence as a reference forecast on perceived forecast skill. Weather Forecast. 23 (5), 1022–1031. https://doi.org/10.1175/2008WAF2007037.1.

[26] Ni, J.Q., 1999. Mechanistic models of ammonia release from liquid manure: a review. J. Agric. Eng. Res. 72 (1), 1–17. https://doi.org/10.1006/jaer.1998.0342.

[27] Nie, Y., Nguyen, N.H., Sinthong, P., Kalagnanam, J., 2023. A time series is worth 64 words: Long-term forecasting with transformers. In: International Conference on Learning Representations (ICLR 2023).

[28] Pagans, E., Barrena, R., Font, X., Sánchez, A., 2006. Ammonia emissions from the composting of different organic wastes. Dependency on process temperature. Chemosphere 62 (9), 1534–1542. https://doi.org/10.1016/j.chemosphere.2005.06.044.

[29] Peng, L., Tang, R., Wang, G., Ma, R., Li, Y., Li, G., Yuan, J., 2023. Effect of aeration rate, aeration pattern, and turning frequency on maturity and gaseous emissions during kitchen waste composting. Environ. Technol. Innovation 29, 102997. https://doi.org/10.1016/j.eti.2022.102997.

[30] Raissi, M., Perdikaris, P., Karniadakis, G.E., 2019. Physics-informed neural networks: A deep learning framework for solving forward and inverse problems involving nonlinear partial differential equations. J. Comput. Phys. 378, 686–707. https://doi.org/10.1016/j.jcp.2018.10.045.

[31] Sander, R., 2015. Compilation of Henry's law constants (version 4.0) for water as solvent. Atmos. Chem. Phys. 15 (8), 4399–4981. https://doi.org/10.5194/acp-15-4399-2015.

[32] Sayara, T., Sánchez, A., 2021. Gaseous emissions from the composting process: Controlling parameters and strategies of mitigation. Processes 9 (10), 1844. https://doi.org/10.3390/pr9101844.

[33] Sole-Mauri, F., Illa, J., Magrí, A., Prenafeta-Boldú, F.X., Flotats, X., 2007. An integrated biochemical and physical model for the composting process. Bioresour. Technol. 98 (17), 3278–3293. https://doi.org/10.1016/j.biortech.2006.07.012.

[34] Thakur, S., Raissi, M., Mitra, H., Ardekani, A.M., 2024. Temporal consistency loss for

physics-informed neural networks. Phys. Fluids 36 (7), 077136. https://doi.org/10.1063/5.0211398.

[35] Vaswani, A., Shazeer, N., Parmar, N., Uszkoreit, J., Jones, L., Gomez, A.N., Kaiser, Ł., Polosukhin, I., 2017. Attention is all you need. In: Adv. Neural Inf. Process. Syst. 30, 5998–6008.

[36] Vovk, V., Gammerman, A., Shafer, G., 2005. Algorithmic Learning in a Random World. Springer-Verlag, New York. https://doi.org/10.1007/b106715.

[37] Wheatcroft, E., 2019. Interpreting the skill score form of forecast performance metrics. Int. J. Forecast. 35 (2), 573–579. https://doi.org/10.1016/j.ijforecast.2018.11.010.

[38] Zhou, T., Li, G., Yuan, J., et al., 2021. Hygienic indicators and organic matter transformation characteristics of livestock and poultry manure aerobic composting. J. Agric. Resour. Environ. 38 (3), 450–458 (in Chinese with English abstract).

[39] Zhou, H., Cheng, Q., Meng, H., et al., 2022. Current situation and bottleneck analysis of solid-liquid manure treatment facilities in national large-scale livestock and poultry farms. J. Agro-Environ. Sci. 41 (6), 1201–1210 (in Chinese with English abstract).

# 物理约束与事件感知的鸡粪好氧发酵反应器氨气排放预测模型及符号关系分析

# Physics-constrained and event-aware forecasting of ammonia emissions in chicken manure aerobic composting reactors: A PINN-PatchTST-GPR framework with symbolic regression analysis

**Lanbin Wei, Haiyong Weng**

College of Mechanical and Electrical Engineering,Fujian Agricultrue and Forestry University,Fuzhou 310002,PR China

Full postal address: College of Mechanical and Electrical Engineering, Fujian Agriculture and Forestry University campus, No. 63, xiyuangong Road, Shangjie Town, Minhou County, Fuzhou City, Fujian Province, PR China.

## Abstract (摘要)

鸡粪好氧发酵 NH3 浓度预测可为曝气决策提供前馈信息，但长周期过程演化与强制曝气脉冲并存，使模型同时面临稳态背景、突发响应、跨步长误差累积和预测风险量化等问题。现有纯数据驱动时序模型通常缺少物理约束，生化反应机理模型难以适应复杂环境变化，二者均难在同一框架内兼顾全局拟合、事件响应与风险量化。为此，本文提出物理信息神经符号时序预测框架 PINN-PatchTST-GPR。基于六个立式反应器的 28 d 发酵过程，组合通道独立 PatchTST、宏观生化反应约束、动态上下界和事件专家。PatchTST 从温度、湿度、CO2 及 NH3 传感器数据中提取物理微分方程、非负约束合理范围，结果符合物理规律；基于分块分割保形预测模块，生成名义 90% 预测区间；PySINDy/PySR 分别将正常与曝气事件表示为低复杂度的候选符号关系。模型总体 RMSE 为 0.8256 ± 0.0132 ppm，$R^2$ 为 0.9817 ± 0.0006，$E_{peak}$ 为 1.6793 ± 0.0725 ppm，$E_{time}$ 为 0.7 ± 0.2 min。总体保形结果为 $PICP_{90}$=99.792 ± 0.047%、MPIW=6.951 ± 0.128 ppm、$IS_{0.1}$=6.968 ± 0.128 ppm。PySINDy RMSE 为 0.4624 ± 0.0758 ppm、$R^2$ 为 0.9896 ± 0.0035，PySR RMSE 为 2.4084 ± 0.0864 ppm、$R^2$ 为 0.9523 ± 0.0044。20 min 曝气组的 $NH_3$ 排放最少。PINN-PatchTST-GPR 在当前六罐、单批次下实现了协同全局预测、事件响应与经验风险量化。

## 1. Introduction (引言)

### 1.1. Background and motivation (背景与动机)

全球集约化畜牧养殖业供给人类蛋白质，同时年均产生数十亿吨高氮畜禽有机废弃物，其中禽类粪污含氮有机质富集最为显著，加剧了氮素向大气与水体扩散，成为全球碳氮循环失衡与畜牧业氮素温室气体减排的核心瓶颈[1, 2]。传统粗放堆沤与粪污末端治理不仅易造成严重养分流失与大肠杆菌、寄生虫等病原体外泄引发的生物安全隐患[3, 4]，而且难以满

足向低碳循环转型的战略要求。工业好氧堆肥生物反应器通过受控的高温发酵环境，加速有机质矿化、腐殖化并彻底灭活病原微生物，是将畜牧业有机废弃物转化为优质有机肥的核心方法[5, 6]。

然而工业好氧堆肥反应器是一个由微生物、固体、液体和气体共同构成的复杂系统，各因素相互影响，并且响应曝气操作常具延迟。在反应微观尺度上，质量含水率、碳氮比（C/N）与温度梯度紧密耦合，共同调控着氨化、硝化与反硝化微生物的代谢速率与种群丰度[7, 8]。当发酵环境进入 >55℃ 的高温期并伴随周期性强制曝气扰动时，更多氮会转化为易挥发的游离氨，并在短时间内进入气相，导致氨气浓度迅速升高，导致氮素以瞬态脉冲的形式剧烈向气相逸散，最高可致 30%~60% 的总氮流失[9, 10]。不仅严重削弱了堆肥产品农用价值，还带来了恶臭二次污染与大气细颗粒前体物排放。

当前工程应用主要依赖经验查表法与事后取样检测来调整参数，缺乏对反应器内部气体动态演化的预测与闭环调控能力[11, 12]。尽管物联网多源传感技术的普及为过程监测提供了高频数据流，但传统纯数据驱动的时序深度学习模型属于“黑箱”架构，在面对剧烈曝气扰动时极易输出违背物料平衡关系的负值或非物理畸变，且无法为现场工艺调优提供透明的生化机理归因[13, 14]。因此，将生化反应常微分方程（ODE）与先进深度时序表征架构深度融合，发展兼具物理约束与高保真预测能力的可解释人工智能（XAI）模型，实现在线预测与生化机理方程逆向提取，已成为突破有机废弃物生物反应器精准减排与自主调瓶颈的迫切需求与核心科学前沿。

**1.2. Problem statement** (问题陈述)

针对生物反应器发酵过程中氨气浓度变化的预测与调控，现有研究主要采用两类技术路线。两类方法各具优势，但在适应复杂工况、预测精度和可解释性方面仍存在局限，难以单独满足工业精准控制的需求：

第一类是以 Monod 型微分方程、阿伦尼乌斯反应动力学及多相质量平衡为核心的白盒机理模型[15, 16]。这类模型以生化反应规律和物料平衡关系为基础，具有良好可解释性。但生物反应器内的物料成分和局部状态差异较大，且运行条件会频繁变化，模型参数难以始终保持适用，预测精度易受影响[17]。实际反应器内的温度和质量含水率存在空间差异，微生物活性及气－液－固多相传质过程会随之变化；曝气事件还会进一步改变气体交换条件。因此，采用固定参数的机理模型难以准确预测气相 $NH_3$ 浓度的突增幅度及其响应时间。对于包含多种快速和缓慢过程的复杂动力学方程时，存在数值求解不稳定与计算耗时较长的问题，从而限制实时应用[11]。

第二类是以循环神经网络（RNN/GRU）、时间卷积网络（TCN）及 Transformer 架构为代表的纯数据驱动黑盒模型[13, 18]。此类模型依托物联网多源传感器数据，展现出捕获长周期时间依赖与非线性特征的优异表征能力。然而纯数据驱动模型缺乏生化反应基本定律与物料平衡边界的先验约束[19]。在面对突发曝气事件时，黑盒模型易输出违背生化反应规律数值，如输出物理边界违背率（PVR）> 0 的负浓度预测，或突破物料平衡关系的畸变数值[14]；其高维复杂的权值矩阵存在“黑箱不可解释性”，无法解构为显式物理方程，致使其既无法为生化反应机理提供透明归因，亦无法在计算资源受限的 MCU 中实现轻量化部署与自主闭环控制[20]。

上述白盒“机理严谨但参数刚性、求解发散”与黑盒“拟合能力强但物理失真、黑箱不可解释”的两极矛盾，构成了当前有机废弃物发酵工程与时序深度学习交叉领域的核心科学知识缺口[21]。如何构建一种能够显式融合生化反应规律和物料平衡关系、抵抗工况瞬态扰动并具备显式方程提取能力的“神经符号物理信息（Gray-box PINN + Symbolic AI）”范式，打破传统机理模型与深度学习的二元壁垒，已成为实现固废生物反应器可信预测与低氮自主控制的核心前沿命题。

### 1.3. Proposed framework (提出框架)

本文提出 PINN-PatchTST-GPR(Ours) 框架，将物理约束、时间序列预测和曝气事件校正结合起来，用于预测鸡粪好氧发酵过程中的 $NH_3$ 浓度变化；符号回归进一步辅助解释不同工况下的预测关系。PatchTST 模型以多变量时间序列的回看窗口为输入，通过分块和通道独立设计长程依赖[13]；主输出为正常时期 $NH_3$ 浓度预测，事件分支只在曝气事件下修正曝气诱发的 $NH_3$ 浓度突增响应。

训练过程中，模型不仅要减小预测 $NH_3$ 浓度与传感器数值之间的数据拟合误差，还要满足两类物理约束：一是减小 ODE 残差，使预测的变化趋势尽量符合设定的反应关系；二是施加非负约束以及正常和曝气工况下的上界约束，避免预测突破物料平衡关系的畸变数值。各项误差按设定权重合并为总损失，其中数据拟合损失还包括曝气事件响应误差。在已知曝气日程的门控下，GPR 专家仅使用训练集的曝气事件样本，学习实际事件响应相对于经验事件倍率 $R_{emp}$ 的对数残差，并将该残差转换为修正后的事件倍率 $R_{pred}$，用于调整曝气事件 $NH_3$ 浓度预测。分块分割保形预测（block-aware split conformal prediction, BASCP）模块使用校准集残差构建名义 90% 的预测区间，用于表示模型在当前测试条件下预测 $NH_3$ 浓度时可能出现的误差范围[22]。PySINDy/PySR 作为事后符号代理，分别对正常期的延迟映射和曝气事件的相对响应进行低复杂度拟合[19,23]。因此在预测起点已知的曝气日程门控下，Ours 生成 $NH_3$ 多步点预测，并由事件分支对曝气响应进行相对修正；分割保形校准模块使用校准集使用构建名义 90% 的预测区间；PySINDy/PySR 则提供独立于深度网络的辅助符号表示。

### 1.4. Contributions (贡献)

本文的贡献集中在预测模型架构、不确定性量化、符号回归、氨气减排四个层面。本文提出 Ours 将 PatchTST 多步 $NH_3$浓度预测、事件触发 GPR 修正以及 ODE 软约束置于同一冻结评估协议。在六反应器上，Ours 取得 $R^2 = 0.9817 \pm 0.0006$、$RMSE = 0.8256 \pm 0.0132$ ppm，Raw PVR 和 Total PVR 均为 0。对曝气事件评估得到 $E_{peak} = 1.6793 \pm 0.0725$ ppm 和 $E_{time} = 0.7 \pm 0.2$ min，相对 Persistence 的滚动 RMSE 技能降低 82.85%。跨反应器 Fold A/B 的 CCC 为 0.9886（95% CI: 0.9874–0.9897）。采用 BASCP 模块输出名义 90% 预测区间，并在当前时间依赖测试协议下报告其经验覆盖率（PICP），用于评估预测不确定性。总体 $PICP_{90}$ 为 99.792%（五个计算种子均值 ± SD：99.792 ± 0.047%）。PySINDy/PySR 进一步将正常与曝气期相对响应压缩为候选符号关系。PySINDy 的 RMSE 和 $R^2$ 分别为 0.4624 ± 0.0758 ppm 和 0.9896 ± 0.0035，PySR 的 RMSE、 $R^2$、峰值误差 $E_{peak}$和达峰时滞 $E_{time}$ 分别为 2.4084 ± 0.0864 ppm、0.9523 ± 0.0044、1.5554 ± 0.2592 ppm 和 2.11 ± 1.17 min。

## 2. Materials and methods (材料与方法)

### 2.1. Experimental setup (实验设置)

#### 2.1.1. Bioreactor operation (反应器运行)

本次数据采集实验构建了一套由 6 个 20 L 的密封不锈钢发酵桶组成的生物反应器，初始装填 16 L。实验划分 3 组，每组设 2 个平行重复。控制系统包含 1 台协同自研控制柜、6 个无线节点箱与 6 组温度、湿度、二氧化碳、氨气传感器。每日 10:00 与 16:00 按组分梯度曝气 10/20/30 分钟，抽气气泵采用 10min 通断循环处理。曝气气泵将空气输入至发酵桶底部，穿透堆体后从顶层通气孔吹入内含 4-6mm 干燥剂与分子筛的干燥装置后，由节点箱内的抽气泵送入 $CO_2$ 传感器。反应堆配置如图 1 所示。

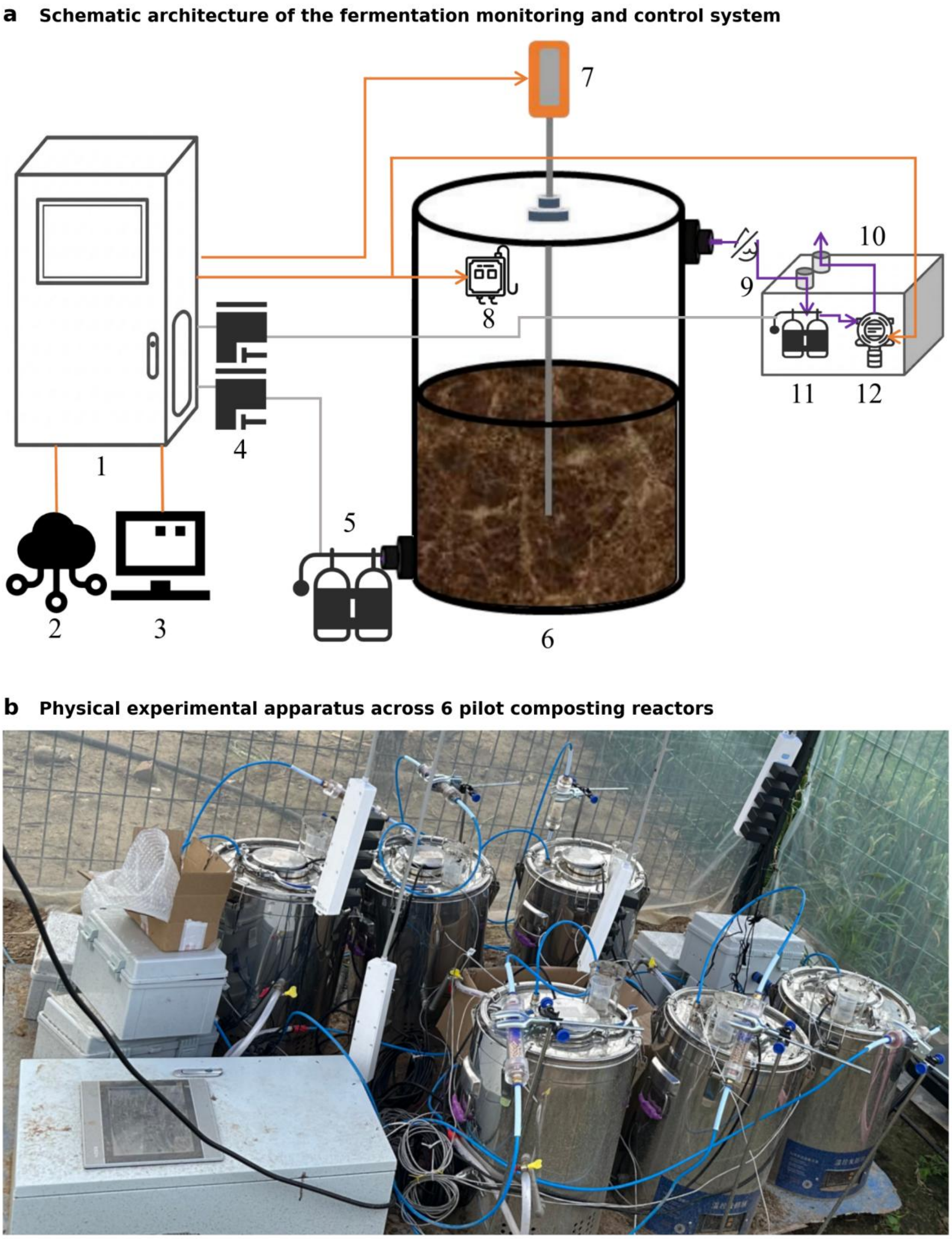

图 1 好氧堆肥实验装置的系统架构。（1）控制柜；（2）云端；（3）显示屏；（4）继电器；（5）曝气气泵；（6）发酵桶；（7）温湿度传感器；（8）氨气传感器；（9）干燥剂、分子筛；（10）防水盒；（11）抽气气泵；（12）二氧化碳传感器

控制柜 MCU 型号为 STM32F407，部署 FreeRTOS 实现监测传感器数据、驱动气泵继电器与多任务微秒级抢占调度；驱动 12 路继电器，控制 6 个曝气速率为 12L/min 的 KZP12-B24 曝气气泵与 6 个抽气速率为 0.5L/min 的 EDLP600-D24B 抽气气泵。各无线节点箱通过 RS-485 与控制柜通信。使用 RS-232 连接昆仑通态 TPC7032Nt 组态屏与 MCU，能够在本地端实时监测，并导出传感器数据。通过 RS-485 连接有人云网端，实现远程监测。使用四氟套管 PT100 温度传感器（精度±0.1℃）与 SN-3000SD-TR-N01 湿度传感器（精度±3% RH）

插入堆体中心，桶内上方空间放置扩散式 SN-3002-NH3-N01-1000P-2 电化学 $NH_3$ 传感器（精度±3%），IP68 级节点箱内安装 RS-485 无线节点与泵吸式 JX-$CO_2$-103-100W 红外 $CO_2$ 传感器（精度±5%F.S,25℃），保障了发酵高温、高湿、腐蚀环境下的数据采集。

在北京市小汤山国家精准农业研究示范基地鸡舍内，采集鸡粪 20kg，筛除羽毛与饲料后粉碎均质化为 4 目颗粒，并购买徐州 U8259 号稻壳，按鸡粪: 干稻壳 : 鲜稻壳 : 水 = 5 : 1 : 1 : 1 质量比混合；在装填及翻堆时用木棍插成 30 个直径 2cm 的垂直通气孔，增大混合物表面积。发酵桶底盘设定 55℃恒温。传感器采样周期设为 10 min/次，在 28 天发酵周期内连续产出 4033 组传感器数据。

**2.1.2. Analytical methods** (分析方法)

本实验严格遵循中国农业行业标准《有机肥料》《Organic fertilizer》（NY/T 525-2026）与粪便无害化卫生要求《Hygienic requirements for harmless disposal of night soil》（GB 7959-2012）。28 天发酵周期中，在第 0、1、7、14、21、28 天采集 250g 鲜样；每 7 日添加 1 次按总重 0.05% 君德复合微生物菌剂，进行人工翻堆，均匀混合。采样后检测干鲜样：取 100 g 鲜样在室内自然风干后，用粉碎机粉碎成 110 目干样,避光密封排气保存，测定总碳（TC）与总氮（TN）；取 125g 鲜样于 4℃避光排气冷藏，测定氨态氮与硝态氮（$NO_3^- - N$）以探究氮素与腐熟进程。使用京研春秋绿 2 号黄瓜种子进行种子发芽指数（GI）实验；取 5g 鲜样加入 50 mL 超纯水，使用厚塑料膜密封后用振荡机在 200r/min 下振荡 120 min，取出后使用离心机在 3000r/min 下离心 15 min，取 20 mL 上清液测定 pH 值与 GI。

**2.1.3. Data preprocessing (数据预处理)**

为保证 PySINDy 与 PySR 提取高精度公式，对传感器数据执行降噪与 $CO_2$ 归一化处理，加入质量补偿乘数，消除人工采样误差。反应器按时间轴划分训练集：验证集：校准集：测试集 =6：1：1：2，生成共 144360 行测试切片。Table3 针对跨发酵桶泛化，设计 1,3,5 前 80%训练，2,4,6 后 20%测试的 FoldA 与 2,4,6 前 80%训练，1,3,5 后 20%测试的 FoldB，研究数据泛化能力。因为本次实验仅翻堆 3 次，且每次翻堆只影响 1 个时刻，故在翻堆时刻进行非线性插值补偿处理。

**2.1.4. Computational environment（计算环境）**

在 Ubuntu 20.04.6 LTS 服务器上完成模型训练。服务器配备 Intel Xeon Silver 4210R @ 2.40 GHz 处理器和四张 NVIDIA GeForce RTX 4090 GPU。NVIDIA 驱动版本为 570.133.07，驱动支持的 CUDA 兼容版本为 12.8；与训练日志对应的 Conda 环境 tsai_env 使用 Python 3.10.20、 PyTorch 2.1.0+cu118、CUDA runtime 11.8 和 cuDNN 8.7.0。模型训练采用批量大小 32、Adam 优化器和 $10^{-4}$ 的初始学习率，最多训练 30 个 epoch，早停耐心值设为 6。实验使用 2021–2025 五个随机种子重复运行，各训练任务绑定单张 GPU。

在统一计算环境下，本文分别运行全流程预测普通与曝气事件配置。各配置共享数据预处理、输入输出长度、随机种子、批量大小和优化器设置；模型间的差异仅由基线架构、事件专家、滚动策略或物理约束开关构成，具体配置及对应评估协议列于 Table 1–6。

**2.2. Biochemical reaction modeling** (生化反应建模)

为了给后续数据驱动的时间序列架构建立严格的物理边界，本研究基于物料平衡关系，对好氧发酵生物反应器顶空区域的气体生化反应机理建模 [24,25]。设置基准自由体积设定为$V_{free}$ = 4.0 L。基于广义连续搅拌釜反应器（Continuous Stirred Tank Reactor, CSTR）质量平衡原理，反应器顶层空间瞬态氨气浓度$C(t)$（单位：ppm）的连续演化受常微分方程（ODE）主导 [25]：

$$V_{free}\frac{dC(t)}{dt} = G_{bio}(T, CO_2) - Q_{air}(t) \cdot C(t)$$

其中$G_{bio}(T,CO_2)$为生物源 $NH_3$ 生成速率（mg/min）；$Q_{air}(t)$为曝气体积流量（L/min）作为阶跃型变量，在曝气和正常阶段分别设为 12.0 和 0 L/min[25]。在 28 天的发酵周期中，固体有机物降解和物料压实导致堆体沉降，顶空自由体积呈现缓慢物理膨胀 [24]。在本研究的质量平衡表达中，气体传输和相间平衡通过浓度与通风流量描述。由于顶层空间的实时变化未被单独测量，令$V_{free}$ = 4.0 L，并由时间序列模块从连续传感器记录中学习可能的短时波动。该处理避免引入无法观测的体积变量，同时保持计算鲁棒性。

在多孔鸡粪基质的高温好氧发酵系统中，$G_{bio}$受堆体中心物料温度 T 与异养微生物呼吸代谢强度共同主导 [25]。基于温度对反应速率的影响规律，修正 Arrhenius 关系，并用 $CO_2$ 浓度反映微生物呼吸活动，模型将生物源 $NH_3$ 生成速率表示为温度与 $CO_2$ 浓度共同作用的非线性函数[24]：

$$G_{bio} = \alpha \cdot exp(k_1(T - T_{base})) \cdot \ln(CO_2)$$

式中：$\alpha = M_{mat} \cdot \alpha_{specific}$（令初始装填物料总重$M_{mat}$ = 10.5 kg， $\alpha_{specific}$为单位生物量对应的产氨参数）；$T_{base}$=31.0℃为计算温度效应时采用的参考温度；$k_1$为产氨速率对温度变化参数；$\ln(CO_2)$表示微生物呼吸增强对产氨反应的促进作用。

深度神经网络在面对强扰动工况时，因缺乏显式物理约束，极易产生负浓度预测，或突破物料平衡关系的畸变数值[14]。本研究根据系统长期运行后趋于稳定的状态，以及 $NH_3$ 在气-液-固相之间交换时的平衡关系，推导出模型允许的动态浓度上下限［$C_{min}$,$C_{max}$］[18,27]。其中物理下限$C_{min}(t)$固定为不能低于零的浓度。

动态浓度上限$C_{max}(t)$根据通风状态$Q_{air}(t)$解耦为两个互斥的物理方程[23,26,27]：

$$C_{max,aerate}(t) = \frac{G_{bio}(T,CO_2)}{Q_{air}(t)}, Q_{air}(t)>0$$

$$C_{max,static}(t) = \beta \cdot exp(k_2(T - 60.0)), Q_{air}(t)=0$$

曝气阶段，持续通风会快速带走顶空中的 $NH_3$，使其浓度接近由通风稀释所决定的模型上限$C_{max,aerate}(t)$，此时可近似认为浓度变化率$\frac{dC(t)}{dt} \approx 0$[26]；在正常阶段，$NH_3$ 极缓被气流持续带出，顶空浓度主要由堆体水分与气体之间的分配平衡决定，并采用 Henry 定律进行描述[27]。本研究以温度 60℃为基准，分别标定出基准饱和经验参数。各实验平行组基准参数按 1-2 号、3-4 号及 5-6 号反应罐组，分别标定为$\beta$ = [ 521.0 , 274.0 , 367.0 ] ppm。

### 2.3. Network architecture (网络架构)

#### 2.3.1. Patch-based representation (基于 Patch 的表征学习)

在每个 10 min 采样时刻构造 8 维过程输入，而非仅使用$M_{\mathrm{raw}} = 4$个原始传感通道。原始观测包含温度 Temp、湿度 Humid、$CO_2$ 和 $NH_3$ 4 个传感器数据。根据每日定时曝气时刻，数据加载程序同步生成事件门控 m、$R_{emp}$、归一化事件相对时间 $\tau$ 和反应器号码 Group_ID 四个辅助变量。这些变量由预测起点可获得的时间信息、反应器身份和曝气日程确定。令长度为 L 的输入窗口为 $Z \in R^{L\times C_{\mathrm{in}}}$，则第 t 个输入向量定义为：

$$\mathbf{z}_t = \left[T_t, \mathrm{Humid}_t, \mathrm{CO}_{2,t}, \mathrm{m}_t, R_{\mathrm{emp},t}, \tau_t, \mathrm{Group_ID}_t, \mathrm{NH}_{3,t}\right]^{\mathrm{T}} \in \mathbb{R}^{C_{\mathrm{in}}}, C_{\mathrm{in}} = 8$$

发酵开始后 3d 内进入高温阶段作为发酵初步成功标志，并据此采用连续 432 个采样点构成历史窗口，既覆盖一个完整日变化，也能够容纳 6 个曝气周期。模型一次输出未来 6 个时间步，即预测起点之后 10－60 min 的 $NH_3$ 浓度轨迹。这一输出面向曝气决策所需的短时前置信息，同时避免将过长预测范围内尚未观测到的工况变化误归因于模型本身。

本研究采用长度为 16 步的 Patch，这一范围足以包含曝气前后的 $NH_3$ 变化。移动步长设为 8 步，形成 50%的重叠，使相邻片段共享部分边界信息，降低短时峰值恰好落在切片

边界时，模型认为事件不连续。因此 432 步历史在每个通道上形成 53 个有效 Patch token，数量由本研究的窗口长度、Patch 长度、移动步长共同决定。

各通道的 Patch 经过共享线性嵌入后映射到 64 维隐空间，再与可学习的位置编码相加并送入 Transformer 编码器。64 维提供了足以表示温度、湿度、气体浓度及事件变量局部模式的中等宽度，同时控制参数量，避免在仅有六个反应器和有限时间跨度的数据上使用过宽表示。位置编码用于保留 Patch 的先后顺序，也是 Transformer 在处理无循环序列表示时恢复时间位置信息的标准做法[28]。编码器设置为 3 层，目的是在有限样本下逐步提取局部到较长范围的时间依赖，而不是无依据地增加网络深度。4 个注意力头可将 64 维表示均分为每头 16 维，使不同头分别关注发酵趋势、滞后、事件邻域数值变化等潜在关系；256 维前馈层约为模型宽度的 4 倍，为非线性变换提供必要容量。dropout 设为 0.1，用于抑制小样本训练中的过拟合。

模型采用通道独立 PatchTST 设计：每个变量先分别完成分块和编码，所有通道共享嵌入层、Transformer 编码器和展平预测头。共享参数可以减少为每个变量单独配置网络所带来的参数膨胀，并使不同变量在相同时间建模规则下获得可比较的表征。使用可逆实例归一化（RevIN）处理窗口内的尺度差异，关闭可学习仿射变换，并以窗口统计量完成归一化和反归一化，减轻不同反应器和不同发酵阶段的量纲漂移。最终，展平预测头汇总各通道的编码结果输出未来 6 步的正常事件 $NH_3$ 浓度轨迹。

**2.3.2. Physics-informed residual learning** (物理信息残差学习)

纯数据损失仅度量预测值与实测值之间的距离，不能直接识别负浓度、突破物料平衡关系的畸变数值。为在训练阶段引入可解释物理约束，本节将 2.2 节定义的浓度边界和气相质量平衡关系作为训练阶段的软约束。物理信息机器学习通过可微残差，将已知方程或不等式引入数据驱动模型，并优化拟合数据[14, 18, 20]。程序将五类物理公式单边惩罚合并为：

$$\mathcal{L}_{\text{boundary}} = \lambda_{\text{boundary}}\left(\mathcal{L}_{\text{lower}} + \mathcal{L}_{\text{upper,static}} + \mathcal{L}_{\text{upper,aerate}} + \mathcal{L}_{\text{upper,event}} + \mathcal{L}_{\text{upper,ratio}}\right)$$

其中$\mathcal{L}_{\text{lower}}$在全时段惩罚负浓度；$\mathcal{L}_{\text{upper,static}}$ 仅在事件门控 $m_t$=0 且不处于过渡边缘时启用；$\mathcal{L}_{\text{upper,aerate}}$在 $Q_{air} > 0$ 的曝气时刻启用；$\mathcal{L}_{\text{upper,event}}$和$\mathcal{L}_{\text{upper,ratio}}$在 $m_t \geq 0.5$ 的曝气事件时刻约束 $NH_3$ 浓度与响应倍率。倍率上限设为 $R_{emp}$+ 0.35。

对于含时间导数的生化反应关系，已有研究采用物理信息神经网络（PINN）将控制方程残差的均方损失加入训练目标，使模型输出在采样点上近似满足给定的物理关系[18]，采用后向 Euler 或有限差分形式构造离散物理残差[29,30]。在此基础上，本文将第 2.2 节的气相质量平衡关系离散到相邻采样间隔$\Delta t$= 10min 的采样网格上，令 t 表示当前预测时刻，t-1 表示其前一个采样时刻；$C_t$ 和 $C_{t-1}$ 均为事件响应组合后、反标准化的 $NH_3$ 预测浓度。使用一阶后向差分近似浓度变化率，并定义离散质量平衡残差：

$$\mathcal{R}_{\text{ode},t} = V_{\text{free}}\frac{C_t - C_{t-1}}{\Delta t} - G_{\text{bio}}(T_t, CO_{2,t}) + Q_{\text{air},t} \cdot C_t$$

式中 $V_{\text{free}}\frac{C_t - C_{t-1}}{\Delta t}$ 表示气相中 $NH_3$ 浓度变化对应的储存通量。$T_t$ 为时刻 t 的温度，先由标准化输入还原为摄氏度；$CO_{2,t}$ 为同一时刻的二氧化碳实测值，归一化后进入 $G_{bio}$。$G_{\text{bio}}(T_t, CO_{2,t})$表示第 2.2 节定义的生物生成项，其函数形式和温度敏感参数由第 2.2 节给出。$Q_{\text{air},t} \cdot C_t$ 表示曝气造成的气相移除通量。因此当氨气存储通量等于生物源通量减去曝气移除通量时，$\mathcal{R}_{\text{ode},t} = 0$，非零残差表示预测轨迹与该质量平衡关系之间存在偏差。

参考 PINN 中对控制方程残差进行均方最小化的标准损失构造[18]，对所有有效差分对进一步计算归一化 ODE 损失：

$$\mathcal{L}_{\text{ode}} = \frac{1}{|\mathcal{M}_{\text{ode}}|}\sum\nolimits_{t\in\mathcal{M}_{\text{ode}}}\left(\frac{\mathcal{R}_{\text{ode},t}}{Scale_{\text{ode}}}\right)^2$$

其中，$M_{ode}$是有效差分对的索引集合，$|M_{ode}|$表示集合中差分对的数量。物理残差仅在相邻两个采样时刻处于相同曝气状态和事件状态时计算。气泵启停以及事件门控切换产生的边界差分不纳入该项损失，避免控制状态突变造成误差。$Scale_{ode}$由储存、生成和移除通量绝对值的均值构成，并在反向传播中停止梯度。归一化后的残差平方使不同发酵时期和工况下的物理项保持相近量级。$\frac{1}{|\mathcal{M}_{ode}|}$为数学简写，实际对这些有效点的归一化 ODE 残差取掩码平均，分母 $+10^{-5}$避免有效点数为零时数值错误，正常情况下不影响结果。若预测序列不足两个时间点，或该批次不存在有效差分对时，则 $L_{ode}=0$，即跳过该批次的 ODE 约束。$L_{ode}$ 为预测轨迹的软非负项，其与简化的 $NH_3$ 生成-去除关系相关。

在数据项与物理残差联合优化的 PINN 框架基础上，本文根据正常期预测、曝气事件响应、$E_{peak}$ 的评估目标，构造全局训练目标为：

$$\mathcal{L}_{\text{total}} = \lambda_{\text{raw}}\mathcal{L}_{\text{raw}} + \mathcal{L}_{\text{normal}} + \lambda_{\text{event,raw}}\mathcal{L}_{\text{event,raw}} + \lambda_{\text{event,ratio}}\mathcal{L}_{\text{event,ratio}} + \lambda_{\text{event,peak}}\mathcal{L}_{\text{event,peak}} + \lambda_{\text{trans}}\mathcal{L}_{\text{trans}} + \gamma(e)\cdot\left(\mathcal{L}_{\text{boundary}} + \lambda_{\text{ode}}\mathcal{L}_{\text{ode}}\right) + \mathcal{L}_{\text{param}}$$

数据项分别覆盖全域数值、正常期数值、曝气事件数值、曝气事件倍率、曝气事件峰值和曝气前后过渡响应。主实验取$\lambda_{\text{boundary}} = \lambda_{\text{ode}} = 0.01$；$\gamma(e)$在前 5 轮取 0，随后 5 轮线性增加至 1 并保持不变，$\mathcal{L}_{\text{param}}$保持学习到的物理参数合理，使公式非负。物理限制预测不违反浓度边界、简化 ODE 变化规律及参数非负条件。各项损失通过权重系数调节相对重要性，物理约束在训练过程中逐步增强。模型不仅实现精准预测，还能够正确描述正常状态、曝气响应、峰值变化和曝气前后过渡过程。

#### 2.3.3. Block conformal prediction（分块保形预测）

分块设计用于减少把相邻且相互关联的 10 min 步长误当成彼此独立的样本。按照随机种子和反应器号码，将校准集数据划分为互不重叠的 1 h 时间块。5 个随机种子分别建立 6 个预测步长和 2 个工况层，共 60 个校准器。每个校准器只使用与其预测步长和工况相对应的数据层,对于每个随机种子，6 个事件步长校准器各包含 66 个时间块；正常期校准器各包含 402 个时间块，曝气期校准器包含 408 个时间块。$PICP_{90}$ 是预先设定的名义覆盖目标，表示按分割保形方法构造的区间能够覆盖 90% 的实测值。有限样本分割保形方法先在校准集数据中计算每个时间块的最大绝对误差，再将这些误差从小到大排列，选取位于目标 PICP 位置内的误差值，并把它作为点预测上下界[22]。本文在每个预测步长和工况内，按照有限样本排名规则计算误差半径：

$$q_{h,r} = s\left(\min\left\{n_{h,r}, \lceil (n_{h,r}+1)(1-\alpha) \rceil\right\}\right)$$

其中 $n_{h,r}$ 是该层的有效校准块数，s(k) 表示排序后的第 k 个误差，$\alpha_{\text{CP}}=0.1$，r 为当前工况。排名被限制在 $[1,n_{h,r}]$ 内，避免程序读取不存在的排序位置。最终预测区间为：

$$C_{t,h}^{(1-\alpha_{\text{CP}})} = \left[y_{t,h} - q_{h,r}, y_{t,h} + q_{h,r}\right]$$

其中 $y_{t,h}$ 是模型给出的 $NH_3$ 点预测值。

#### 2.3.4. Symbolic regression (符号回归)

本节将符号回归作为事后可解释性分析。正常与曝气期的 $NH_3$ 浓度变化方式不同：前者极缓慢变化，后者在短时间内出现明显突增与下降。因此不能用同一套符号回归公式评价两个阶段。所以点预测模型固定为正常期使用 PINN–PatchTST，曝气期使用 GPR 事件专家；两者按曝气时刻进行拼接。

正常期 PySINDy 以冻结 PINN－PatchTST 的单步输出作为回归目标，查看当前时刻之前连续 432 个步长的历史信息，随后在验证集中比较多种设置，选择最能复现 PINN－PatchTST 输出、同时公式尽量简洁稳定的参数组合。PySINDy 的离散延迟映射属于 SINDy

类数据驱动符号建模方法[19]，其一般形式为：

$$\dot{x}(t)\Theta(x(t))\Xi$$

其中，$x(t)$ 是系统状态，$\dot{x}(t)$ 是状态变化率，$\Theta(x)$ 是由候选变量和非线性函数组成的函数库，$\Xi$ 是稀疏系数矩阵。PySINDy 通过筛选少量非零系数，得到较为简洁的候选状态演化关系。在本研究中，该方法被用于拟合冻结 PINN–PatchTST 的离散延迟映射；因此所得表达式是离散时间下的候选预测关系。PySINDy 衡量的是公式对冻结教师输出的单步复现，其作用是把正常期预测关系写成可逐项检查的候选公式，而不是重新确认一条普适的生化反应机理模型。曝气期 PySR 的输入向量定义为：

$$\mathbf{x}_i = \left[g_{1,i},\ g_{2,i},\ g_{3,i},\ \tau_i,\ T_i,\ \mathrm{CO}_{2,i},\ R_{\mathrm{emp},i},\ \frac{h_i}{6}\right]^{\mathsf{T}} \in \mathbb{R}^8$$

其中 g1,g2,g3 为反应器组别的二值识别码，若当前反应器属于该组则对应变量为 1，否则为 0，三者满足 $g1 + g2 + g3 = 1$；$h \in 1,\dots,6$ 为预测步长，其中 $h_i/6$ 为归一化预测步长。PySR 将符号表达式发现视为预测误差与表达式复杂度之间的权衡问题[31]。本研究据此论文内的方法公式，在验证集上选择能够较好复现教师模型、同时保持结构简洁和数值稳定的候选表达式。PySR 得到的表达式仅用于近似 GPR 的事件响应并提供可审查的符号表示，最终曝气预测仍由 GPR 完成。

上述公式表示离散时间下的延迟映射或正常与曝气期响应近似关系。逐步预测过程中，模型输出会被恢复到实测值尺度，并将负预测值限制为零，这两步属于数值处理。在五个随机种子下分别保存表达式及其系数，用于评估随机种子引起的计算变异。由于不同种子可能得到不同的表达式结构，正文将这些结果表述为候选符号关系。符号回归的指标不与完整流程的最终预测指标直接等价比较。

### 2.4. Evaluation metrics (评估指标)

所有评估指标均先固定实测值及其时间划分，再将每个反应器在各目标时刻的预测值与实测值逐一配对，分别报告完整测试集中正常与曝气期结果。五个种子的均值 ± 标准差反映计算过程中的随机变异。

本文以 RMSE 作为主要误差指标，对较大的预测偏差给予更高惩罚。$R^2$ 作为补充指标，用于说明模型能够解释测试数据波动的程度。MAE 对少数 $E_{peak}$ 的额外惩罚较弱，且当 $NH_3$ 浓度接近零时，MAPE、sMAPE 和 WAPE 的分母过小，结果不稳定，因此不适用于该模型[32]。MASE 用于比较不同反应器和浓度范围下的预测误差，其基准误差仅根据训练集内相隔 24 h 的观测值计算，MASE < 1 表示模型的平均误差小于该基准[29]。全时段样本中正常期占比超 93%，直接计算总体误差导致掩盖曝气后短时间内的 $NH_3$ 浓度变化。曝气事件增量指标公式如下：

$$R^2_{\Delta,\mathrm{event}} = 1 - \frac{\sum_{i\in E}\left[\left(\hat{y}_i^{\mathrm{event}} - \hat{y}_i^{\mathrm{base}}\right) - \left(y_i^{\mathrm{raw}} - y_i^{\mathrm{clean}}\right)\right]^2}{\sum_{i\in E}\left[\left(y_i^{\mathrm{raw}} - y_i^{\mathrm{clean}}\right) - \overline{\Delta y}_E\right]^2}$$

在事件样本集合 E 内，本文将第 i 个样本的曝气事件模型预测值 $\hat{y}_i^{\mathrm{event}}$ 与正常期预测值 $\hat{y}_i^{\mathrm{base}}$ 之差定义为预测曝气增量，将实测值 $y_i^{\mathrm{raw}}$ 与非线性插值得到的参考浓度 $y_i^{\mathrm{clean}}$ 之差定义为观测曝气增量，并对两组增量计算标准决定系数 $R^2_{\Delta,\mathrm{event}}$。该决定系数以事件集合中观测曝气增量的平均值 $\overline{\Delta y}_E$ 为基线，比较预测增量误差与观测增量总变异，从而衡量模型对曝气浓度变化的解释能力。

$E_{peak}$ 和 $E_{time}$ 分别从两个方面评价模型对连续曝气事件的预测：$E_{peak}$ 比较模型预测的最高 $NH_3$ 浓度与实际最高浓度差；$E_{time}$ 比较预测峰值出现时间与实际峰值出现时间相差多少分钟。因此前者衡量浓度幅度失真，后者衡量时间位置失真。模型可能准确预测峰值大小

但提前或滞后出现，也可能峰值时间正确但浓度过高或过低，这两个指标可以将这两类误差区分开。

Wheatcroft 的综述讨论了“模型误差/参考误差”的 skill-score 结构[33];Mittermaier 专门讨论了以 Persistence 作为参考预报[34]；Mayer 和 Yang 则针对 RMSE 提出了相应的 RMSE skill-score 形式[35]。相对 Persistence 的 RMSE Skill 采用预报验证中常用的相对误差改进形式，并以百分数报告：

$$\text{Skill} = 100\left(1 - \frac{\text{RMSE}model}{\text{RMSE}Persistence}\right)$$

PVR 参考预测可行性[14,18]，在输出投影或截断前计算：

$$\text{Raw PVR} = 100N^{-1}\sum_t I\,(y_t < 0)$$

$$\text{Total PVR} = 100N^{-1}\sum_t I\,(y_t < Cmin,t \vee y_t > Cmax,t)$$

其中 $y_t$ 表示模型在第 t 个步长给出的预测值，N 为预测数量，I 为违反次数；Raw PVR 只检测非负，Total PVR 是适用下界、正常期/曝气期上界的并集，重复越界只计一次。校准器按“步长×工况”优先使用 60 min 非重叠残差块最大值 [36]。因为单报 PICP 会偏向过宽区间，单报 MPIW 会忽略覆盖，$IS_{0.1}$ 联合惩罚两者 [37,38]，所以同时报告 $PICP_{90}$、MPIW 和 $IS_{0.1}$。

## 3. Results and discussion (结果与讨论)

### 3.1. Predictive performance and ablation (预测性能与消融实验)

为检验模型能否同时应对正常和曝气期预测，本研究在六个反应器上比较了 10 种模型架构。表 1 汇总了各模型的总体预测误差、正常和曝气期的预测表现、曝气峰值及时滞误差以及 PVR 比例。

**Table 1**

表 1 基于严格时间顺序测试集的 10 种模型氨气排放预测性能综合对比（5 个随机种子均值 ± 标准差）。加粗表示最优结果，下划线表示次优结果。

| Model | Overall RMSE | Overall $R^2$ | $R^2_{normal}$ | $R^2_{event}$ | $E_{peak}$ | $E_{time}$ | MASE | Total PVR | Raw PVR |
|---|---|---|---|---|---|---|---|---|---|
| **PINN-PatchTST-GPR(Ours)** | **0.8256 ± 0.0132** | **0.9817 ± 0.0006** | **0.9744 ± 0.0011** | **0.9711 ± 0.0020** | **1.6924 ± 0.0837** | 0.8 ± 0.2 | **0.0124 ± 0.0001** | **0.0000 ± 0.0000** | **0.0000 ± 0.0000** |
| NLinear | <u>1.0217 ± 0.0128</u> | <u>0.9720 ± 0.0007</u> | <u>0.9622 ± 0.0011</u> | <u>0.9663 ± 0.0034</u> | <u>2.0224 ± 0.0961</u> | <u>0.7 ± 0.4</u> | <u>0.0155 ± 0.0002</u> | **0.0000 ± 0.0000** | **0.0000 ± 0.0000** |
| DLinear | 1.2613 ± 0.1565 | 0.9568 ± 0.0106 | 0.9355 ± 0.0195 | 0.9618 ± 0.0039 | 2.2889 ± 0.3536 | 0.9 ± 0.6 | 0.0205 ± 0.0037 | **0.0000 ± 0.0000** | **0.0000 ± 0.0000** |
| Linear | 1.5886 ± 0.3895 | 0.9290 ± 0.0314 | 0.8841 ± 0.0568 | 0.9579 ± 0.0087 | 2.3214 ± 0.4779 | **0.6 ± 0.0** | 0.0283 ± 0.0091 | 0.0007 ± 0.0015 | 0.0007 ± 0.0015 |
| PatchTST | 2.3797 ± 0.6682 | 0.8384 ± 0.0907 | 0.7641 ± 0.1497 | 0.8469 ± 0.0680 | 4.6632 ± 0.9769 | **0.6 ± 0.5** | 0.0246 ± 0.0033 | 0.1462 ± 0.1258 | 0.1462 ± 0.1258 |
| Transformer | 2.1742 ± 0.3931 | 0.8698 ± 0.0464 | 0.9020 ± 0.0439 | 0.6807 ± 0.1146 | 8.3061 ± 1.8652 | 5.6 ± 3.8 | 0.0304 ± 0.0075 | **0.0000 ± 0.0000** | **0.0000 ± 0.0000** |
| Persistence | 4.1089 ± 0.0000 | 0.5469 ± 0.0000 | 0.8672 ± 0.0000 | -0.2310 ± 0.0000 | 11.7500 ± 0.0000 | 10.4 ± 0.0 | 0.0352 ± 0.0000 | **0.0000 ± 0.0000** | **0.0000 ± 0.0000** |
| iTransformer | 12.3561 ± 1.2078 | -3.1283 ± 0.8150 | -4.1959 ± 1.1169 | -4.9720 ± 1.1122 | 40.1687 ± 3.3201 | 1.0 ± 1.2 | 0.2271 ± 0.0262 | 0.0887 ± 0.0933 | 0.0887 ± 0.0933 |
| Autoformer | 37.4413 ± 6.7746 | -37.6042 ± 13.3197 | -68.3513 ± 25.0614 | -10.8965 ± 1.8076 | 41.8262 ± 5.9536 | 13.1 ± 2.0 | 0.6766 ± 0.1365 | 7.5222 ± 3.3906 | 7.4072 ± 3.4127 |
| Informer | 56.9387 ± 8.7926 | -87.6596 ± 26.4911 | -160.7484 ± 48.6345 | -16.0002 ± 5.1563 | 42.5578 ± 7.5668 | 8.0 ± 2.4 | 1.2047 ± 0.1863 | 0.0048 ± 0.0072 | **0.0000 ± 0.0000** |

由表 1 可见，Ours 的总体 RMSE 为 0.8256 ± 0.0132 ppm，$R^2$为 0.9817 ± 0.0006，MASE 为 0.0124 ± 0.0001，均为当前比较中的最优值。相较次优的 NLinear，其 RMSE 按均值降低 19.2%；Persistence 的 RMSE 为 4.1089 ppm。曝气事件的 $E_{peak}$ 和 $E_{time}$ 分别为 1.6924 ± 0.0837 ppm 和 0.8 ± 0.2 min，为所有模型最优与次优。虽然 PatchTST 的 $E_{time}$ 最优，但是其存在 PVR 问题，且其他指标均不如 Ours。该优势在曝气期与全局 $R^2$的联合分布以及曝气峰值和时滞误差对比中进一步显现，如图 2 所示。

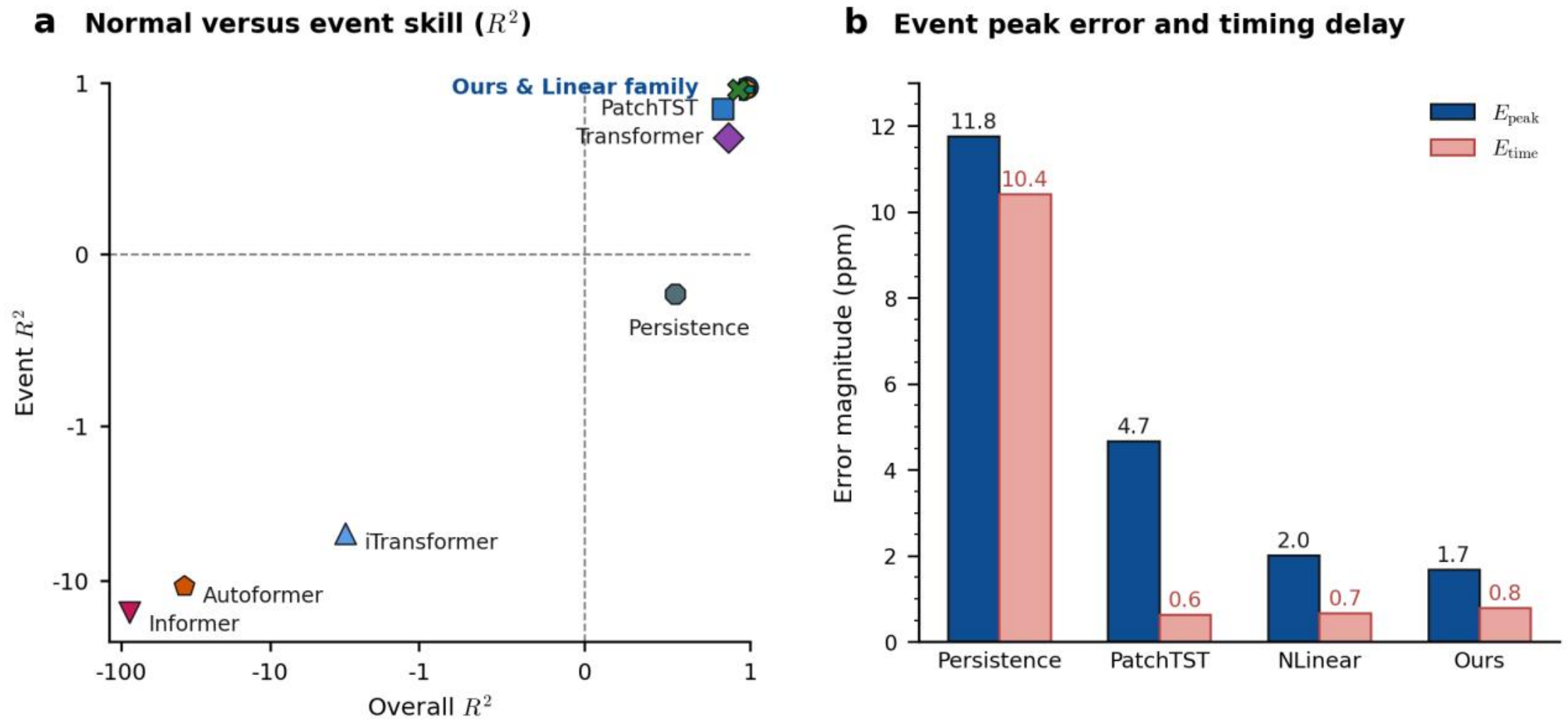

【Figure 2】

图 2a 显示，Ours 与线性模型位于总体和事件 $R^2$均在正区域内，而 Persistence 及部分 Transformer 系列模型在曝气期出现明显性能衰减。图 2b 进一步表明，Ours 的 $E_{peak}$ 约为 1.7 ppm、$E_{time}$ 为 0.8 min；Persistence 的对应误差分别为 11.8 ppm 和 10.4 min。图 2 说明总体精度差异主要受曝气扰动影响。在事件前单次预测评估中，模型在曝气开始前一次性预测未来 6 个时间步，但只评价第 1 个时间步精度。在滚动预测评估中，模型每到新的预测起点，都会根据当时可获得的信息重新预测未来 6 个时间步。两种评估的结果见表 2。

**Table 2**

表 2 各模型在曝气突变工况下的曝气前预测与滚动 h=1 步预测性能对比（均值 ± 标准差）。加粗表示最优结果，下划线表示次优结果。

| Model | Pre-aeration RMSE | Pre-aeration R² | Rolling RMSE | Rolling R² |
|---|---|---|---|---|
| **PINN-PatchTST-GPR(Ours)** | **2.1716 ± 0.0813** | **0.9697 ± 0.0023** | **2.1609 ± 0.0812** | **0.9700 ± 0.0022** |
| NLinear | 2.2973 ± 0.0514 | 0.9661 ± 0.0015 | 2.6236 ± 0.3261 | 0.9552 ± 0.0118 |
| DLinear | 3.0155 ± 0.2504 | 0.9412 ± 0.0098 | 3.0514 ± 0.5245 | 0.9387 ± 0.0221 |
| Linear | 4.0388 ± 0.8624 | 0.8913 ± 0.0463 | 5.0439 ± 1.7719 | 0.8204 ± 0.1116 |
| PatchTST | 4.3828 ± 1.0367 | 0.8710 ± 0.0613 | 5.7495 ± 2.3468 | 0.7592 ± 0.2130 |
| Transformer | 7.6547 ± 3.0101 | 0.5769 ± 0.3560 | 7.9368 ± 2.9748 | 0.5497 ± 0.3318 |
| Persistence | 16.1858 ± 0.0000 | -0.6836 ± 0.0000 | 12.6006 ± 0.0000 | -0.0204 ± 0.0000 |
| iTransformer | 31.3047 ± 4.6277 | -5.4081 ± 1.9004 | 30.1972 ± 4.9724 | -4.9873 ± 1.8844 |
| Autoformer | 45.4688 ± 9.4181 | -12.7424 ± 5.7017 | 39.7601 ± 3.8670 | -9.2365 ± 1.9018 |
| Informer | 52.5096 ± 7.7304 | -17.0270 ± 5.1860 | 50.2274 ± 7.9323 | -15.5365 ± 4.9948 |

【Table 2】

表 2 显示，Ours 在单次曝气前预测中的 RMSE 为 2.1716 ± 0.0813 ppm，在滚动预测中的 RMSE 为 2.1609 ± 0.0812 ppm；对应的 Persistence RMSE 分别为 16.1858 ± 0.0000 ppm 和 12.6006 ± 0.0000 ppm。NLinear 在单次曝气前预测中接近 Ours，其 RMSE 为 2.2973 ± 0.0514 ppm，但在滚动预测中升至 2.6236 ± 0.3261 ppm。为区分预测性能变化与物理损失项的作用，本研究固定 PatchTST 主干、GPR 事件专家及全部非物理损失，仅改变 ODE 残差项和动态边界项的权重，结果如表 3 所示。

**Table 3**

表 3 固定事件专家下，物理信息非负的消融实验对比（均值 ± 标准差）。加粗表示最优结果，下划线表示次优结果。

| Physics Configuration | Overall RMSE | $R^2_{normal}$ | Pre-event RMSE | $R^2$ Δ,event |
|---|---|---|---|---|
| Soft bounds only | **0.8239 ± 0.0120** | **0.9785 ± 0.0006** | **2.1625 ± 0.0766** | **0.9602 ± 0.0033** |
| **Full PINN (Ours)** | <u>0.8256 ± 0.0132</u> | <u>0.9744 ± 0.0011</u> | <u>2.1716 ± 0.0813</u> | <u>0.9600 ± 0.0037</u> |
| No physics | 0.8276 ± 0.0134 | 0.9783 ± 0.0005 | 2.1840 ± 0.0822 | 0.9598 ± 0.0038 |
| ODE only | 0.8277 ± 0.0134 | 0.9783 ± 0.0005 | 2.1846 ± 0.0829 | 0.9598 ± 0.0038 |

【Table 3】

由表 3 可见，Soft bounds only、Full PINN、No physics 和 ODE only 的总体 RMSE 分别为 0.8239 ± 0.0120、0.8256 ± 0.0132、0.8276 ± 0.0134 和 0.8277 ± 0.0134 ppm。四种配置的均值差异最大为 0.0038 ppm，处于五个计算种子离散程度的同一量级。因此在当前工况和权重设定下，物理项尚未显示出可单独归因的点预测精度增益，但完整配置也未表现出明显的精度损失。尽管 Full PINN 的 RMSE 不是最佳值，本文仍将其作为主配置，因为它同时保留 ODE 动态约束和浓度边界约束，能够在预测性能基本相当的条件下提供更完整的物理接口；这一选择表示模型设计取舍，而不是已证明 ODE 项带来了更佳预测。PVR 不参与本表配置排序，相关指标保留在 表 1 中。为检验模型能否将在部分反应器中学到的关系迁移到未参与训练的反应器，本研究另设按反应器划分的 Fold A/B 迁移测试，结果如表 4 所示。

**Table 4**

表 4 针对 Fold A/B 划分下的模型性能对比（5 个随机种子均值 ± 标准差）。加粗表示最优结果，下划线表示次优结果。

| Model | Fold A Overall RMSE | Fold B Overall RMSE | Equal Fold Mean RMSE | Fold A Overall $R^2$ | Fold B Overall $R^2$ | Equal Fold Mean $R^2$ |
|---|---|---|---|---|---|---|
| **PINN-PatchTST-GPR(Ours)** | **0.9671 ± 0.0403** | **0.8318 ± 0.0164** | **0.8994** | **0.9753 ± 0.0021** | **0.9811 ± 0.0007** | **0.9782** |
| NLinear | 1.0201 ± 0.0131 | 0.9974 ± 0.0051 | 1.0088 | 0.9725 ± 0.0007 | 0.9728 ± 0.0003 | 0.9727 |
| PatchTST | 1.8258 ± 0.2252 | 1.0623 ± 0.0347 | 1.4441 | 0.9108 ± 0.0230 | 0.9692 ± 0.0020 | 0.9400 |
| Linear | 1.6991 ± 0.4068 | 1.4470 ± 0.2391 | 1.5730 | 0.9202 ± 0.0352 | 0.9416 ± 0.0182 | 0.9309 |
| DLinear | 1.9151 ± 0.5818 | 1.3819 ± 0.3630 | 1.6485 | 0.8959 ± 0.0606 | 0.9450 ± 0.0317 | 0.9204 |
| Transformer | 3.1986 ± 1.1016 | 2.9059 ± 1.0236 | 3.0522 | 0.7040 ± 0.1764 | 0.7465 ± 0.1660 | 0.7253 |
| Persistence | 4.1377 ± 0.0000 | 4.0799 ± 0.0000 | 4.1088 | 0.5476 ± 0.0000 | 0.5454 ± 0.0000 | 0.5465 |
| iTransformer | 11.9115 ± 1.3486 | 11.2142 ± 2.5261 | 11.5628 | -2.7876 ± 0.8484 | -2.5737 ± 1.5589 | -2.6807 |
| Autoformer | 30.2303 ± 6.9972 | 45.4855 ± 13.7168 | 37.8579 | -24.1836 ± 11.6013 | -59.6110 ± 33.8273 | -41.8973 |
| Informer | 43.9853 ± 12.9064 | 49.5995 ± 15.2669 | 46.7924 | -53.6450 ± 31.2373 | -71.2750 ± 41.6181 | -62.4600 |

【Table 4】

表 4 显示，Ours 在两折中均取得最低的总体 RMSE，分别为 0.9671 ± 0.0403 ppm 和 0.8318 ± 0.0164 ppm；对应的总体 $R^2$ 分别为 0.9753 ± 0.0021 和 0.9811 ± 0.0007。两个折次等权平均后，Ours 的 RMSE 为 0.8994 ppm，$R^2$ 为 0.9782，均优于 NLinear 的 1.0088 ppm 和 0.9727。PatchTST、Linear 和 DLinear 的平均 RMSE 为 1.4441–1.6485 ppm，$R^2$ 为 0.9204–0.9400；Transformer 和 Persistence 的误差进一步升高，RMSE 分别为 3.0522 ppm 和 4.1088 ppm。iTransformer、Autoformer 和 Informer 的 $R^2$ 分别降至 -2.6807、-41.8973 和 -62.4600，表明这些模型在反应器更换后难以进行 $NH_3$ 浓度预测。该结果说明 Ours 在本研究的两折跨反应器测试中保持了较高的一致性。物理项消融的随机种子变化及两折迁移结果进一步汇总于图 3。

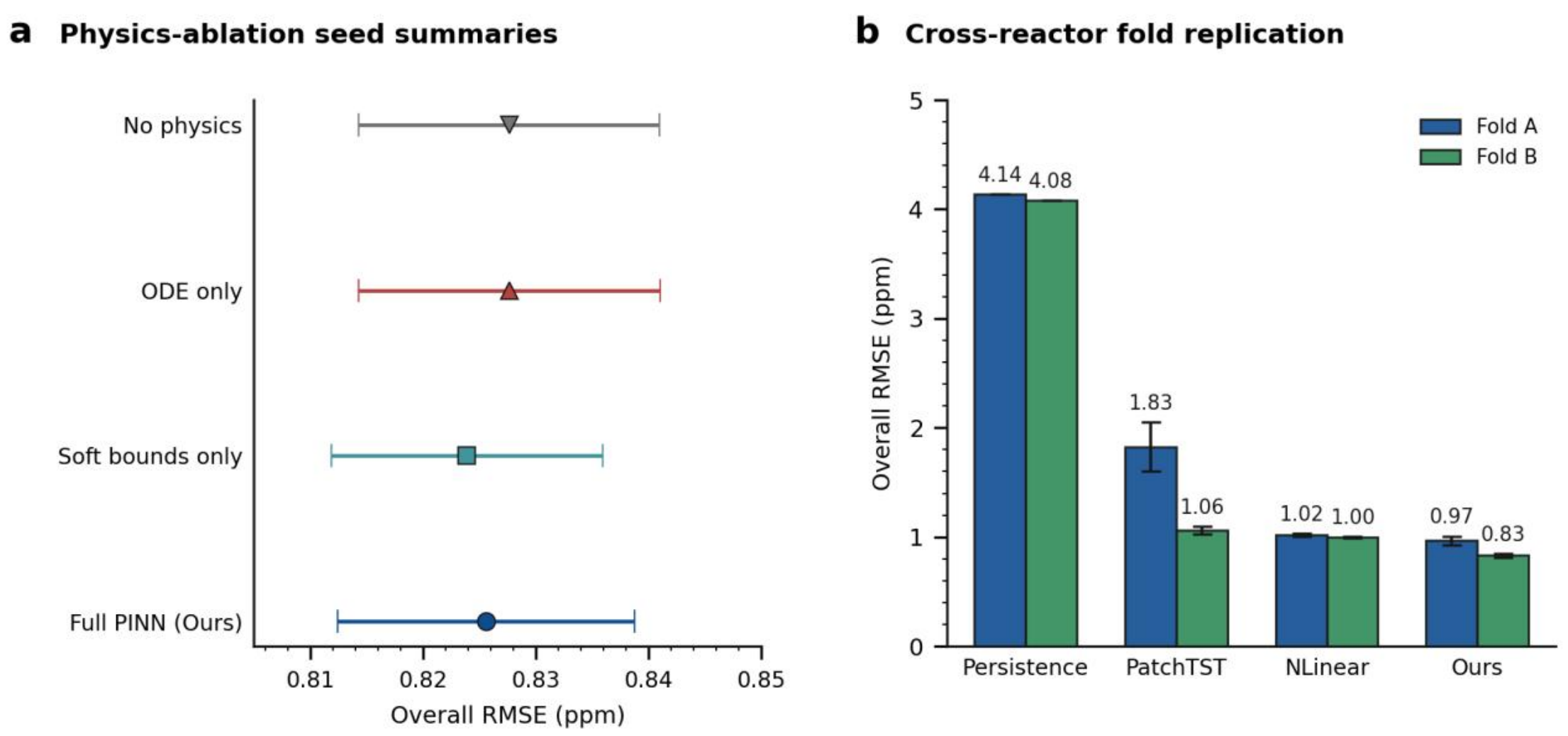

【Figure 3】

图 3a 显示四种消融物理配置的 RMSE 区间高度重叠，与表 3 中极小精度增益的结果一致，误差线表示五个计算种子的样本标准差。图 3b 显示 Ours 在 Fold A 和 Fold B 中的 RMSE 均低于图示基线，但 Fold 间差异仍然存在。总体而言，当前证据支持该模型在给定反应器和曝气时刻下的预测模型优势。

**3.2. Uncertainty quantification** (不确定性量化)

为客观评估不同滚动预测策略在瞬态 $NH_3$ 浓度外推中的鲁棒性，本研究构建了多阶段时序测试协议。鉴于曝气期具有显著的 $NH_3$ 浓度突增特征，单一的全周期全局指标容易掩盖模型在曝气期的鲁棒性。为此，数据集被分为三个指标区间：全局 RMSE 与 $R^2$、浓度极值前后 ± 10 min 的峰期及回落到正常期的恢复期。各策略在不同阶段的对比评估结果如表 5 所示。

**Table 5**

表 5 各滚动预测策略在曝气期的峰值期与恢复期的预测性能及相对技能增效对比（均值 ± 标准差）。加粗表示最优结果，下划线表示次优结果。

| Rolling Strategy | Overall RMSE | Overall R² | Skill vs Persistence | Peak Phase RMSE | Recovery Phase RMSE |
|---|---|---|---|---|---|
| **Frozen PINN (Ours)** | **1.8547 ± 0.0660** | **0.9771 ± 0.0016** | **86.37 ± 0.49 %** | **2.2450 ± 0.1754** | **0.7204 ± 0.0566** |
| Re-anchored PINN | <u>1.8622 ± 0.0643</u> | <u>0.9769 ± 0.0016</u> | <u>86.31 ± 0.47 %</u> | <u>2.2584 ± 0.1691</u> | <u>0.7228 ± 0.0486</u> |
| Persistence | 13.6039 ± 0.0000 | -0.2328 ± 0.0000 | 0.00 ± 0.00 % | 22.3150 ± 0.0000 | 1.0897 ± 0.0000 |
| ODE Rollout | 50.3355 ± 0.2969 | -15.8782 ± 0.1991 | -270.01 ± 2.18 % | 9.8661 ± 0.0380 | 69.2380 ± 0.4616 |

【table5】

如表 5 汇总的多阶段滚动评估结果所示，本研究提出的 Frozen PINN 在所有评价维度上均取得了最优表现。Frozen PINN 的总体 RMSE 为 1.8547 ± 0.0660 ppm，Re-anchored PINN 为 1.8622 ± 0.0643 ppm；相应的峰期 RMSE 为 2.2450 ± 0.1754 与 2.2584 ± 0.1691 ppm，恢复期 RMSE 为 0.7204 ± 0.0566 与 0.7228 ± 0.0486 ppm。这表明在当前测量噪声与物理约束下，重锚定机制引入的残差扰动超过了其校准收益。相较于 Persistence 基准实

现了高 86.37 ± 0.49 %的预测技能得分。相比之下，纯机理驱动的 ODE Rollout 出现灾难性的开环积分发散，尤其在恢复期漂移至 69.2380 ppm，证实了脱离残差补偿的纯微分方程无法维持稳定；而 Persistence 基线在曝气峰期崩溃，其 RMSE = 22.3150 ppm，凸显了物理归纳偏置对捕捉快速浓度跃迁的决定性价值。

为判断预测区间能否覆盖真实值，同时避免区间过宽，因为区间越宽通常越容易获得较高覆盖率，因此本研究同时报告 90%覆盖率($PICP_{90}$)、各反应罐的区间宽度（MPIW）与综合考虑 MPIW 和漏覆盖惩罚的 $IS_{0.1}$[37,39]。图 5 从三个层面展示了六个反应罐的结果：曝气事件中的预测轨迹、不同反应罐和预测步长的 $PICP_{90}$、MPIW 与 $IS_{0.1}$。

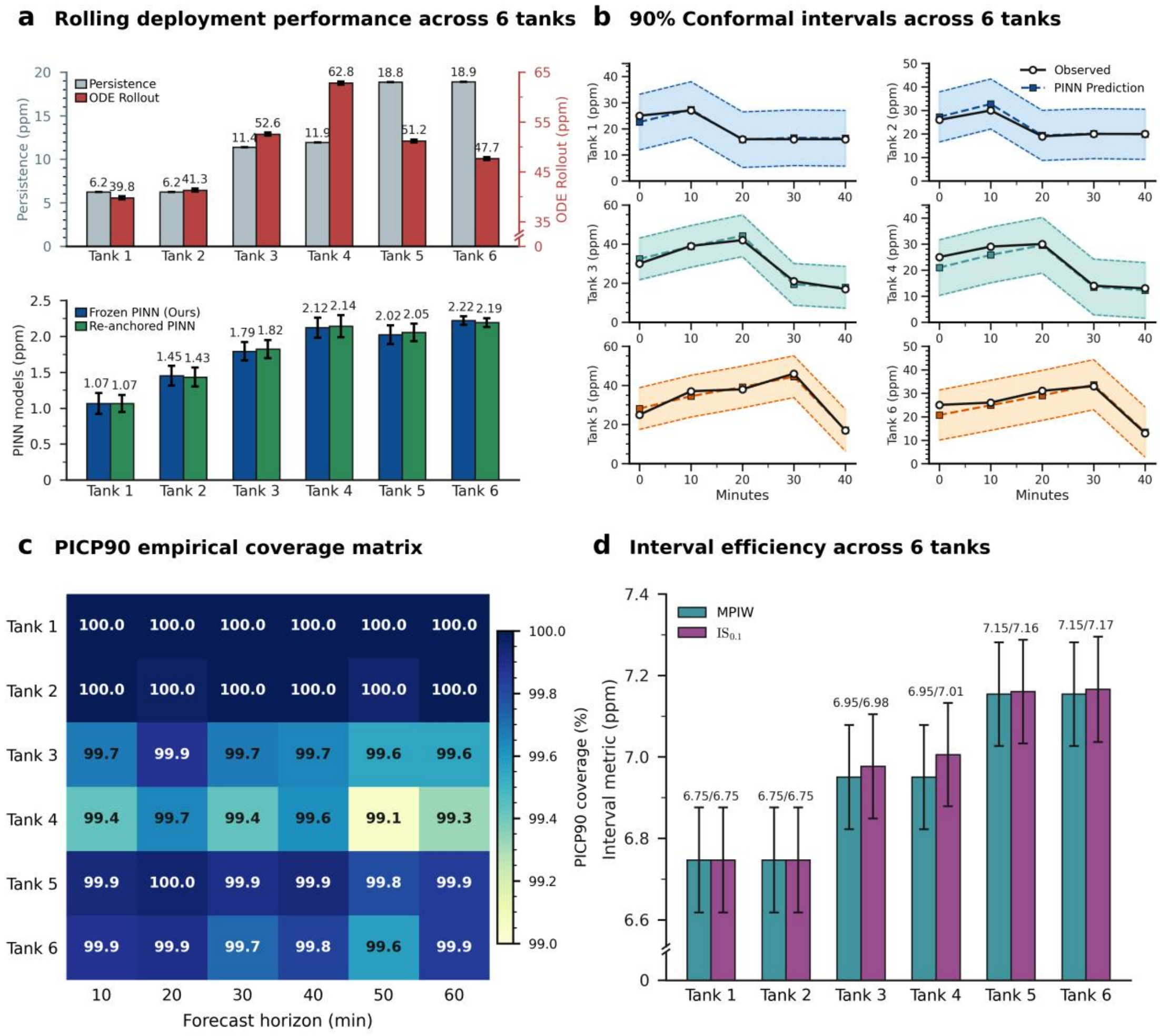

【figure5】

在校准集内，本研究将连续 60 min 的数据划分为互不重叠的时间块，以每个时间块中的最大绝对预测误差为依据，再分别针对不同预测步长和正常/曝气工况确定预测区间，遵循 BASCP 的校准逻辑[22]。五个计算种子的全反应器与全预测步长平均结果为：$PICP_{90}$ 为 99.792 ± 0.047%，MPIW 为 6.951 ± 0.128 ppm，$IS_{0.1}$ 为 6.968 ± 0.128 ppm，每个种子均包含 28,872 个测试点，平均宽度约为 7 ppm。为避免总体平均值掩盖个别反应罐或预测步长的低覆盖问题，图 5c 进一步列出了“反应罐×预测步长”的 PICP，图 5d 比较了各反应罐的 MPIW 和综合评分。

在曝气期发生时，模型给出的预测范围更宽，更不易漏实测值。曝气期的平均 $PICP_{90}$ 为 99.961 ± 0.086%，MPIW 为 19.349 ± 0.630 ppm；正常期则分别为 99.775 ± 0.046% 和 5.728 ± 0.147 ppm。即曝气期的预测区间为正常期的 3.38 倍。较宽的区间使几乎所有曝气期 $NH_3$ 数值都落在预测范围内，体现该方法能够应对曝气扰动。过宽的区间会使模型难

以准确判断 $NH_3$ 浓度可能达到的水平，进而影响系统决策的精细程度。因此，本研究同时采用 MPIW 和综合区间评分 $IS_{0.1}$ 评价兼顾 PICP 与 MPIW 指标，而不预设缺乏数据依据的 3 ppm 宽度标准。

按不同预测步长和不同反应器分别检查后，未发现总体的高覆盖率主要来自某一个步长或某一个反应器。按六个预测步长区分的平均对应范围分别为 $PICP_{90}$ 为 99.655% – 99.892%，MPIW 为 6.711 – 7.339 ppm；按六个反应器区分的平均对应范围分别为 99.418% – 100.000% 和 6.747 – 7.155 ppm。这说明在不同预测提前时间和不同反应器之间一致为高覆盖率，而不是由少数局部结果支撑[22]。总体而言，在曝气期强扰动下仍覆盖了大多数实测值，但曝气期区间会明显变宽，但是模型对 $NH_3$ 浓度范围的判断也足够精细。

3.3. **Symbolic extraction of kinetics** (动力学的符号化提取)

按照 SINDy 通过候选函数库和稀疏回归提取低复杂度关系的方法[19]，当前冻结工件中 seed=2023 得到的代表性候选式为：

$$\begin{aligned}\hat{\tilde{C}}_{t+1}^{(2023)} &= 0.016015\tilde{C}_{t-3} + 0.024875\tilde{C}_{t-11} + 0.053337\tilde{C}_{t-17} + 0.030142\tilde{C}_{t-35} \\ &+ 0.021957\tilde{C}_{t-71} - 0.039474\tilde{C}_{t-215} + 0.192784\tilde{M}_{t,0:5} + 0.549972\tilde{M}_{t,0:17} \\ &+ 0.331500\tilde{M}_{t,0:71} - 0.195308\tilde{M}_{t,0:143}\end{aligned}$$

其中，$\hat{\tilde{C}}_{t+1}$是下一个步长的标准化 $NH_3$ 浓度预测值；$\tilde{C}_{t-k}$ 表示向前 k 个采样点的标准化 $NH_3$ 浓度，分别对应前 3、11、17、35、71 和 215 个时间步的历史信息。$\tilde{M}_{t,0:k}$ 表示从当前时刻向前 k 个步长取平均浓度，并经标准化处理。波浪号表示变量已按均值和标准差缩放。该 seed 的稀疏结果未保留温度、湿度、$CO_2$ 与常数项。输出还原为 ppm 值，并执行非负约束。这些指标表示符号式对教师模型输出的复现程度，不是相对于实测 $NH_3$ 的独立预测误差。由于不同计算种子得到的保留项并不完全相同，该公式应视为当前数据范围内的代表性候选关系，而不是普适动力学定律。PINN-PatchTST→PySINDy 的 RMSE 为 0.4624 ± 0.0758 ppm，$R^2$ 为 0.9896 ±0.0035。 表 6 汇总了五个计算种子的平均结果及其波动范围，用于比较不同事件响应方法在曝气期的表现。结果见表 6：

**Table 6**

表 6 曝气事件专家分支消融与神经符号蒸馏保真度（5 个随机种子均值 ± 标准差）。加粗表示最优结果，下划线表示次优结果。

| **Model** | **Event RMSE** | **$R^2$ Δ,event** | **$E_{peak}$** | **$E_{time}$** |
|---|---|---|---|---|
| No Specialist | 11.5943 ± 0.2294 | -1.5534 ± 0.0000 | **13.2902 ± 0.1843** | 10.44 ± 1.19 |
| GPR→PySR | 2.4084 ± 0.0864 | 0.9523 ± 0.0044 | **1.5554 ± 0.2592** | 2.11 ± 1.17 |
| Lookup | 2.2735 ± 0.0351 | <u>0.9607 ± 0.0009</u> | <u>1.6468 ± 0.0499</u> | 0.94 ± 0.15 |
| GAMM | <u>2.2559 ± 0.0890</u> | **0.9608 ± 0.0011** | 1.7053 ± 0.1126 | <u>0.92 ± 0.16</u> |
| PySR | 3.1794 ± 1.1887 | 0.9357 ± 0.0194 | 2.3002 ± 0.8536 | 1.08 ± 0.21 |
| GPR | **2.1716 ± 0.0813** | 0.9600 ± 0.0037 | 1.6924 ± 0.0837 | **0.81 ± 0.15** |

【Table 6】

未加入事件专家的模型，其曝气事件 RMSE 为 11.5943 ± 0.2294 ppm，$R^2$ Δ,event 为 -1.5534，$E_{peak}$ 为 13.2902 ± 0.1843 ppm，$E_{time}$ 为 10.44 ± 1.19 min。加入 GPR 事件专家后，事件 RMSE 降至 2.1716 ± 0.0813 ppm，$E_{time}$ 降至 0.81 ± 0.15 min，这说明专门处理曝气突变的机器学习模型能够更好地跟踪浓度峰值及其出现时间。

为检验 GPR 事件响应能否被压缩为较短且可检查的表达式，PySR 按照符号回归中权衡“预测误差与表达式复杂度”的原则进行拟合[31]。seed =2023 得到的公式为：

$$f_{\mathrm{PySR}}^{(2023)}(\mathbf{x}_t) = g_{2,t} \cdot R_{\mathrm{emp},t}\left(\frac{h_t}{6} - 0.53659093\right)\left(R_{\mathrm{emp},t} - 1.7477332\right)$$

其中 0.53659093 和 1.7477332 是由训练数据拟合得到的无量纲常数，不具有独立的生物学含义。该公式输出表示事件对数响应的候选修正，评估时再将其转换为乘性事件响应。该表达式仅用于近似 GPR 的事件响应，不替代 GPR 事件专家。

在冻结测试中，PySR 近似 GPR 响应后的事件 RMSE 为 2.4084 ± 0.0864 ppm，$R^2_{\Delta,event}$ 为 0.9523 ± 0.0044，$E_{peak}$ 为 1.5554 ± 0.2592 ppm，$E_{time}$ 为 2.11 ± 1.17 min。虽然五个计算种子得到的公式并不完全一致，但是这些公式均符合作为 GPR 事件响应的低复杂度、可检查范式。为比较不同事件专家对曝气后 $NH_3$ 浓度快速变化的捕捉能力，在保持正常期预测分支、数据切分和评价方法不变的条件下，本研究比较了六种曝气期配置，如 Figure 4 所示：

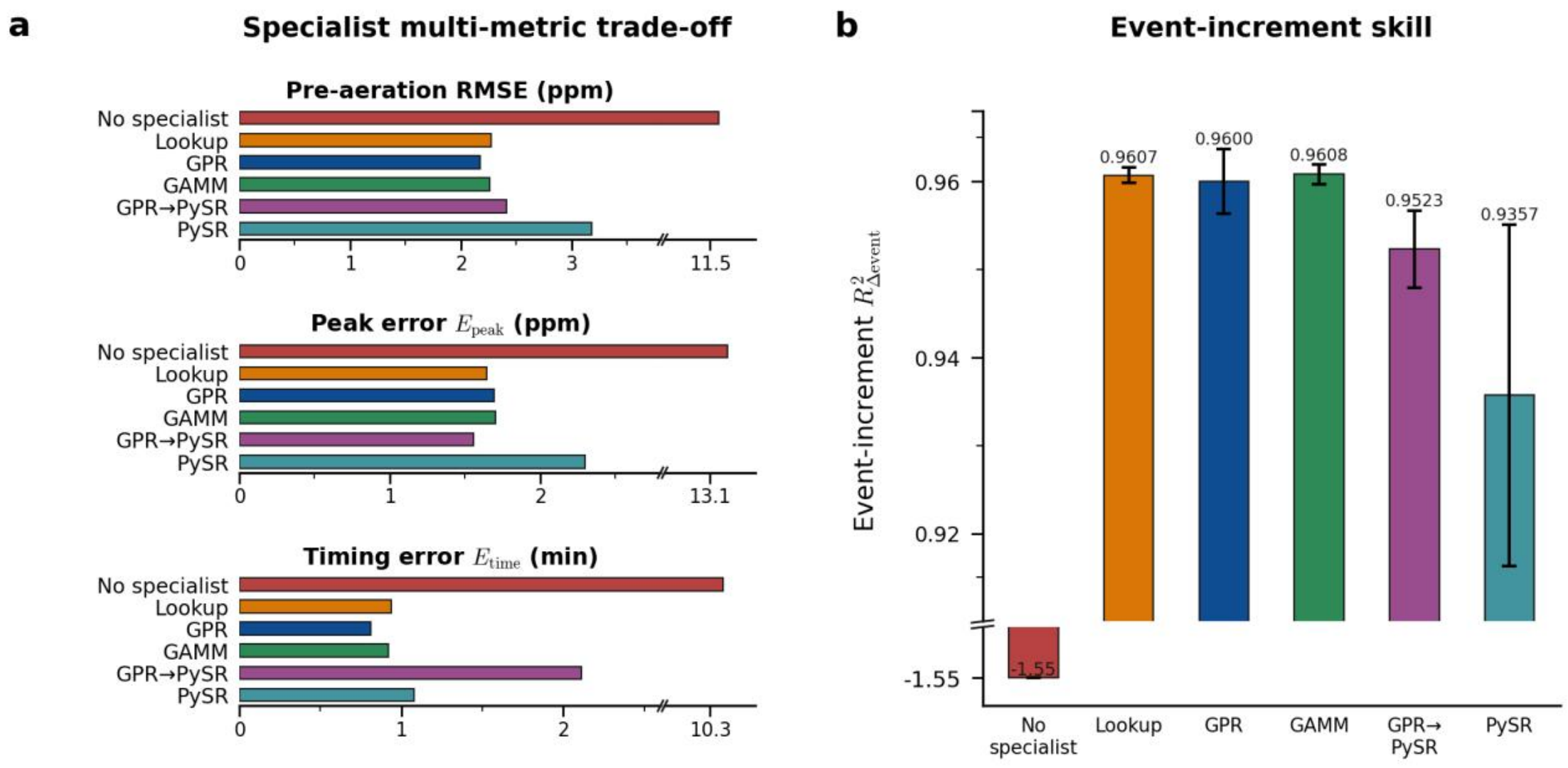

【Figure 4】

从 Figure 4a 可以看出，未使用事件专家时，曝气前 RMSE、$E_{peak}$ 和 $E_{time}$ 分别达到 11.5943 ± 0.2294 ppm、13.2902 ± 0.1843 ppm 和 10.44 ± 1.19 min，明显高于事件专家。加入事件专家后，这三项误差分别降至约 2.17 – 3.18 ppm、1.56 – 2.30 ppm 和 0.81 – 2.11 min。GPR 的曝气前 RMSE 和 $E_{time}$ 最低，分别为 2.1716 ± 0.0813 ppm 和 0.81 ± 0.15 min；GPR→PySR 的 $E_{peak}$ 最低，为 1.5554 ± 0.2592 ppm，但其 $E_{time}$ 增加至 2.11 ± 1.17 min。Figure 4b 显示 No specialist 的 $R^2_{\Delta,event}$ 为 -1.55，而 Lookup、GPR 和 GAMM 分别达到 0.9607、0.9600 和 0.9608；GPR→PySR 和 PySR 分别为 0.9523 和 0.9357。由此可见，事件专家能够显著改善曝气期精度，GPR 为曝气期最优事件专家。

双符号分支均由显式有限算子组成。在采用滚动窗口、离线合并标准化参数并将事件步长预先归一化的实现中，每次预测一步 PySINDy 公式需 14 次乘法、12 次加法和 6 次减法；PySR 公式需 5 次乘法、2 次减法和 1 次指数运算。该实现仅需保存最多 216 个历史 $NH_3$ 步长，预计占用约 0.8 – 1.8 KB 的 SRAM，具体取决于采用定点数还是 float32。因此推荐 8 位 MCU 运行定点版本，32 位 MCU 运行软件浮点版本，带 FPU 的 MCU 直接运行浮点计算。据此推断其计算与存储开销显著低于 Ours，且不依赖深度学习推理框架或 GPU，因而具明确的边缘部署可行性。

### **3.4. Aeration strategy optimization** (曝气策略优化)

尽管各罐的峰值高度和降温速度存在差异，六条温度曲线均表现出“升温期—高温期—降温期”的共同过程。六个反应器在 28 d 发酵过程中均呈现出升温期、高温期与降温期过程，同时伴随含水率逐步下降、$NH_3$、$CO_2$浓度以及 pH、GI、TC、TN、氨态氮、硝态氮的阶段性变化，如图 6 所示：

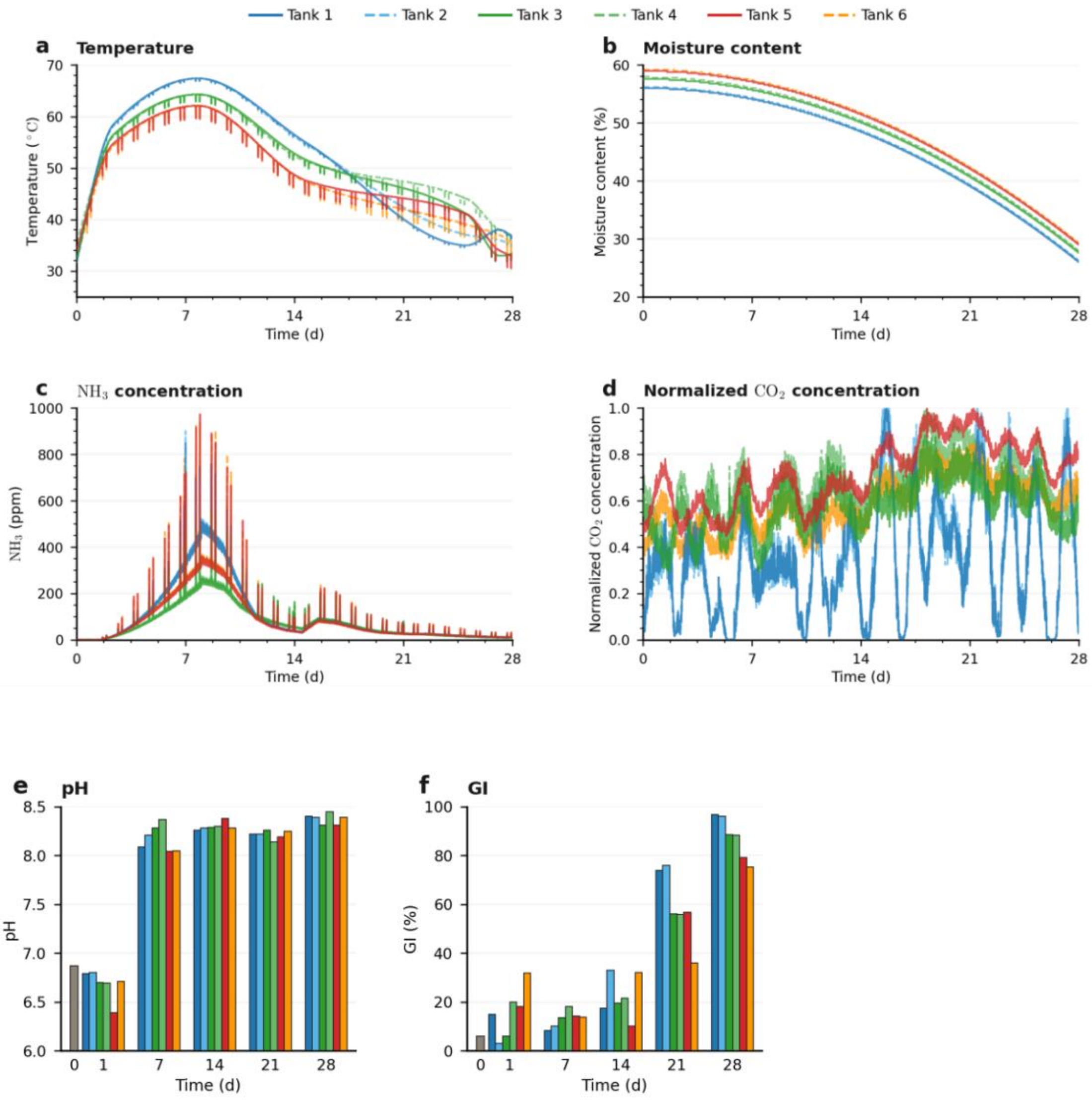

【Figure 6】

温度在 0~3 天内快速升高，并在第 8 天达到峰值。10 min 曝气组的峰值为 67.4 ℃，20 min 组为 64.3 ℃，30 min 组为 62.1 ℃；随后各罐温度逐步缓慢下降，在第 28 天降至 33-37 ℃。曝气时温度下降 0.5-2.7℃，曝气结束后温度缓慢回升至曝气前数值。含水率则从初始约 56-59%持续下降至第 28 天的约 26-29%。$NH_3$浓度在试验前 2 天内为零，第 8 天出现峰值，其中 30 min 组的第一阶段峰值最高，为 938 – 973 ppm；10 min 组为 849 – 901ppm；20 min 组最低，为 563 – 643 ppm。第 14 天附近仍可观察到幅度较小的次级波动，此后逐渐回落。20 min 组的排放响应最低，而 10 min 组最高。归一化 $CO_2$浓度则呈现周期性波动，并在发酵后半段整体维持较高水平；其中 10 min 组的低谷更明显，30 min 组在后期的基线相对较高。

pH 在第 0 天为 6.87，第 1 天短暂下降至 6.39-6.80，随后在第 7 天升至 8.04-8.37，并在第 14-28 天维持在 8.14-8.45 之间。GI 在前 14 d 内处于未腐熟水平并有波动，第 21 天后快

速升高，第 28 天达到 75.17%-96.84%。终点 GI 的组间差异表现为 10 min 最好；pH 则在三组间保持相近范围。由此，Figure 6 同时呈现了不同曝气时长对应的温湿、气相、腐熟变化和曝气事件响应。

**3.5. Implications and limitations** (意义与局限性)

Ours 的工程价值在于为曝气控制提供预测信息，但目前无法替控制器直接下达曝气动作。曝气事件分支可在预设曝气时刻之前提示 $NH_3$ 浓度可能发生的变化，作为前馈信号；名义 90% 的 BASCP 区间可帮助控制系统选择更保守或更积极的曝气动作；PySINDy/PySR 得到的简化符号关系则提供便于检查的快速近似。Table 5 显示，在连续滚动预测中，直接使用已训练模型、不再更新参数的 Frozen PINN 与每获得一批新观测后重新校正预测的 Re-anchored PINN 整体表现相近。因此在本研究的离线前馈评估中，Frozen PINN 已能提供稳定预测，无需逐步调整模型。但是该结果仅适用于当前六个反应罐和既定测试条件；当发酵底物、外界环境或曝气农艺发生变化时，Re-anchored PINN 仍可能有助于修正预测偏差。保形区间为当前测试协议下的经验预测风险范围，而不是设备的绝对安全边界，也不能保证在新批次中保持相同覆盖率[22,37]。同样 Table 6 中的符号公式只是事件专家的可解释近似，不能替代 GPR，也不是已被验证的曝气控制方程[23]。符号回归将复杂模型的响应关系压缩为简单公式，可部署在资源受限的边缘设备上，支持近实时预测；但这一部署可行性仍需通过实际嵌入式运行和闭环曝气试验验证。

本研究的鲁棒性还受到实验布置的限制。三种曝气方案各由两个反应罐作平行重复，因此组间差异不能证明氨气减排处理已达稳定。Table 3 中各模型均未出现负浓度，说明当前测试样本没有观察到这类违规；但 PVR 为零并不能单独证明模型在其他农艺下始终满足物料平衡公式。

## 4. Conclusion

本研究针对鸡粪好氧发酵过程中平稳背景与曝气后快速变化同时存在的 $NH_3$ 浓度预测问题，提出了 Ours 框架，结合长时序传感器数据、物料平衡关系约束和曝气事件专家。在六个反应罐连续 28 d 的发酵中，Ours 的总体 RMSE 为 0.8256 ± 0.0132 ppm，$R^2$ 为 0.9817 ± 0.0006；曝气期的 $R^2_{event}$ 为 0.9711 ± 0.0020，$E_{peak}$ 为 1.6924 ± 0.0837 ppm，$E_{time}$ 为 0.8 ± 0.2 min。冻结滚动预测评估中，Ours 采用 Frozen PINN 的 RMSE 为 1.8547 ± 0.0660 ppm，相比 Persistence 基线的 RMSE 降低了 86.37 ± 0.49%。

BASCP 的总体 $PICP_{90}$ = 99.792 ± 0.047%，MPIW = 6.951 ± 0.128 ppm，$IS_{0.1}$ = 6.968 ± 0.128 ppm。曝气期的 $PICP_{90}$ 为 99.961 ± 0.086%，但 MPIW 增至 19.349 ± 0.630 ppm，说明快速变化阶段需要更宽的预测范围才能维持较高覆盖率。物理项消融结果表明，Soft bounds only 的点预测略优于完整 PINN，因此 ODE 项为限制不合理输出的约束。Table 6 中的 PySINDy/PySR 结果属于事件响应的辅助近似[19,23]。在当前六反应器农艺下，Ours 可在预设曝气时刻前提供 $NH_3$ 浓度预测，为后续利用预测风险区间和符号代理进行控制决策提供基础。Figure 6 显示，在 $NH_3$ 减排方面，20 min 曝气组最优。未来的研究应在中试规模设施中进一步融合完整的碳氮比 C/N、总氮、铵态氮和硝态氮固相化学腐熟度演化曲线，以构建兼顾气态氮减排最大化与堆肥有机物腐殖化最优控制策略。

CRediT 作者贡献声明 (CRediT authorship contribution statement)

Lanbin Wei：研究方法（Methodology）、软件开发与算法实现（Software）、实验验证（Validation）、形式化分析（Formal analysis）、实验实施与调查（Investigation）、数据管护与整理（Data curation）、可视化绘图（Visualization）、初稿撰写（Writing – original draft）、论文审阅与编辑（Writing – review & editing）。

Haiyong Weng：经费获取（Funding acquisition）、资源保障（Resources）、项目管理（Project administration）、监督指导（Supervision）、论文审阅与编辑（Writing – review & editing）。

利益冲突声明 (Declaration of competing interest)
作者声明不存在任何可能影响本论文研究工作的已知竞争性经济利益或人际关系。

致谢 (Acknowledgements)

本研究得到了福建省农业信息感知技术重点实验室建设引导性项目（项目编号：KJG22052A）的资助。

数据可用性声明 (Data availability)

支持本研究结论的数据与代码可根据合理要求，由通讯作者与第一作者都同意后，向读者提供。

参考文献：

[1] FAO. (2023). The State of Food and Agriculture 2023: Revealing the true cost of food to transform agrifood systems. Rome: Food and Agriculture Organization of the United Nations. DOI: 10.4060/cc7724en.
[2] 刘振涛, 董红敏, 陶秀萍, 等. (2022). 中国畜禽粪污资源化利用政策演进与减排路径. 农业工程学报, 38(14), 215–224.
[3] Galgani, P., van der Voet, E., & Korevaar, G. (2014). Composting, anaerobic digestion and biochar production in Ghana: Environmental-economic assessment in the context of voluntary carbon markets. Waste Management, 34(12), 2454-2465. DOI: 10.1016/j.wasman.2014.07.027.
[4] 周海宾, 程琼仪, 孟海波, 等. (2022). 全国规模化畜禽养殖场固液粪污处理设施建设现状与短板分析. 农业环境科学学报, 41(6), 1201–1210.
[5] 丁祥. (2022). 高效好氧发酵生物反应器热湿与气体调控机理研究. 农业机械学报, 53(9), 310–318.
[6] Girón-Rojas, C., Gil, E., Garcia-Ruiz, A., Iglesias, N., & López, M. (2020). Assessment of biowaste composting process for industrial support tool development through macro data approach. Waste Management, 105, 364-372. DOI: 10.1016/ j.wasman.2020.02.019.
[7] 周桐, 李国学, 袁京, 等. (2021). 畜禽粪便好氧堆肥卫生学指标与有机质转化特征. 农业资源与环境学报, 38(3), 450–458.
[8] 马若男, 李国学, 袁京, 等.(2021). 初始C/N比与调理剂对鸡粪高温堆肥腐熟度及臭气排放的影响. 环境科学, 42(5), 2510–2519.
[9] Sayara, T., & Sánchez, A. (2021). Gaseous emissions from the composting process: Controlling parameters and strategies of mitigation. Processes, 9(10), 1844. DOI: 10.3390/pr9101844.
[10] Kupper, T., Häni, C., Neftel, A., Kincaid, C., Büchler, M., Amon, B., & VanderZaag, A. (2020). Ammonia and greenhouse gas emissions from slurry storage

– A review. Agriculture, Ecosystems & Environment, 300, 106963. DOI: 10.1016/ j.agee.2020.106963.
[11]Bernal, M. P., Alburquerque, J. A., & Moral, R. (2009). Composting of animal manures and chemical criteria for compost maturity assessment: A review. Bioresource Technology, 100(22), 5444 – 5453.DOI : 10.1016/j.biortech.2008.11.027
[12]Bari, Q. H., Koenig, A., & Guihe, T. (2000). Kinetic analysis of forced aeration composting – I. Reaction rates and temperature. Waste Management & Research, 18(4), 303-312. DOI:10.1177/0734242X0001800402.
[13]Nie, Y., Nguyen, N. H., Sinthong, P., & Kalagnanam, J. (2023).A time series is worth 64 words: Long-term forecasting with transformers. In International Conference on Learning Representations (ICLR 2023).
[14]Karniadakis, G. E., Kevrekidis, I. G., Lu, L., Perdikaris, P., Wang, S., & Yang, L. (2021). Physics-informed machine learning. Nature Reviews Physics, 3(6), 422 – 440.
[15]Mason, I. G. (2006). Mathematical modelling of the composting process: A review. Waste Management, 26(1), 3 – 21. DOI: 10.1016/ j.wasman.2005.01.021.
[16]Sole-Mauri, F., Illa, J., Magrí, A., Prenafeta-Boldú, F. X., &Flotats, X. (2007). An integrated biochemical and physical model for the composting process. Bioresource Technology, 98(17), 3278 – 3293. DOI: 10.1016/j.biortech.2006.07.012.
[17]Peng, L., Tang, R., Wang, G., Ma, R., Li, Y., Li, G., & Yuan, J. (2023). Effect of aeration rate, aeration pattern, and turning frequency on maturity and gaseous emissions during kitchen waste composting. Environmental Technology & Innovation, 29, 102997. DOI: 10.1016/j.eti.2022.102997.
[18]Raissi, M., Perdikaris, P., & Karniadakis, G. E. (2019). Physics-informed neural networks: A deep learning framework for solving forward and inverse problems involving nonlinear partial differential equations. Journal of Computational Physics, 378, 686 – 707.
[19]Brunton, S. L., Proctor, J. L., & Kutz, J. N. (2016). Discovering governing equations from data by sparse identification of nonlinear dynamical systems (SINDy). Proceedings of the National Academy of Sciences (PNAS), 113(15), 3932 – 3937.
[20]Cuomo, S., Di Cola, V. S., Giampaolo, F., et al. (2022). Scientific machine learning through physics-informed neural networks: Where we are and what's next. Journal of Scientific Computing, 92(3), 88.
[21]Haug, R. T. (1993). The Practical Handbook of Compost Engineering. Boca Raton: CRC Press.
[22]Angelopoulos, A. N., & Bates, S. (2023). Conformal Prediction: A Gentle Introduction. Foundations and Trends in Machine Learning, 16(4), 494 – 591. DOI: 10.1561/2200000101.
[23]Cranmer, M., Sanchez-Gonzalez, A., Battaglia, P., Xu, R., Cranmer, K., Spergel, D., & Ho, S. (2020). Discovering symbolic models from deep learning with inductive biases. In Advances in Neural Information Processing Systems, 33,

17429 - 17442.
[24]Ni, J. Q. (1999). Mechanistic models of ammonia release from liquid manure: a review. Journal of Agricultural Engineering Research, 72(1), 1 - 17.
[25]Pagans, E., Barrena, R., Font, X., & Sánchez, A. (2006). Ammonia emissions from the composting of different organic wastes. Dependency on process temperature. Chemosphere, 62(9), 1534 - 1542.
[26]Levenspiel, O. (1999). Chemical Reaction Engineering (3rd Edition). New York: John Wiley & Sons.
[27]Sander, R. (2015). Compilation of Henry's law constants (version 4.0) for water as solvent. Atmospheric Chemistry and Physics, 15(8), 4399 - 4981.
[28]Vaswani A, Shazeer N, Parmar N, et al. Attention is all you need[J]. Advances in neural information processing systems, 2017, 30.
[29]Lim, K. L.; Dutta, R.; Rotaru, M. Physics Informed Neural Network using Finite Difference Method. In: 2022 IEEE International Conference on Systems, Man, and Cybernetics (SMC). 2022, pp. 1828-1833. DOI: 10.1109/SMC53654.2022.9945171
[30]Thakur, S.; Raissi, M.; Mitra, H.; Ardekani, A. M. Temporal consistency loss for physics-informed neural networks. Physics of Fluids 36, 077136 (2024). DOI: 10.1063/5.0211398
[31]Cranmer, M. (2023). Interpretable Machine Learning for Science with PySR and SymbolicRegression.jl. arXiv:2305.01582. DOI: 10.48550/arXiv.2305.01582.
[32]Hyndman, R. J.; Koehler, A. B. Another look at measures of forecast accuracy. International Journal of Forecasting, 2006, 22(4), 679 - 688. DOI: 10.1016/j.ijforecast.2006.03.001
[33]Wheatcroft, E., 2019. Interpreting the skill score form of forecast performance metrics. International Journal of Forecasting 35(2), 573 - 579. https://doi.org/10.1016/j.ijforecast.2018.11.010.
[34]Mittermaier, M.P., 2008. The potential impact of using persistence as a reference forecast on perceived forecast skill. Weather and Forecasting 23(5), 1022 - 1031. https://doi.org/10.1175/2008WAF2007037.1.
[35]Mayer, M.J., Yang, D., 2024. Potential root mean square error skill score. Journal of Renewable and Sustainable Energy 16(1), 016501. https://doi.org/10.1063/5.0187044
[36]Vovk, V., Gammerman, A., Shafer, G., 2005. Algorithmic learning in a random world. Springer-Verlag, New York, NY. https://doi.org/10.1007/b106715.
[37]Gneiting, T., & Raftery, A. E. (2007). Strictly proper scoring rules, prediction, and estimation. Journal of the American Statistical Association, 102(477), 359 - 378. https://doi.org/10.1198/016214506000001437
[38]Kuleshov, V., Fenner, N., Ermon, S., 2018. Accurate uncertainties for deep learning using calibrated regression. In: Dy, J., Krause, A. (Eds.), Proceedings of the 35th International Conference on Machine Learning. Proceedings of Machine Learning Research, vol. 80, pp. 2796 - 2804. https://proceedings.mlr.press/v80/kuleshov18a.html.
[39]Khosravi, A., Nahavandi, S., Creighton, D., & Atiya, A. F. (2011). Comprehensive review of neural network-based prediction intervals and new advances. IEEE

Transactions on Neural Networks, 22(9), 1341 - 1356. https://doi.org/10.1109/TNN.2011.2162110